\documentclass{article}

\usepackage[preprint,nonatbib]{neurips_data_2023}

\usepackage[utf8]{inputenc} 
\usepackage[T1]{fontenc}    
\usepackage{hyperref}       
\usepackage{url}            
\usepackage{booktabs}       
\usepackage{amsfonts}       
\usepackage{nicefrac}       
\usepackage{xr}
\usepackage{microtype}      
\usepackage[dvipsnames,svgnames]{xcolor}
\usepackage{soul}         
\usepackage{amsmath,bm}
\usepackage{multirow}
\usepackage{array}
\usepackage{makecell}
\usepackage{graphicx}
\usepackage[small]{caption}
\usepackage{diagbox}
\usepackage{tabularx}
\usepackage{adjustbox}
\usepackage{bbding}
\usepackage{transparent}
\usepackage{subcaption}
\usepackage{amsfonts}

\usepackage{ulem}

\usepackage[noabbrev,nameinlink]{cleveref}
\crefname{property}{property}{Property}
\creflabelformat{property}{(#1)#2#3}
\creflabelformat{equation}{#1#2#3}

\newcolumntype{x}[1]{>{\centering\arraybackslash\hspace{0pt}}p{#1}}

\makeatletter
\newcommand{\thickhline}{%
    \noalign {\ifnum 0=`}\fi \hrule height 1.3pt
    \futurelet \reserved@a \@xhline
}

\title{Micro-seismic Elastic Reflection Full Waveform Inversion with An Equivalent Source}

\author{%
  Hanchen Wang$^{1}$\\
  \And 
  Qiang Guo$^{2}$ \\
  \And
  Tariq Alkhalifah$^{3}$\\
   \And
\textnormal{$^1$Los Alamos National Laboratory, US \ ~~~
$^2$SLB, UK \
} \\
$^3$King Abdullah University of Science and Technology, KSA \\
\texttt{hanchen.wang@lanl.gov}\\
\texttt{qiang.guo@kaust.edu.sa}\\
\texttt{tariq.alkhalifah@kaust.edu.sa}
}

\begin{document}
\bibliographystyle{unsrt}  

\maketitle
\begin{abstract}
In micro-seismic event measurements, pinpointing the passive source's exact spatial and temporal location is paramount. This research advocates for the combined use of both P- and S-wave data, captured by geophone monitoring systems, to improve source inversion accuracy. Drawing inspiration from the secondary source concept in Elastic Reflection Full Waveform Inversion (ERFWI), we introduce an equivalent source term. This term combines source functions and source images. Our optimization strategy iteratively refines the spatial locations of the source, its temporal functions, and associated velocities using a full waveform inversion framework. Under the premise of an isotropic medium with consistent density, the source is defined by two spatial and three temporal components. This offers a nuanced source representation in contrast to the conventional seismic moment tensor. To address gradient computation, we employ the adjoint-state method. However, we encountered pronounced non-linearity in waveform inversion of micro-seismic events, primarily due to the unknown source origin time, resulting in cycle skipping challenges. To counteract this, we devised an objective function that is decoupled from the source origin time. This function is formulated by convolving reference traces with both observed and predicted data. Through the concurrent inversion of the source image, source time function, and velocity model, our method offers precise estimations of these parameters, as validated by a synthetic 2D example based on a modified Marmousi model. This nested inversion approach promises enhanced accuracy in determining the source image, time function, and velocity model.
\end{abstract}
\section{Introduction}
In recent years, the field of micro-seismic events has been a hot topic in the exploration geophysics community, due mainly to the fact that hydraulic fracturing is becoming a widely used process in unconventional exploration in the oil and gas reservoirs with dense rocks such as shale. Clouds of micro-seismic events indicate the growth of faults and the occurrence of fractures, which can help the engineers reactivate failed regions or discover new ones. The hydraulic stimulation will often trigger or induce micro-seismic events \cite{duncan2005there}. A high pressure condition is created in the reservoir area by hydraulic injections, eventually cracking the dense rock in order for the gas and oil to flow freely. Such cracks can be monitored by seismic sensors, either in a well or on the earth's surface, detected by passive seismic event detection techniques \cite{chen2018fast}. Therefore, petroleum engineers are looking for more accurate methods to locate these micro-seismic events with the recorded signals, as knowing the accurate locations of micro-seismic events is critical for engineering the hydraulic process \cite{warpinski2009microseismic}.
\newline
The most common method of finding micro-seismic source locations or passive earthquakes is based on traveltime picks of both acoustic and shear wave arrivals \cite{eisner2009uncertainties,waldhauser2000double}. Such techniques require the manual determination of micro-seismic event arrival times in the recorded data. With the help of a simple relationship between traveltime, velocity and distance, each event can be located by smearing traveltimes over isochrones; ideally, several isochrones of different traces will intersect at the same point, granted the accurate velocity, referred to as the source location. This manual picking process, however, can hardly be done accurately, due to the obvious reason that micro-seismic data naturally have a poor signal-to-noise ratio (S/N). \cite{bose2009automatic,kummerow2010using,song2010improved}.
\newline
Alternatively, one can use the migration based methods \cite{droujinine2005generalized,larmat2006time,larmat2010time,lu2008locating,steiner2008time,artman2010source,haldorsen2012full,nakata2016reverse,wang2017time}, among which using the time-reversal imaging (TRI) to migrate the micro-seismic signals to the source locations is a popular one. With TRI, less user input is required compared to the traveltime based methods. It also provides more accuracy than the traveltime based methods by utilizing not only the traveltime, but also the waveform information. We can back propagate the time-reversed recorded traces, guided by the wave equation, and focus the signals at the sources, in both time and space. Not having to pick arrival times is the biggest advantage of TRI methods, especially when imaging with low S/N seismic data. If a good enough acquisition system is available, provided we have an accurate model and physics representation, the correct source locations will be imaged. If, however, only a limited aperture or a sparse spatial sampled receiver system is available, TRI is obviously not the optimal method to locate the passive sources. Much research has been published on handling poor acquisition systems and sparsely sampled wavefields \cite{fink2000time,tanter2000time,tanter2001optimal,aubry2001optimal,jonsson2004retrofocusing,sava2011micro,gallot2012passive}.
All TRI methods to locate the micro-seismic source may suffer from spatially defocused images, which is caused by the errors between the true and the imaging velocity models. Migration velocity analysis (MVA) may be a solution to minimize the velocity error \cite{sun2018automatic,song2019microseismic}.
\newline
Another approach of micro-seismic imaging is based on the full waveform inversion (FWI) technique \cite{kaderli2015microseismic,behura2015expedited,sun2016full,kamei2017full,song2019passive}. This technique makes it possible to generally update the micro-seismic event locations and the velocity models by minimizing the objective function that is usually described by the differences between the true and predicted data as any other FWI. The update is terminated when the data misfit is sufficiently small. The inverted source location is considered as the final located source position. Most existing FWI methods, however, assume an acoustic case and sufficiently good initial conditions are also essential to overcome the cycle skipping problem caused by the unknown source origin time, which makes it challenging in practice.
\newline
On the other hand, recent development of machine learning methods shows great potential of efficient and accurate micro-seismic event relocation and velocity model estimation \cite{alkhalifah2022mlreal,wang2021data,deng2022openfwi,wang2021direct}. The machine learning methods take seismic observations as input images and predict corresponding seismic parameters as outputs, such as velocity models and passive seismic event locations. However, such data-driven methods require high demanding labeled or unlabelled datasets for the network training stage, and without enough high quality data, these methods will not perform at the desired level.
\newline
Besides the source location and function, the source mechanism inversion is also critical for passive seismic data. A frequency domain method is proposed to localize micro-seismic events blind to the source time function \cite{jamali2014research}. However, they confine their search to the hypo-centers of the sources and the normalized amplitudes of the moment tensors. Other methods were introduced that include simultaneous determination of the micro-seismic event location and the source mechanism given the input velocity model \cite{droujinine2011elastic,zhebel2012simultaneous,anikiev2014joint}. The key requirement for these methods is the knowledge of the correct velocity model. When the velocity is inaccurate, the source location and mechanism are affected.
\newline
The work presented here builds on the work of \cite{wang2018microseismic}, based on an acoustic FWI method, and further introduces an equivalent source term as the force function in the elastic wave equation. This term is extended from the concept of the equivalent sources in elastic reflection waveform inversion \cite{guo2017elastic}. The introduced source term is constructed by separating the source term in space and time, named as the source images and the source functions, respectively. An FWI approach measuring the L2-norm data misfit is used to update the spatial and the temporal components of the source. An FWI scheme with a source time function independent objective function is used to mitigate the cycle skipping problem caused by the unknown source origin time while updating the velocity models. This source function independent FWI approach works well with any given source function, because it is based on convolved wavefields between the observed and modeled ones to mitigate the role played by the source function \cite{choi2011source}. After all the parameters are inverted, the proposed method can accurately estimate the source with initial velocity models, which are quite different from the true ones so that the key requirement of accurate migration velocity can be relaxed. We implement the proposed method to a modified Marmousi model to show the novelty of inverting for the source and the velocity models with a nested approach.

\section{Equivalent Source In Elastic Micro-seismic Source Case}
In this section, we introduce an equivalent source term for the elastic micro-seismic case. We borrow the concept of the scatters (or perturbations) in Elastic Reflection Full Waveform Inversion (ERFWI) to represent the micro-seismic source term in our inversion method. Thus, we first review the concept of the ERFWI and then we derive the equivalent source term and its gradients.
\subsection{Scatters In Elastic Reflection Waveform Inversion}
Full waveform inversion is highly non-linear, and the non-linearity increases in the elastic cases because of the complicated wave propagation in the subsurface medium. ERFWI is one way to overcome this issue, by separating the model parameters into low-wavenumber components, which controls the kinematics, and the high-wavenumber scattering \cite{wang2013reflection,alkhalifah2015natural}.
\newline
In an isotropic medium with a constant density, it is assumed that the squared $V_p$ and $V_s$ velocities (denoted by $\alpha^2 \, \mbox{and} \, \beta^2$, respectively) can be divided into two parts,
\begin{equation}
\alpha^2 = \alpha_0^2 + \delta \alpha^2 \, \mbox{and} \, \beta^2 = \beta_0^2 + \delta \beta^2
\label{eq:m}
\end{equation}
where $\alpha_0$ and $\beta_0$ refer to the low-wavenumber (background) parts; $\delta \alpha$ and $\delta \beta$ are the high-wavenumber scattering of the P- and S- velocities, respectively. The total wavefield, denoted by $\bf u$, can thus be separated correspondingly into two parts,
\begin{equation}
\bf{u} = \bf{u}_0 + \delta \bf{u},
\label{eq:u}
\end{equation}
where $\bf u_0$ is the wavefield propagating in the smooth background medium, and $\delta \bf u$ is the scattered wavefield simulated by the scatters. The perturbations in the stress tensors ($\delta \tau$) can be derived from Hooke's law, giving us:
\begin{equation}
\begin{array}{cl}
& \displaystyle \delta \tau _{ xx }=\left( \alpha^2 _{ 0 }+\delta \alpha^2 \right) \frac { \partial \delta u_{ x } }{ \partial x } +\left( \alpha^2 _{ 0 }+\delta \alpha^2 - 2\beta^2 _{ 0 } - 2\delta \beta^2 \right) \frac { \partial \delta u_{ z } }{ \partial z } +\widetilde { \tau  } _{ xx }; \\ &  \displaystyle \delta \tau _{ zz }=\left( \alpha^2 _{ 0 }+\delta \alpha^2 - 2\beta^2 _{ 0 } - 2\delta \beta^2 \right) \frac { \partial \delta u_{ x } }{ \partial x } +\left( \alpha^2 _{ 0 }+\delta \alpha^2  \right) \frac { \partial \delta u_{ z } }{ \partial z } +\widetilde { \tau  } _{ zz }; \\ &  \displaystyle \delta \tau _{ xz }=\left( \beta^2 _{ 0 }+\delta \beta^2  \right) \left( \frac { \partial \delta u_{ x } }{ \partial z } +\frac { \partial \delta u_{ z } }{ \partial x }  \right) +\widetilde { \tau  } _{ xz } ,
\end{array}
\label{eq:delta_tau}
\end{equation}
where $\widetilde{\tau}_{ij}$ can be described as
\begin{equation}
\begin{array}{cl}
& \displaystyle \widetilde { \tau  } _{ xx }=\delta \alpha^2 \left( \frac { \partial u_{ x } }{ \partial x } +\frac { \partial u_{ z } }{ \partial z }  \right) -2\delta \beta^2 \frac { \partial u_{ z } }{ \partial z }; \\ & \displaystyle \widetilde { \tau  } _{ zz }=\delta \alpha^2 \left( \frac { \partial u_{ x } }{ \partial x } +\frac { \partial u_{ z } }{ \partial z }  \right) -2\delta \beta^2 \frac { \partial u_{ x } }{ \partial x }; \\ & \displaystyle \widetilde { \tau  } _{ xz }=\delta \beta^2 \left( \frac { \partial u_{ x } }{ \partial z } +\frac { \partial u_{ z } }{ \partial x }  \right).
\end{array}
\label{eq:tau_tilde}
\end{equation}
The stress functions of equation~\ref{eq:tau_tilde} are substituted into equation~\ref{eq:delta_tau} as forces in order to generate the scattered waves at the model perturbation locations. It is considered as a de-migration procedure. It is critical to estimate the scatter image and the incident wavefield at the scatter points as source wavelets in the de-migration procedure \cite{wu2014non,guo2017elastic}.
\newline
Displacement field in the background medium ($\alpha_0$ and $\beta_0$) are then taken into account for estimating the model perturbations ($\delta \alpha$ and $\delta \beta$). 
\subsection{Equivalent Source Term In Micro-seismic Case And Its Gradient}
We extend the concept of the scatters in ERFWI to the micro-seismic case. The passive seismic events triggered from the deep subsurface can be considered as separated single scatters in the de-migration procedure. Now we re-write the micro-seismic source term by a form of the equivalent source, which produces the same micro-seismic seismic events.
\newline
If we consider the 2D case, the elastic wave propagation is granted by the following wave equation:
\begin{equation}
\rho \left( \frac { { \partial  }^{ 2 }{ u }_{ i } }{ { \partial  }t^{ 2 } }  \right) -\frac { \partial  }{ \partial { x }_{ j } } \left( { c }_{ ijkl }\frac { \partial { u }_{ k } }{ \partial { x }_{ l } }  \right) =-M_{ij} \frac{\partial [\delta(\textbf{x}-\textbf{x}^s)]}{\partial x_j}S(t),
\label{eq:equivalent_source}
\end{equation}
where ${\bf u}$ is the elastic displacement field with two components, $c_{ijkl}$ are components of the stiffness tensor; $\rho$ is the density; the r.h.s. source term is an equivalent force term \cite{aki2002quantitative,dahlen1998theoretical}; $\textbf{x}^s$ is the source location coordinate and $S(t)$ is the source time function. $M_{ij}$ represents the components of the full seismic moment tensor. One of the limitations of this description for inversion applications of the source is that the moment tensor $\bf{M}$, the source location $x^s$, and the source function are separate variables, with accurate estimation of these variables requiring the estimation of the source location.
\newline
We now modify equation~\ref{eq:equivalent_source} using an equivalent source term to mimic the source mechanism in the micro-seismic case, which is derived from the ERFWI, and such that the r.h.s. of the elastic wave equation becomes:
\begin{equation}
{ f }_{ i }=\partial _j \tau ^0 _{ij},
\label{eq:sou_term}
\end{equation}
where each component of ${ \tau  }_{ ij } ^0$ can be written as:
\begin{equation}
\begin{array}{cl}
&  \displaystyle { \tau }_{ 11 }^0= \delta \alpha^2 \left( { w }_{ 11 }+{ w }_{ 22 } \right) -2\delta \beta^2 { w }_{ 22 } ; \\ & \displaystyle { \tau }_{ 22 }^0=\delta \alpha^2 \left( { w }_{ 11 }+{ w }_{ 22 } \right) -2\delta \beta^2 { w }_{ 11 } ; \\ & \displaystyle  { \tau }_{ 12 }^0={ \tau }_{ 21 }^0=\delta \beta^2 \left( { w }_{ 12 }+{ w }_{ 21 } \right).
 \end{array}
\label{eq:tau}
\end{equation}
Here, $\delta \alpha$ and $\delta \beta$ are denoted as source images and $\left[ w_{ij} \right]$ is the multi-components source time function. The stress tensors in equation~\ref{eq:tau} and equation~\ref{eq:tau_tilde} have the same form, but the incident wavefield at the scatter points replaced by $\left[ w_{ij} \right]$, that $\frac{\partial u_x}{\partial x}$ is replaced by $w_{11}$, $\frac{\partial u_z}{\partial z}$ is replaced by $w_{22}$, $\frac{\partial u_x}{\partial z}$ and $\frac{\partial u_z}{\partial x}$ are replaced by $w_{12}$ and $w_{21}$, respectively. In ERFWI, $\delta \alpha$ and $\delta \beta$ represent the model perturbations. It is expected that all the possible scatters in the model are described by the two parameters. The back-propagated wavefield $\bf u$ is saved at each model point as the source term at the scatter point. In the micro-seismic case, $\delta \alpha$ and $\delta \beta$ represent the actual micro-seismic sources in space, expressed in terms of perturbations in $P-$ and $S-$ waves velocities, respectively. Compared to the model perturbations in the ERFWI case, instead of having scatters all over the model, it is expected to have only one single scatter at the source location. Thus, the back-propagated wavefield is only needed at the source location as its source time function. In this case, the source will also have inherent representation of the radiation patterns besides the P- and S- wave velocity perturbations, which we will further discuss in the following sections. The source time function $\left[ w_{ij} \right]$ corresponds to the perturbation locations. Later we will estimate the coefficients $\left[ w_{ij} \right]$ by the back-propagated wavefield at the source location. Thus, the source spatial location $\bf x^s$ and the seismic moment tensor $\textbf{M}$ are embedded into these source images. This decomposition of the source assumes that the time and space dependency are decoupled, which leads to a more manageable space of unknown that is spatially equal to the model space. 
\newline
If we ignore the density variable, by considering it as constant for example, the wave equation then becomes:
\begin{equation}
 \left( \frac { { \partial  }^{ 2 }{ u }_{ i } }{ { \partial  }t^{ 2 } }  \right) -\frac { \partial  }{ \partial { x }_{ j } } \left( { a }_{ ijkl }\frac { \partial { u }_{ k } }{ \partial { x }_{ l } }  \right) =\nabla \cdot { \tau  }_{ ij } ^0.
\label{eq:eqv_sou}
\end{equation}
\newline
where $a_{ijkl}$ is the density normalized elastic coefficients. In reality the density may have some influences on the passive seismic data inversion, but density is not easy to recover. To simplify the inversion, we here choose to ignore the density in this study. Considering the data corresponding to the passive events being acquired on the Earth surface (or in a well), $\textbf d(x_r,t)$, we can devise an iterative approach by defining the optimization problem using the classic least-square minimization of
\begin{equation}
{ E }\left( m \right) =\frac { 1 }{ 2 } \sum _{ r }{ \int { \left\| \textbf u\left( { x }_{ r },t;m \right) -\textbf {d} \left( { x }_{ r },t \right)  \right\|^2 _2 dt }  },
\label{eq:obj_func}
\end{equation}
with respect to $\delta \alpha$, $\delta \beta$ and $\left[ w_{ij} \right]$, where $\textbf u(x_r,t;m)$ is the modeled waveform recorded by the receivers at $x_r$ determined by the current parameters, including $\delta \alpha$, $\delta \beta$, $ w_{ij} $ and velocities. 
The parameters $\delta \alpha$ and $\delta \beta$ can be retrieved iteratively using the following gradients:
\begin{equation}
\nabla_{ \delta \alpha  } { E }=-{ \int _{ t }{ dt\left( { w }_{ 11 }+{ w }_{ 22 } \right) \left( \frac { \partial \widehat { { u } } _{ x } }{ \partial x } +\frac { \partial \widehat { { u } } _{ z } }{ \partial z }  \right)  }  },
\label{eq:alpha_grd}
\end{equation}
\begin{equation}
\label{eq:beta_grd}
 \nabla_{ \delta \beta  } { E }= - \int _{ t } dt \, 2{ w }_{ 11 }\frac { \partial \widehat { { u }} _{ x }  }{ \partial x } +2{ w }_{ 22 }\frac { \partial \widehat { { u } }_{ z }  }{ \partial z }  + ({ w }_{ 12 }+{ w }_{ 21 } )(\frac { \partial \widehat { { u } } _{ x } } {\partial z } + \frac { \partial \widehat { { u }}_{ z }   }{ \partial x }),
\end{equation}
where $\hat{\bf u}$ refers to the adjoint wavefield. Besides the source spatial location, $\delta \alpha$ and $\delta \beta$ represent the source mechanism in a more compact form other than the seismic moment tensor.
\newline
Although the objective function we use to update the velocity models, which will be introduced in the following section, frees us from knowing the source time function prior to the inversion, we prefer to apply a simultaneous inversion for the source time functions, in order to get a more accurate estimate than the case of ignoring the time functions:
\begin{equation}
\begin{array}{cl}
& \displaystyle \nabla {E}  _ {{ w}_{ 11 }}=\int_x\int_z \, dxdz \delta \alpha^2 \left( \frac{\partial \widehat { { u } }_{ x }}{\partial x} +\frac{\partial \widehat { { u } }_{ z }}{\partial z} \right) -2\delta \beta^2 \frac{\partial \widehat { { u } }_{ z }}{\partial z} ; \\& \displaystyle \nabla {E}  _ {{ w}_{ 22 }}=\int_x\int_z dxdz \, \delta \alpha^2 \left( \frac{\partial \widehat { { u } }_{ x }}{\partial x}+\frac{\partial \widehat { { u } }_{ z }}{\partial z} \right) -2\delta \beta^2 \frac{\partial \widehat { { u }}_{ x } }{\partial x}; \\& \displaystyle  \nabla {E}  _ {{ w}_{ 12/21 }}=\int_x\int_z dxdz \, \delta \beta^2 \left( \frac{\partial \widehat { {  u }}_{ x } }{\partial z} + \frac{\partial \widehat { { u } }_{ z }}{\partial x} \right).
\end{array}
\label{eq:E_wt}
\end{equation}
\newline
More details on the gradient derivations are provided in Appendix $A$.
\newline
In multi-parameter optimization, there will always be the potential for trade-off. In this case, we need to wisely choose the step-length of each parameter or use a small enough fixed step-length to mitigate the coupling issue.
\section{Source Image Radiation Patterns Analysis}
\begin{figure}
\begin{subfigure}{0.16\textwidth}
  \centering
  \includegraphics[width=1\linewidth]{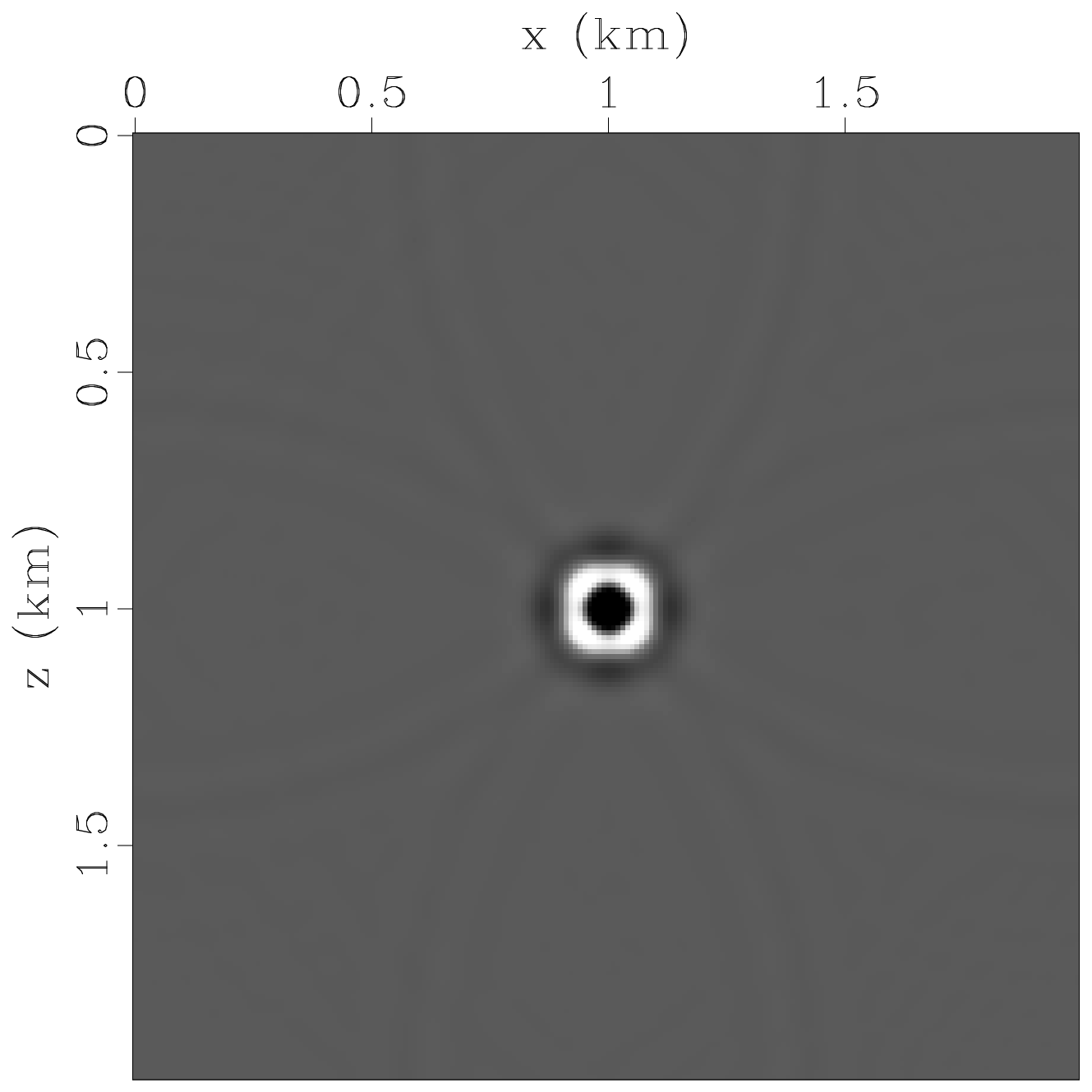}
  \caption{$\delta \alpha, \left( 1,1,0 \right) $ }
  \label{fig:a1}
\end{subfigure}
\begin{subfigure}{0.16\textwidth}
  \centering
  \includegraphics[width=1\linewidth]{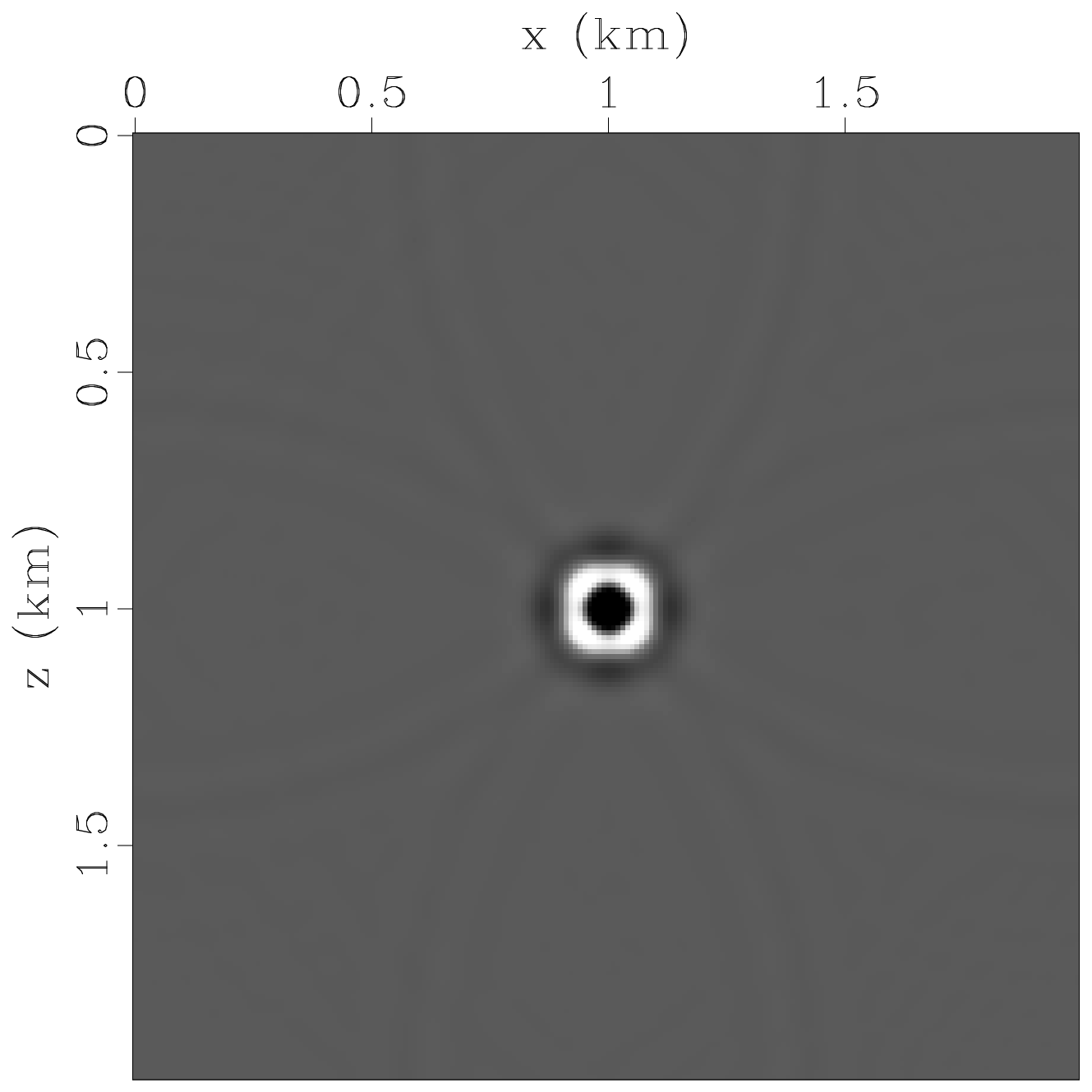}
  \caption{$\delta \beta, \left( 1,1,0 \right) $}
  \label{fig:b1}
\end{subfigure}
\begin{subfigure}{0.16\textwidth}
  \centering
  \includegraphics[width=1\linewidth]{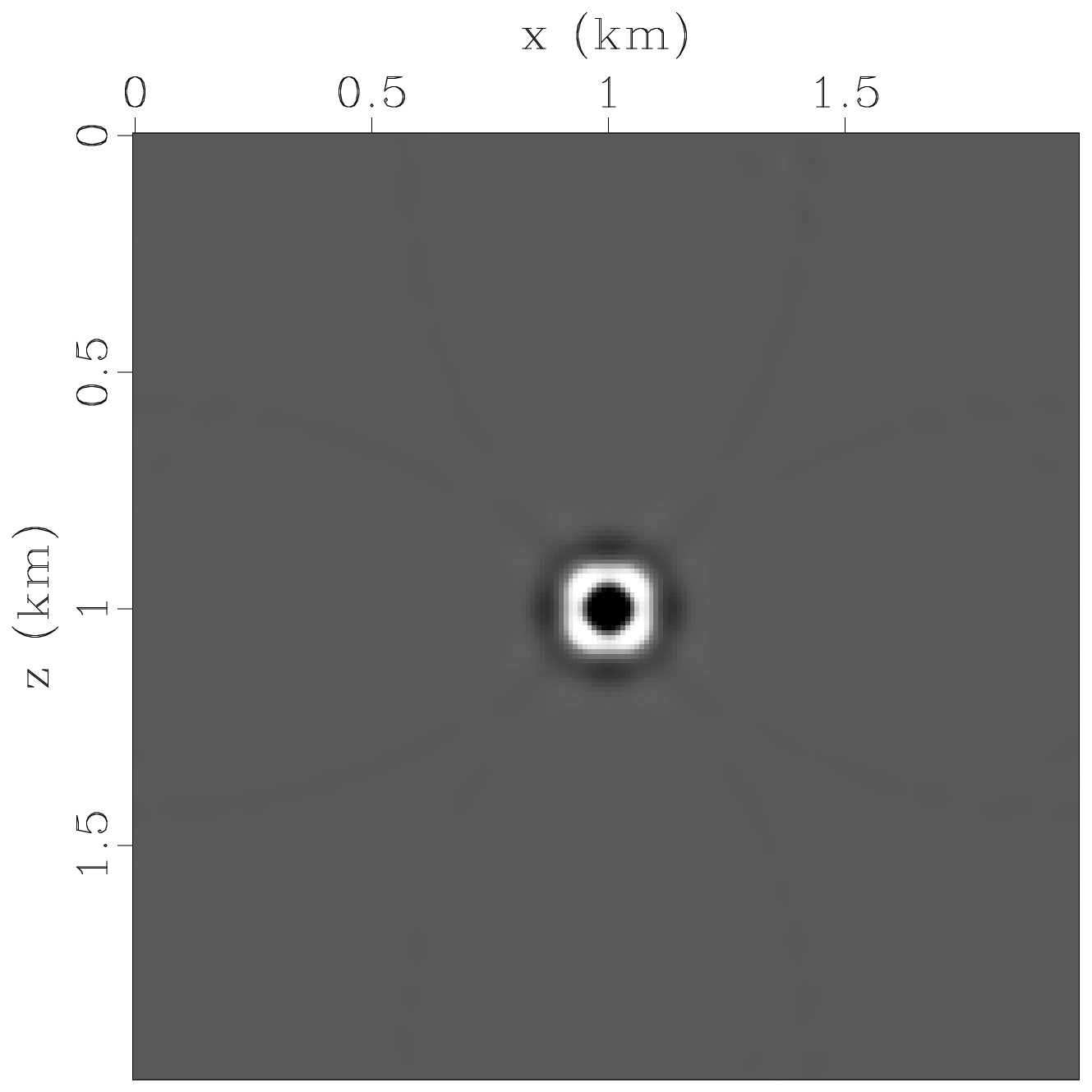}
  \caption{$\delta \alpha, \left( -1,-1,0 \right) $}
  \label{fig:a2}
\end{subfigure}
\begin{subfigure}{0.16\textwidth}
  \centering
  \includegraphics[width=1\linewidth]{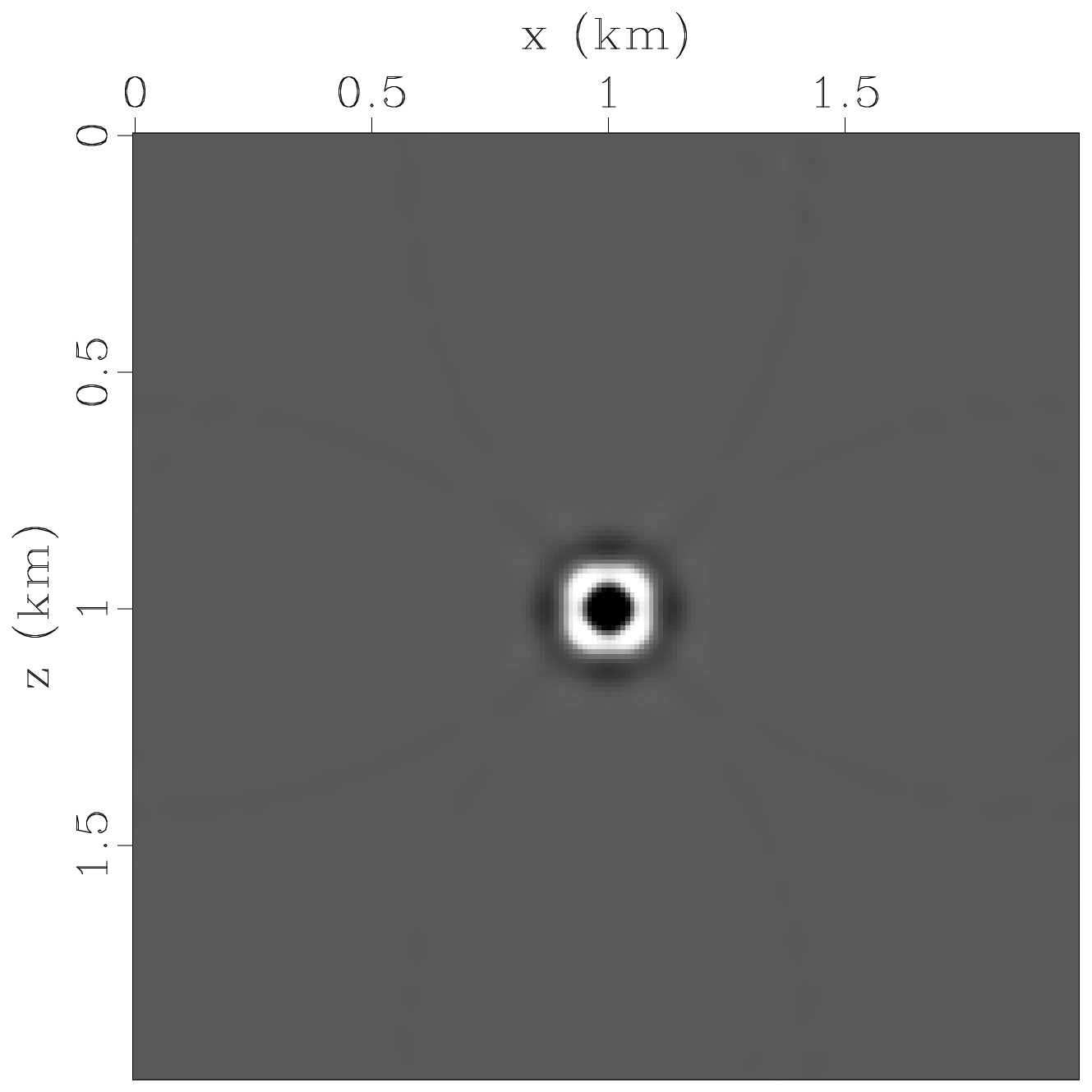}
  \caption{$\delta \beta, \left( -1,-1,0 \right) $}
  \label{fig:b2}  
\end{subfigure}
\begin{subfigure}{0.16\textwidth}
  \centering
  \includegraphics[width=1\linewidth]{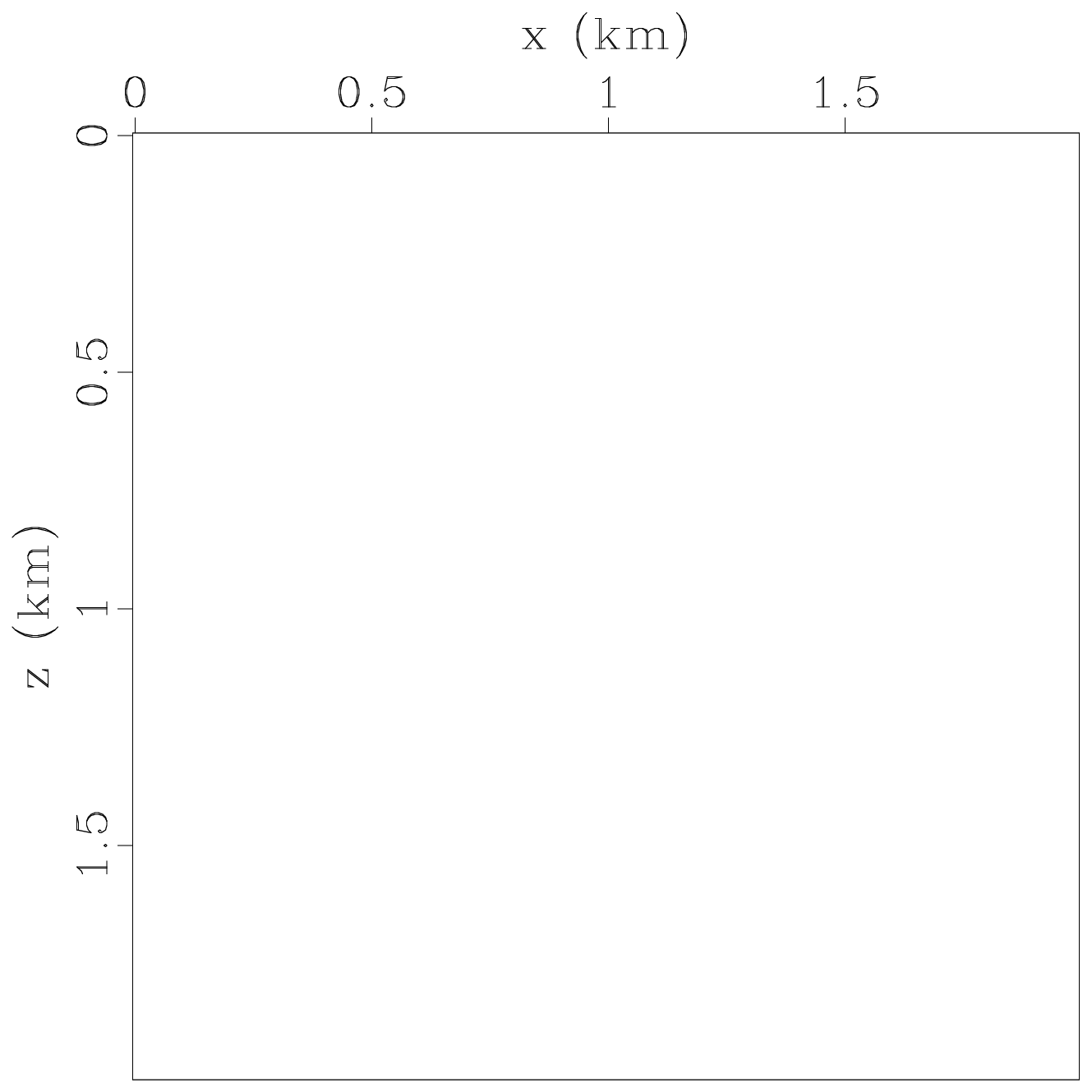}
  \caption{$\delta \alpha, \left(0,0,1\right) $}
  \label{fig:a3}
\end{subfigure}
\begin{subfigure}{0.16\textwidth}
  \centering
  \includegraphics[width=1\linewidth]{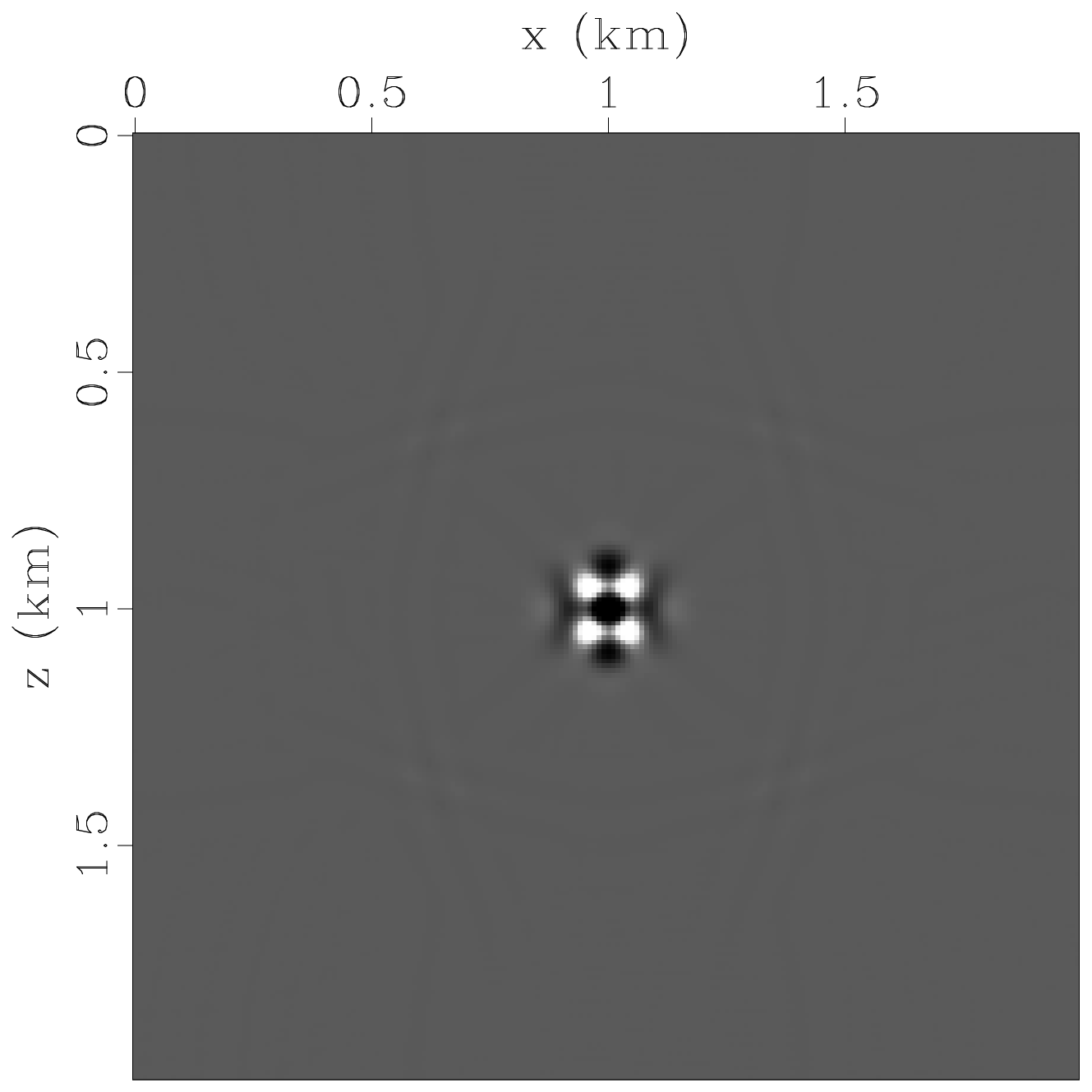}
  \caption{$\delta \beta, \left(0,0,1\right)$}
  \label{fig:b3}  
\end{subfigure}
\begin{subfigure}{0.16\textwidth}
  \centering
  \includegraphics[width=1\linewidth]{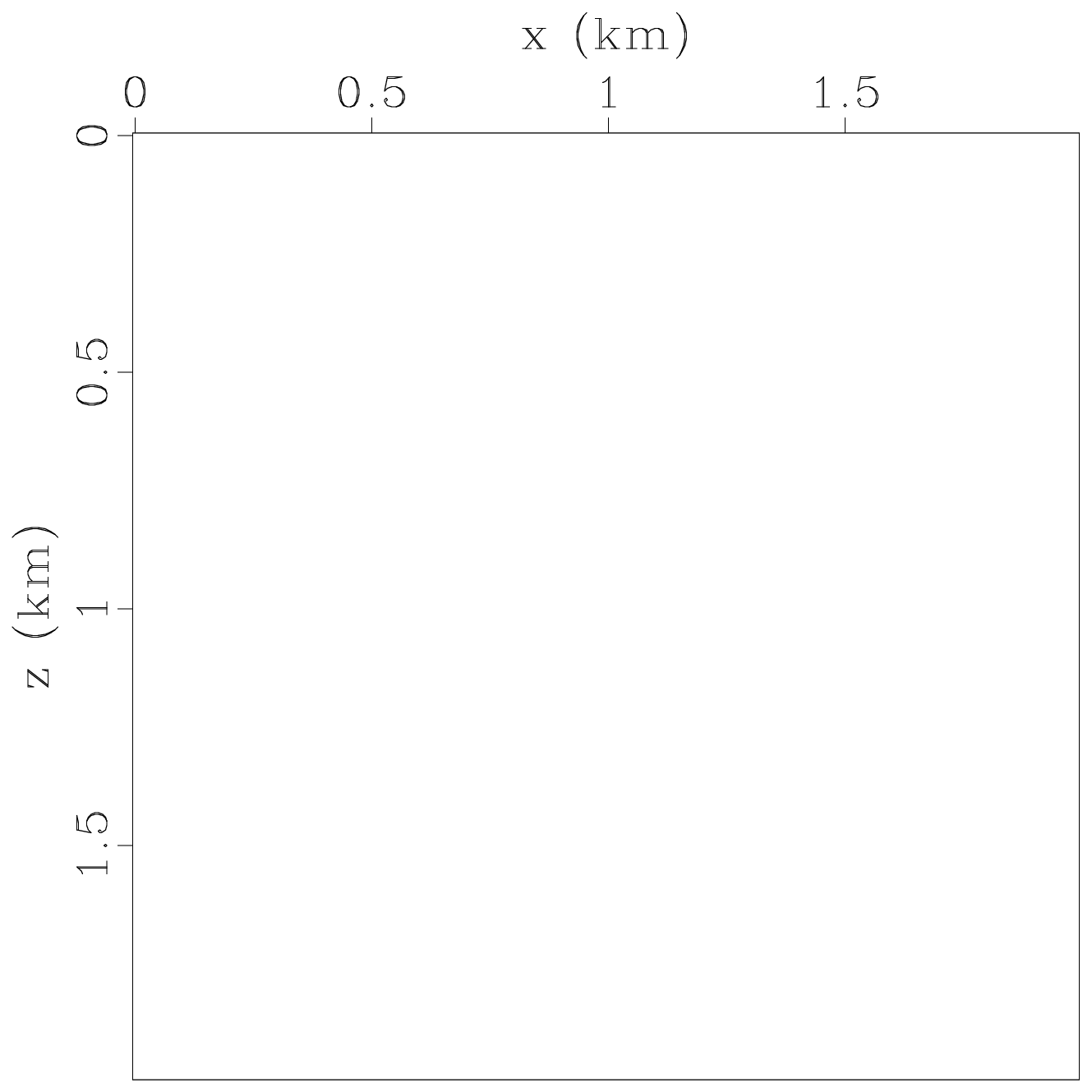}
  \caption{$\delta \alpha, \left(1,-1,0\right)$}
  \label{fig:a4}
\end{subfigure}
\begin{subfigure}{0.16\textwidth}
  \centering
  \includegraphics[width=1\linewidth]{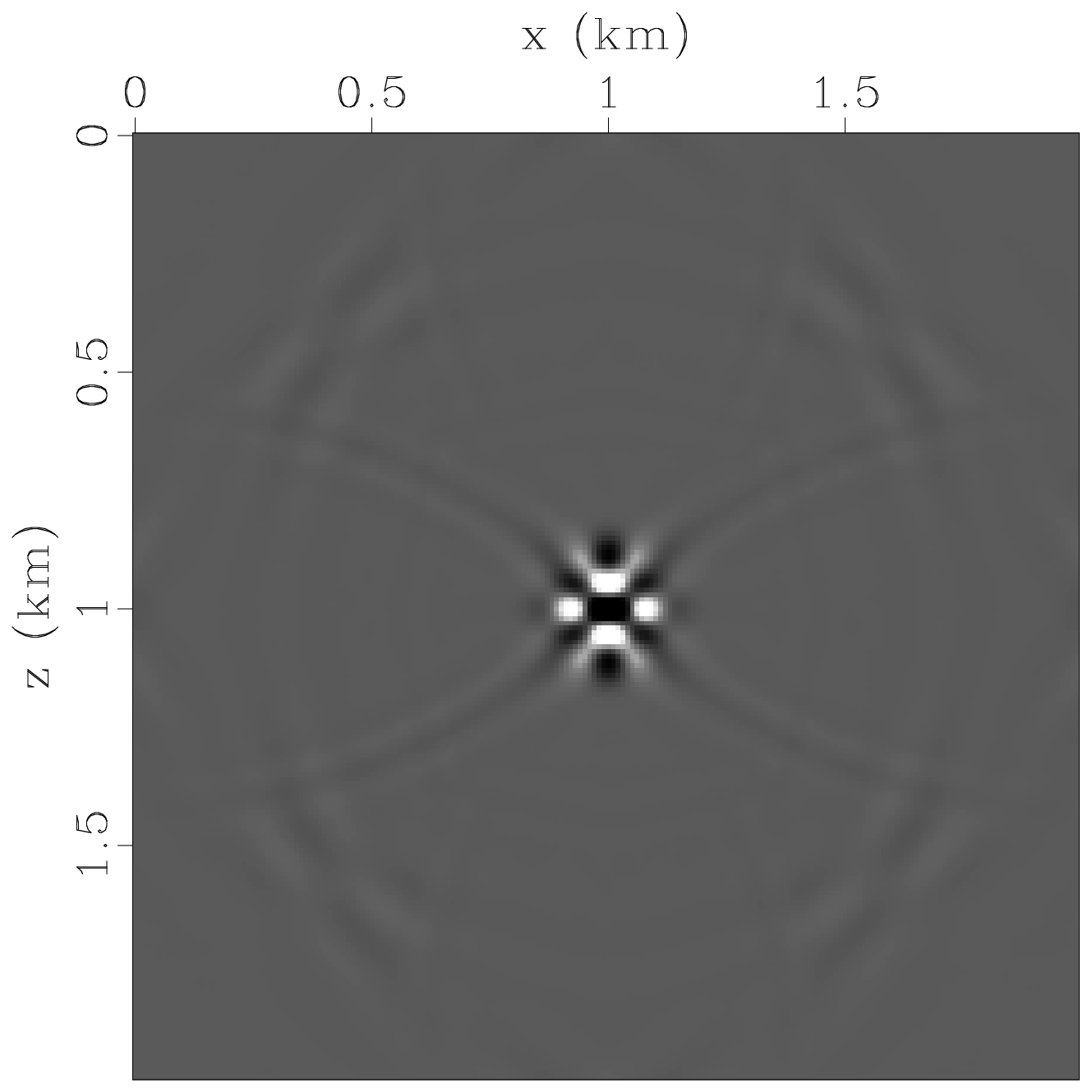}
  \caption{$\delta \beta, \left(1,-1,0\right)$}
  \label{fig:b4}  
\end{subfigure}
\begin{subfigure}{0.16\textwidth}
  \centering
  \includegraphics[width=1\linewidth]{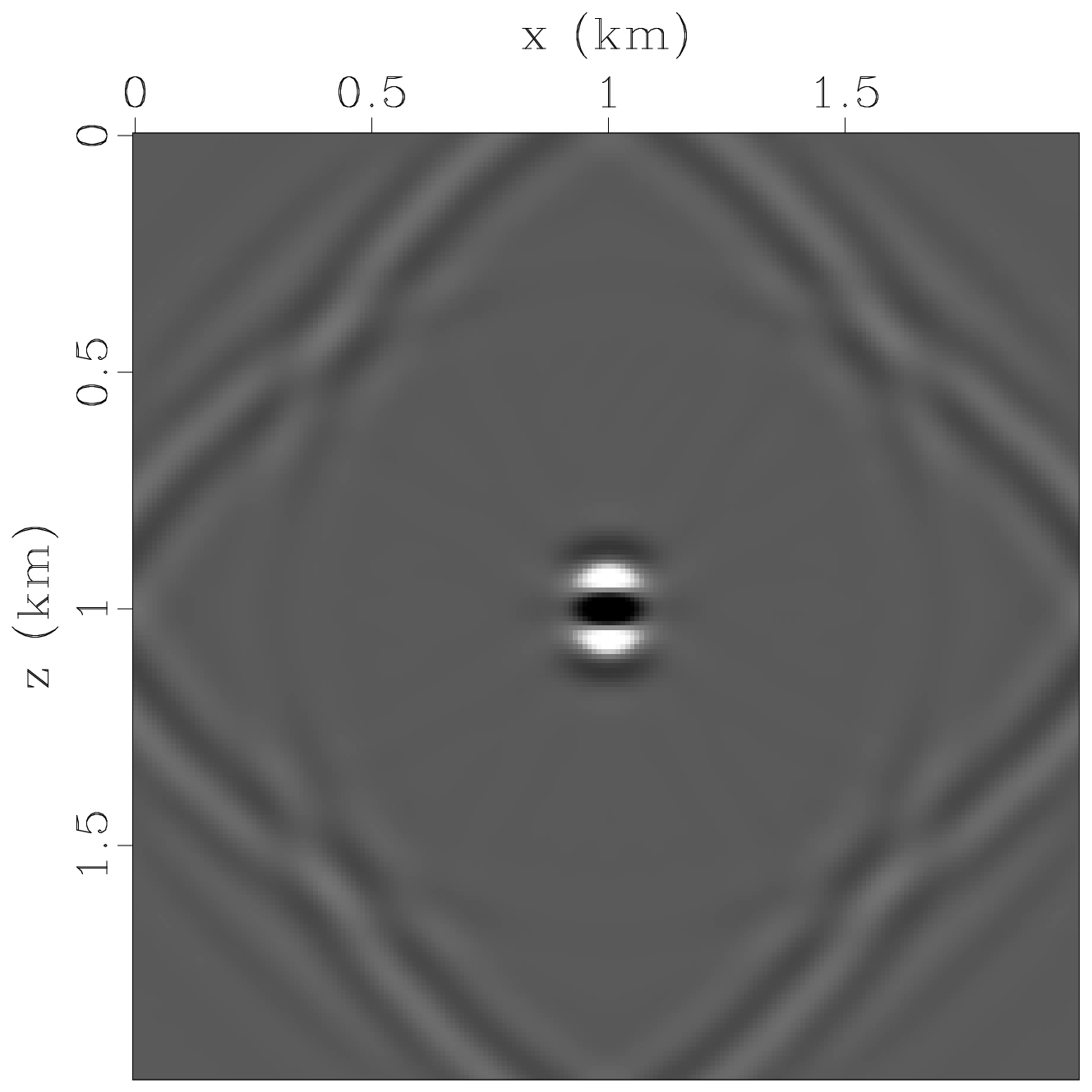}
  \caption{$\delta \alpha, \left(1,0,0\right)$}
  \label{fig:a5}
\end{subfigure}
\begin{subfigure}{0.16\textwidth}
  \centering
  \includegraphics[width=1\linewidth]{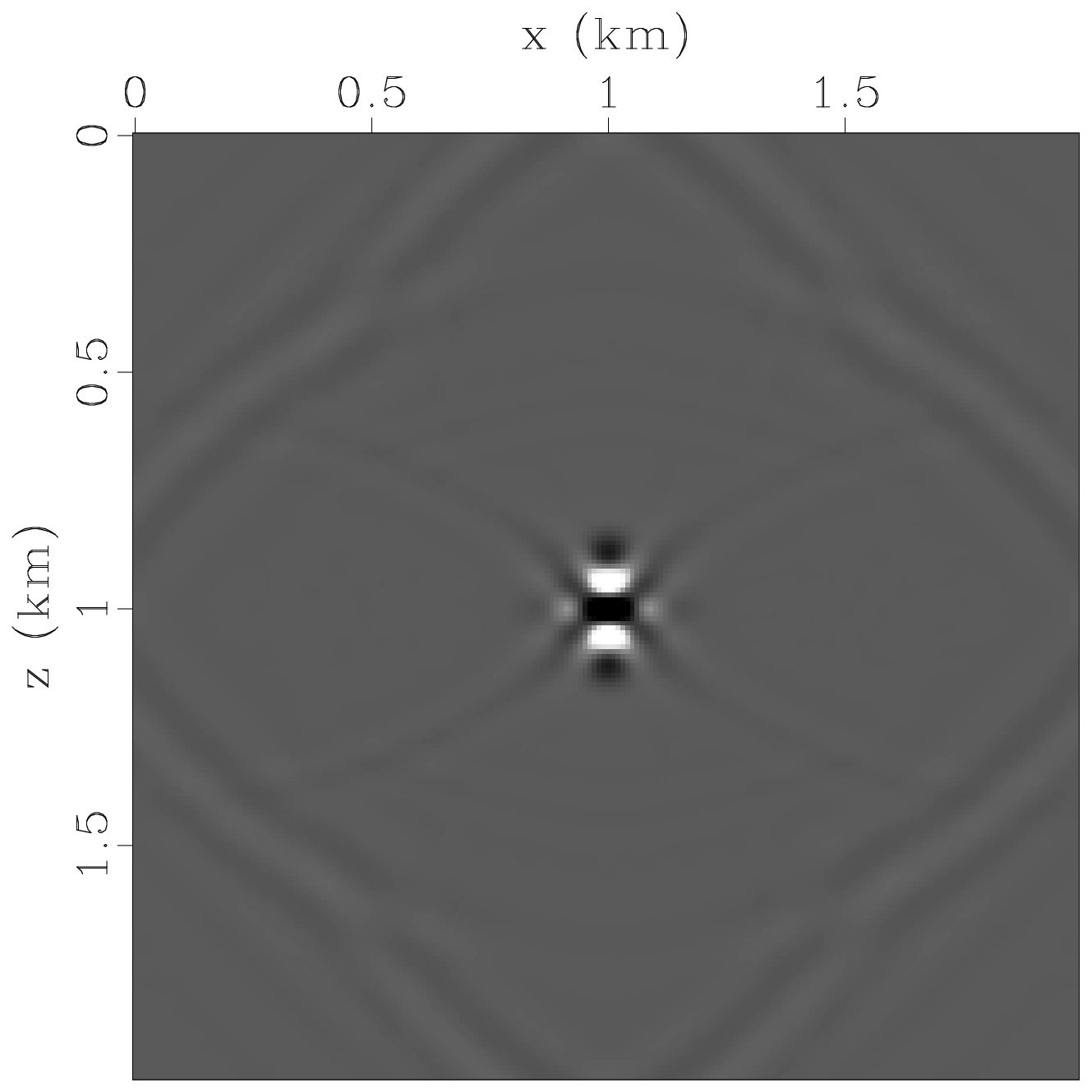}
  \caption{$\delta \beta, \left(1,0,0\right)$}
  \label{fig:b5}  
\end{subfigure}
\begin{subfigure}{0.16\textwidth}
  \centering
  \includegraphics[width=1\linewidth]{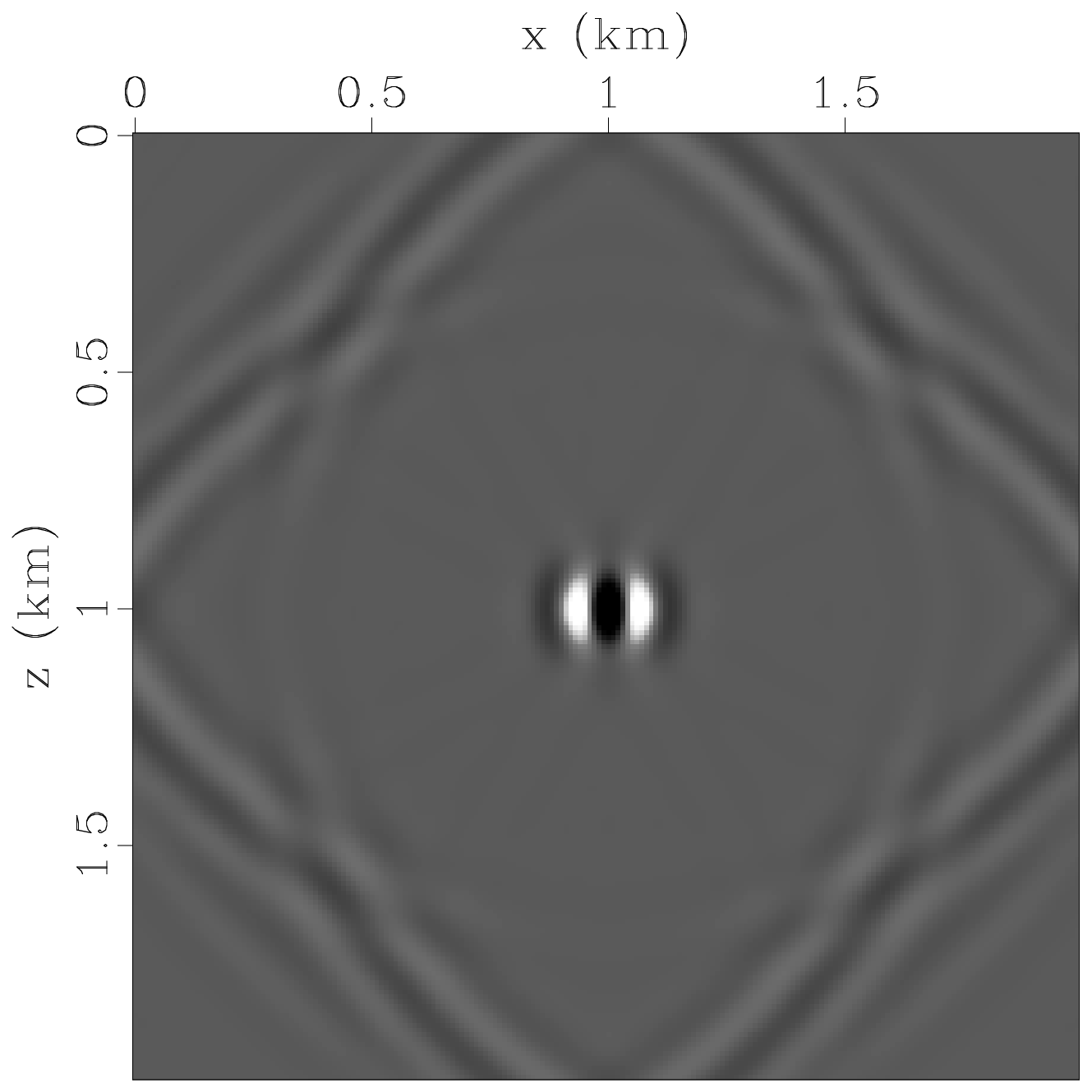}
  \caption{$\delta \alpha, \left(0,1,0\right)$}
  \label{fig:a6}
\end{subfigure}
\begin{subfigure}{0.16\textwidth}
  \centering
  \includegraphics[width=1\linewidth]{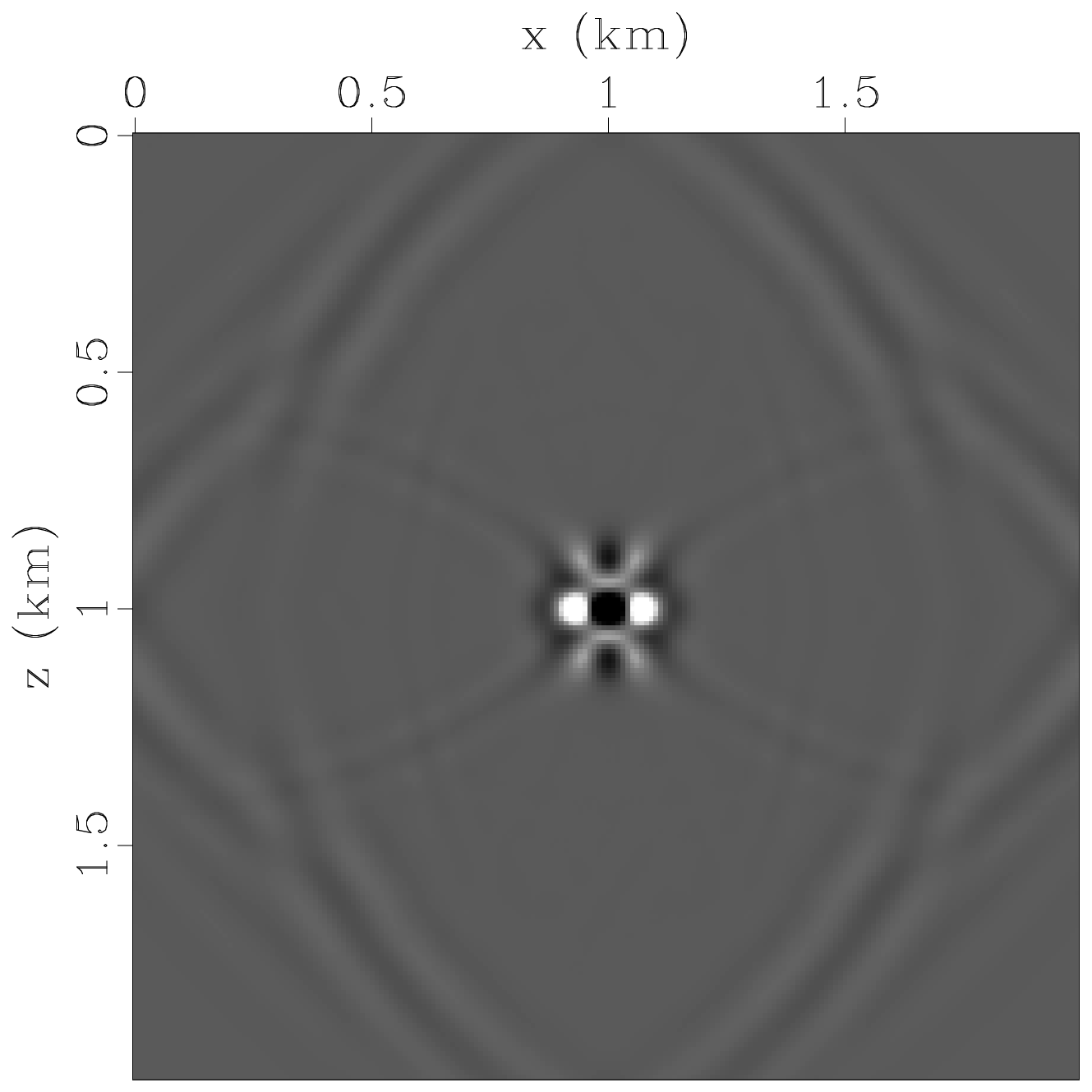}
  \caption{$\delta \beta, \left(0,1,0\right)$}
  \label{fig:b6}  
\end{subfigure}
\begin{subfigure}{0.16\textwidth}
  \centering
  \includegraphics[width=1\linewidth]{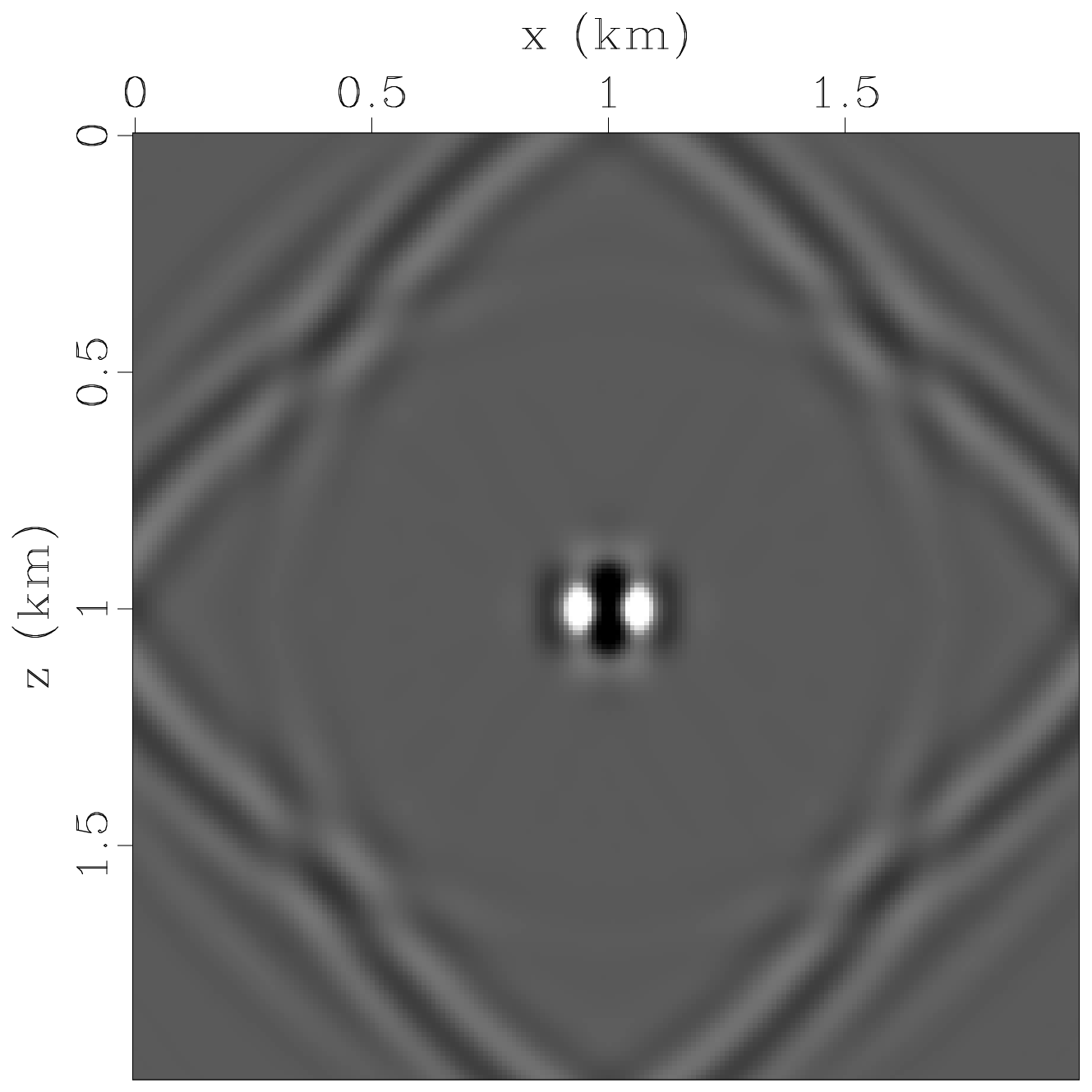}
  \caption{$\delta \alpha, \left(1,-2,0\right)$}
  \label{fig:a7}
\end{subfigure}
\begin{subfigure}{0.16\textwidth}
  \centering
  \includegraphics[width=1\linewidth]{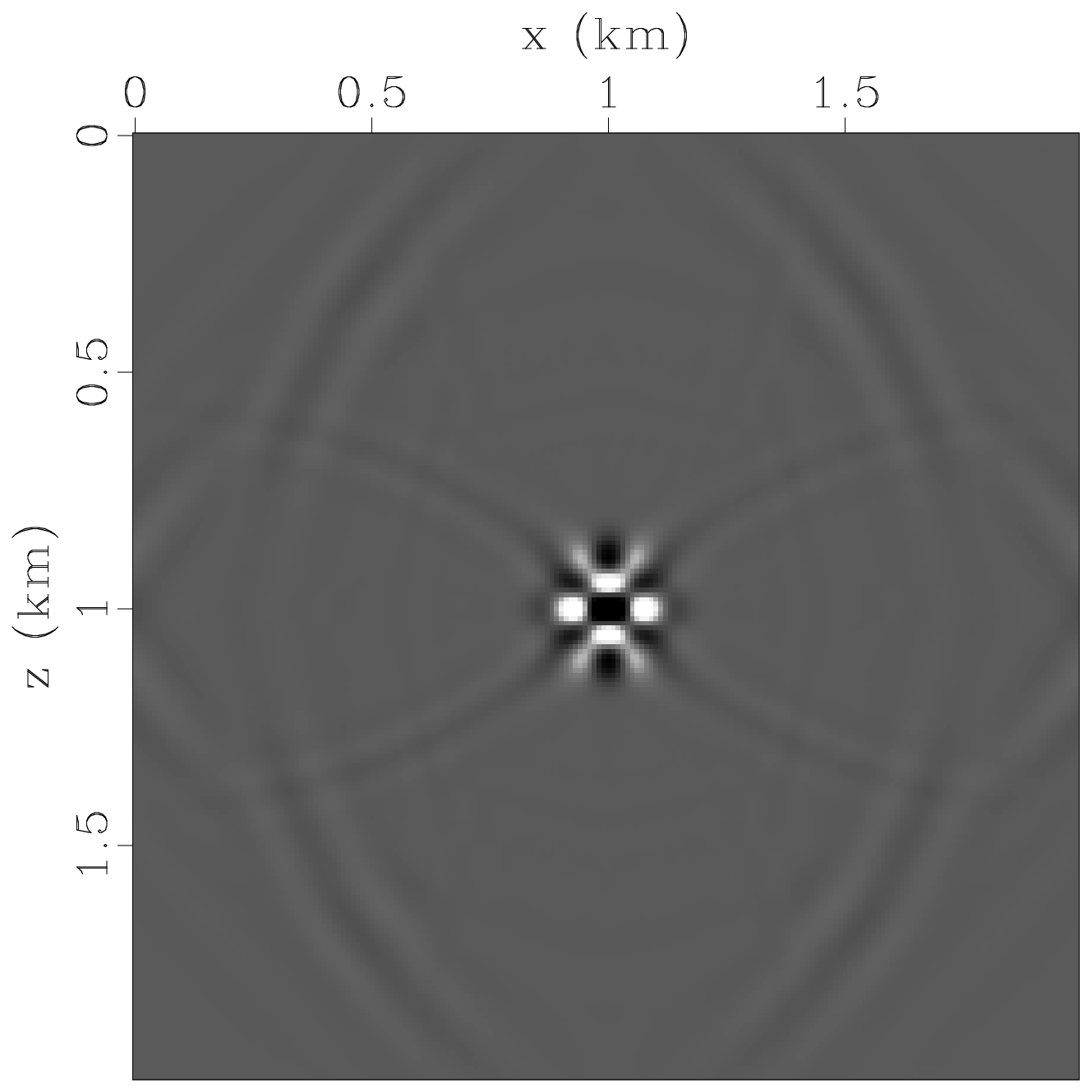}
  \caption{$\delta \beta, \left(1,-2,0\right)$}
  \label{fig:b7}  
\end{subfigure}
\begin{subfigure}{0.16\textwidth}
  \centering
  \includegraphics[width=1\linewidth]{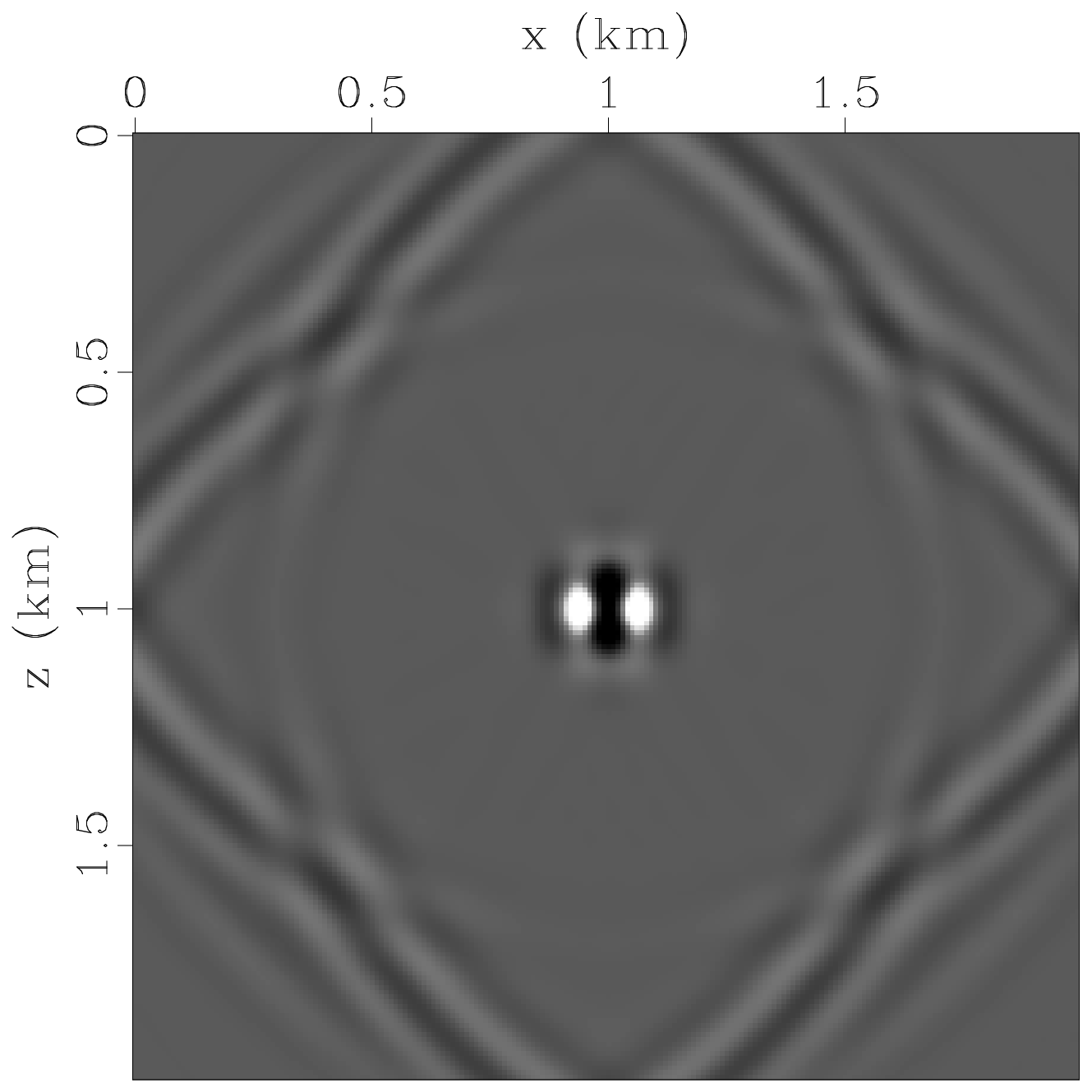}
  \caption{$\delta \alpha, \left(-1,2,0\right)$}
  \label{fig:a8}
\end{subfigure}
\begin{subfigure}{0.16\textwidth}
  \centering
  \includegraphics[width=1\linewidth]{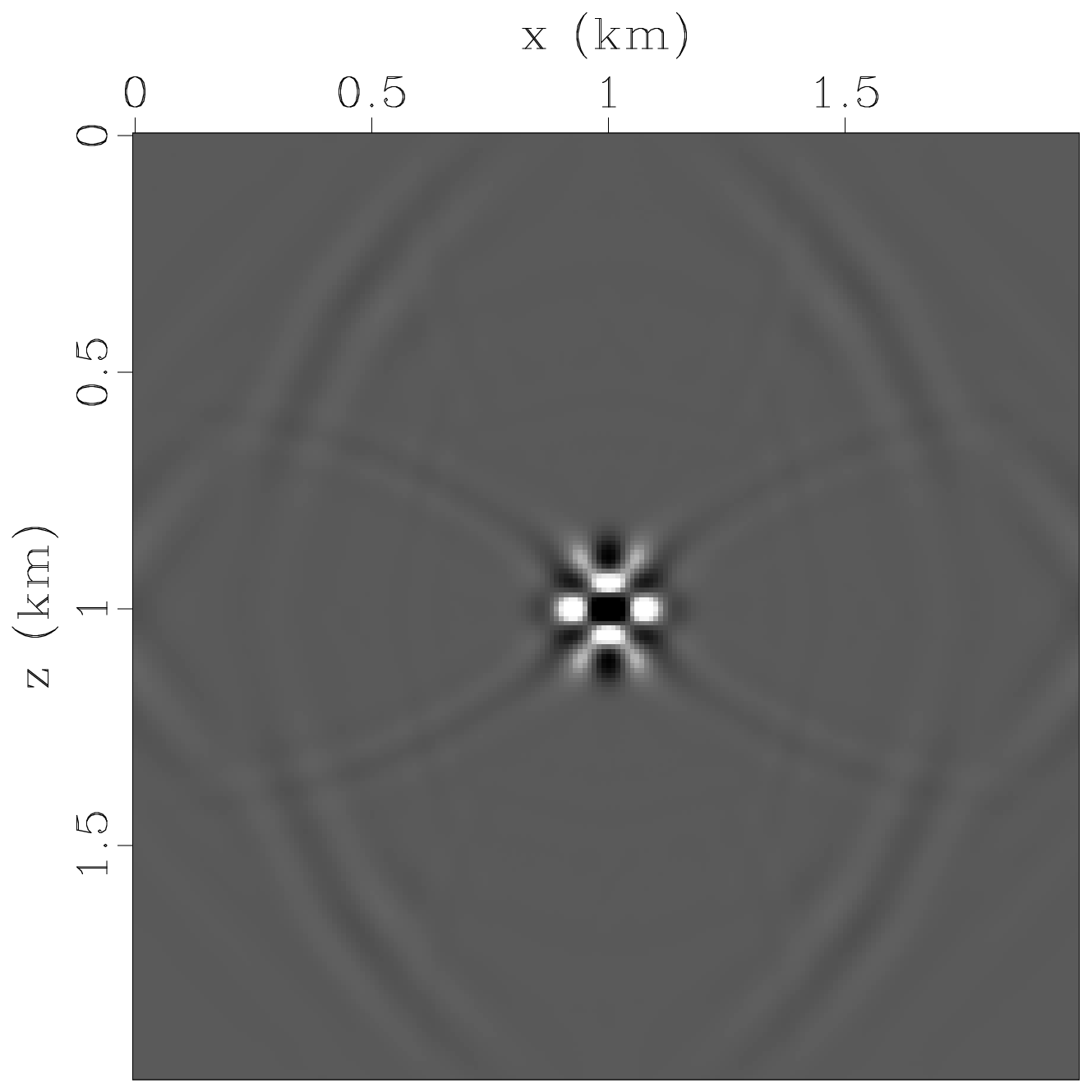}
  \caption{$\delta \beta, \left(-1,2,0\right)$}
  \label{fig:b8}  
\end{subfigure}
\begin{subfigure}{0.16\textwidth}
  \centering
  \includegraphics[width=1\linewidth]{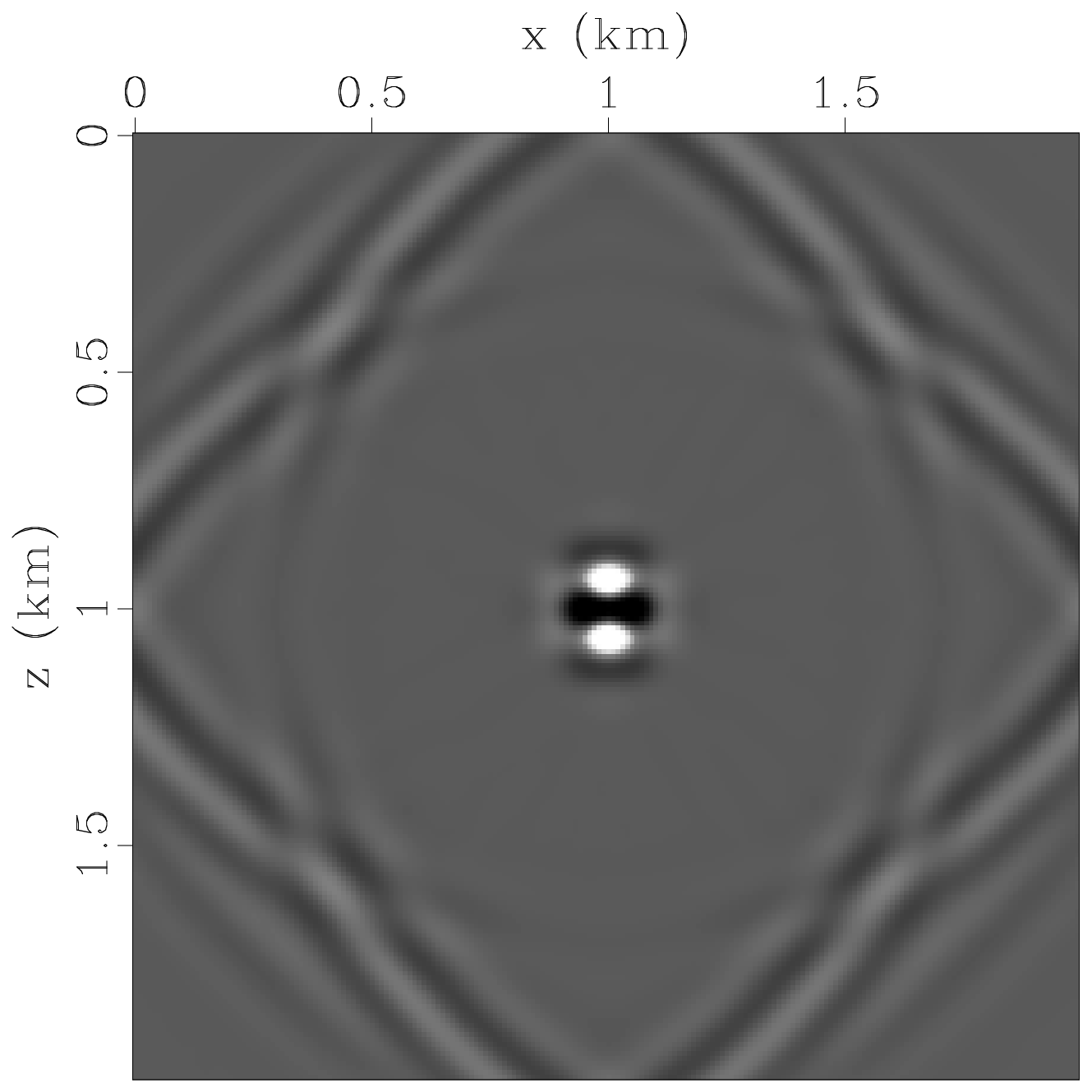}
  \caption{$\delta \alpha, \left(-2,1,0\right)$}
  \label{fig:a9}
\end{subfigure}
\begin{subfigure}{0.16\textwidth}
  \centering
  \includegraphics[width=1\linewidth]{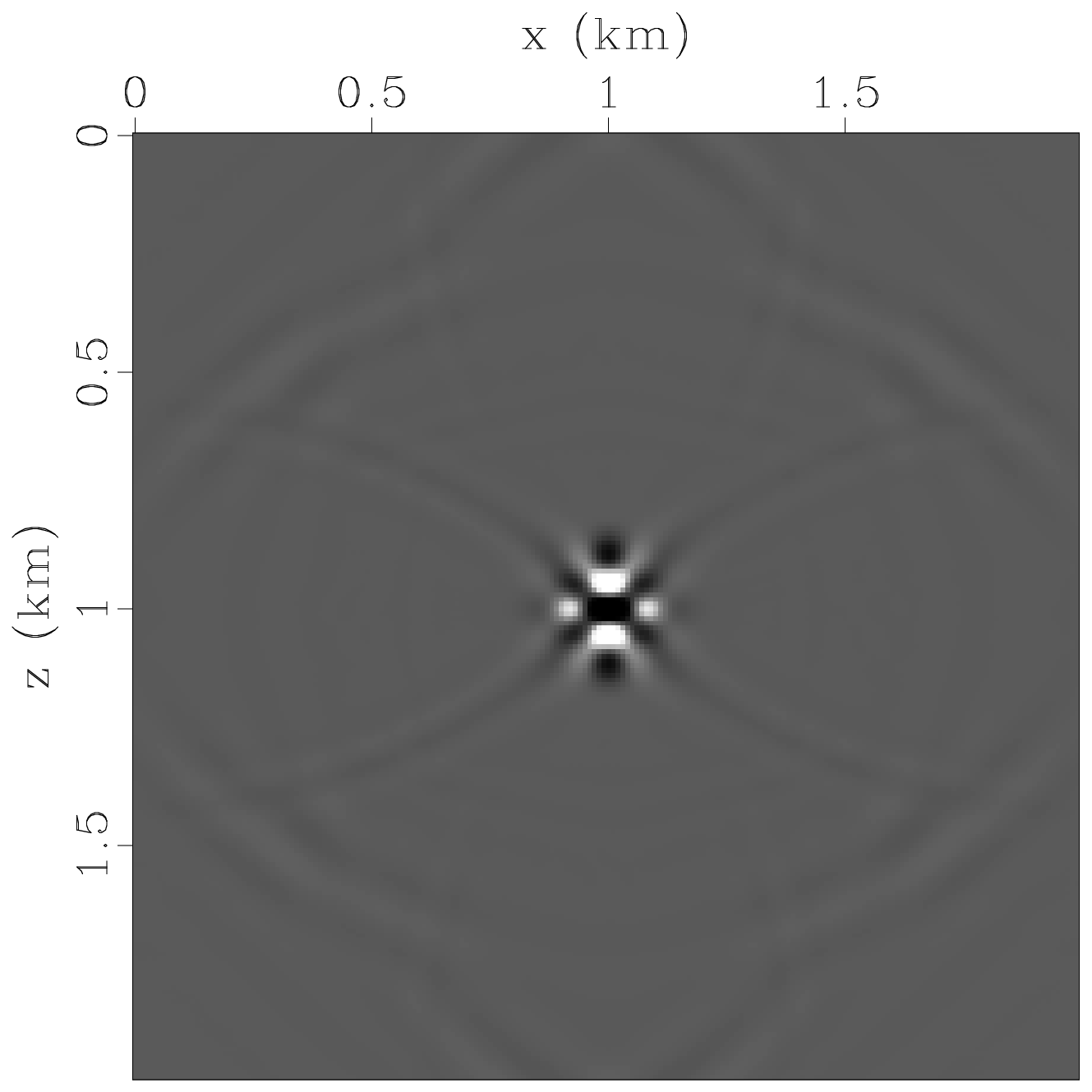}
  \caption{$\delta \beta, \left(-2,1,0\right)$}
  \label{fig:b9}  
\end{subfigure}
\caption{Inverted source images with receivers on all four boundaries. Moment tensors $M_i={M_{11},M_{22},M_{12}}$ used for forward modeling: (a)(b) $M_1={1,1,0}$; (c)(d) $M_2={-1,-1,0}$; (e)(f) $M_3={0,0,1}$; (g)(h) $M_4={0,-1,0}$; (i)(j) $M_5={1,0,0}$; (k)(l) $M_6={0,1,0}$; (m)(n) $M_7={1,-2,0}$; (o)(p) $M_8={-1,2,0}$; (q)(r) $M_9={-2,1,0}$.}
\label{fig:images_all}
\end{figure}

\begin{figure}
\begin{subfigure}{0.16\textwidth}
  \centering
  \includegraphics[width=1\linewidth]{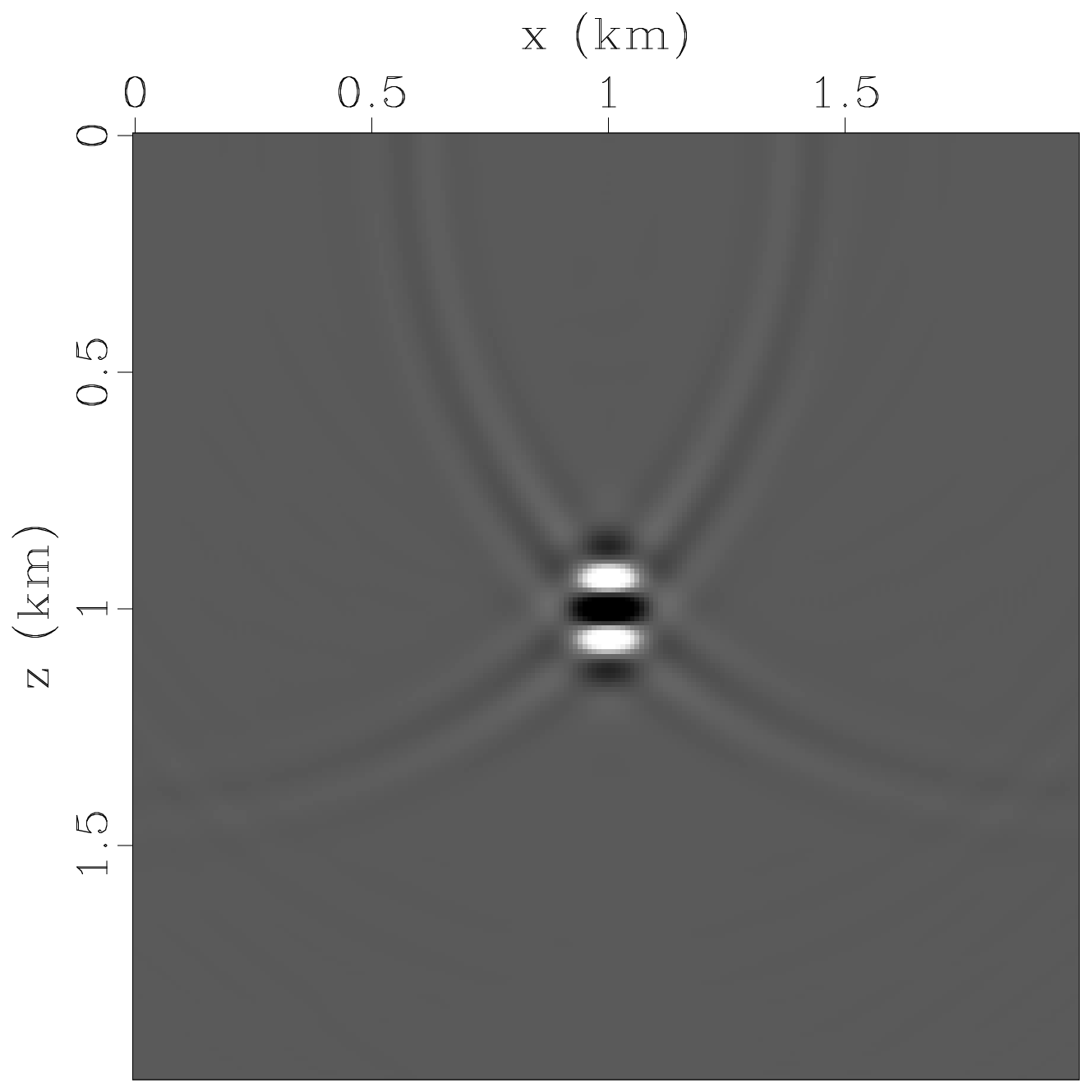}
  \caption{$\delta \alpha, \left( 1,1,0 \right) $ }
  \label{fig:as1}
\end{subfigure}
\begin{subfigure}{0.16\textwidth}
  \centering
  \includegraphics[width=1\linewidth]{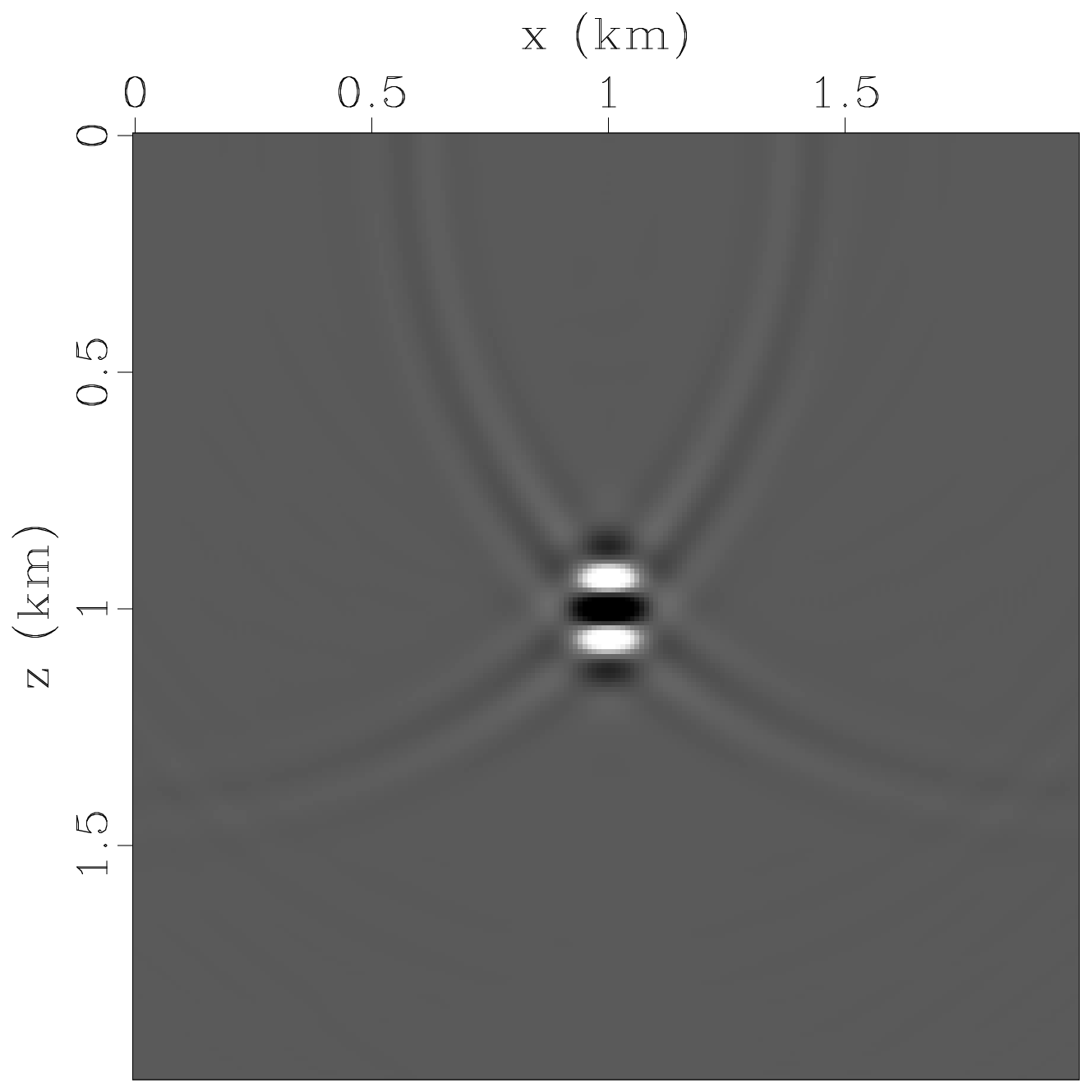}
  \caption{$\delta \beta, \left( 1,1,0 \right) $}
  \label{fig:bs1}
\end{subfigure}
\begin{subfigure}{0.16\textwidth}
  \centering
  \includegraphics[width=1\linewidth]{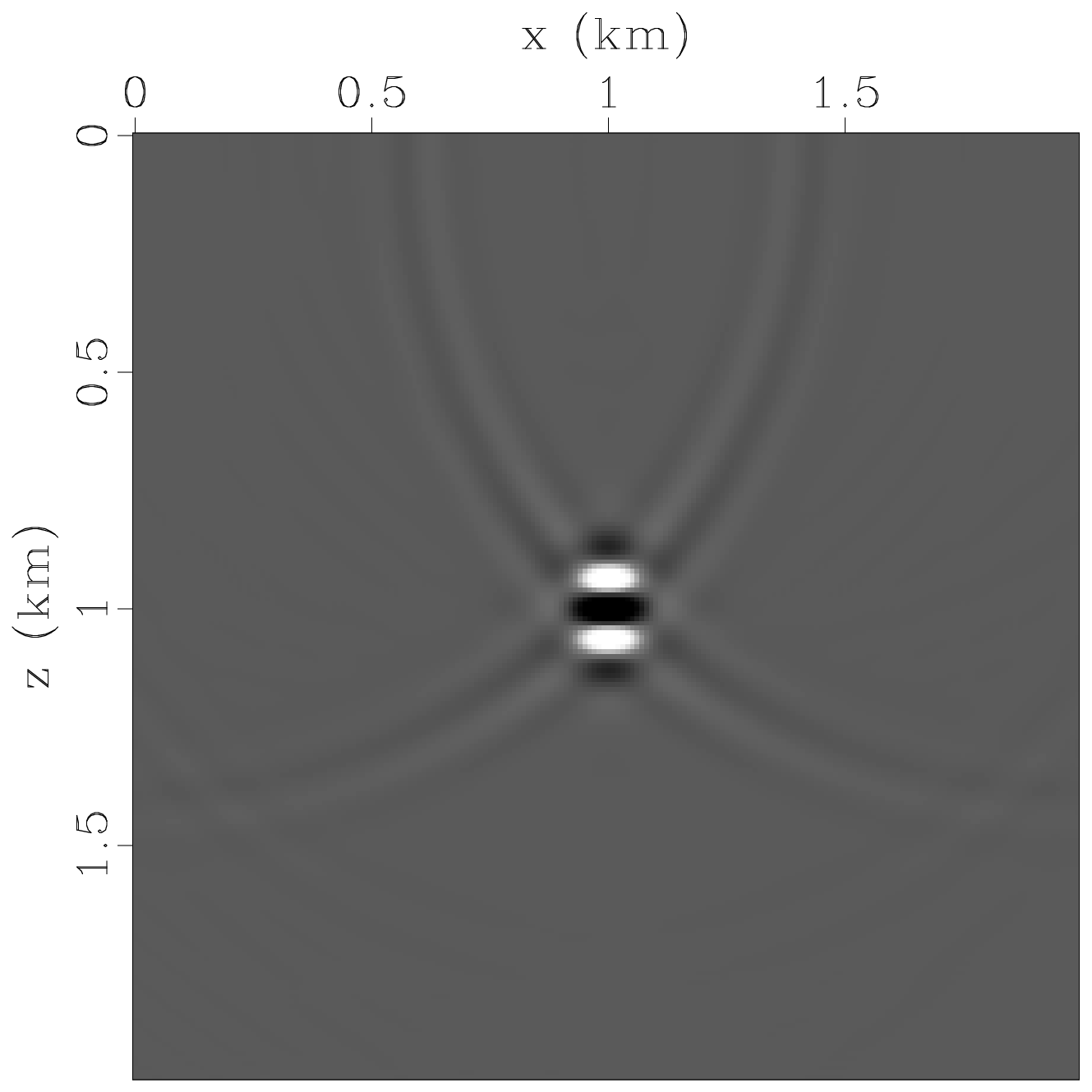}
  \caption{$\delta \alpha, \left( -1,-1,0 \right) $}
  \label{fig:as2}
\end{subfigure}
\begin{subfigure}{0.16\textwidth}
  \centering
  \includegraphics[width=1\linewidth]{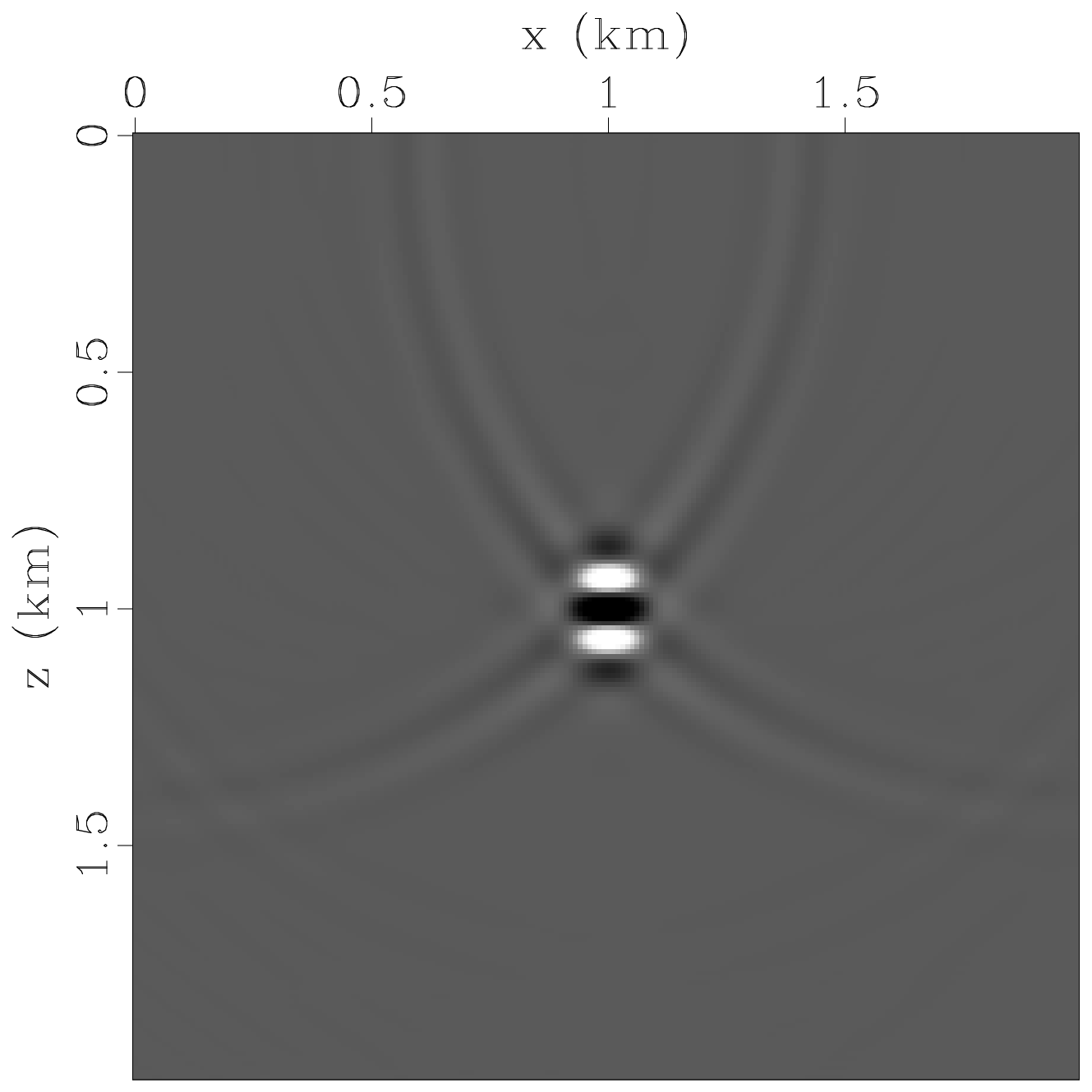}
  \caption{$\delta \beta, \left( -1,-1,0 \right) $}
  \label{fig:bs2}  
\end{subfigure}
\begin{subfigure}{0.16\textwidth}
  \centering
  \includegraphics[width=1\linewidth]{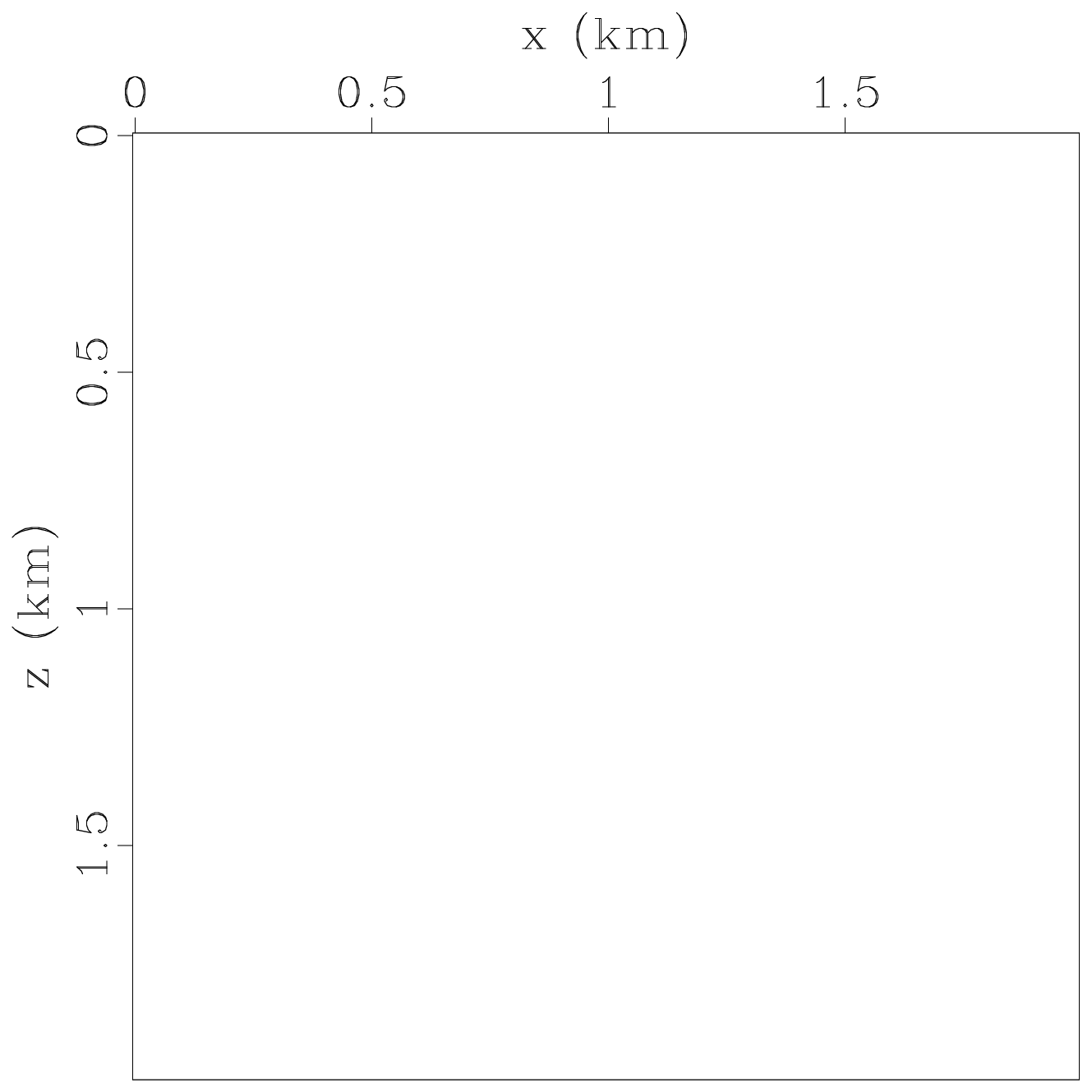}
  \caption{$\delta \alpha, \left(0,0,1\right) $}
  \label{fig:as3}
\end{subfigure}
\begin{subfigure}{0.16\textwidth}
  \centering
  \includegraphics[width=1\linewidth]{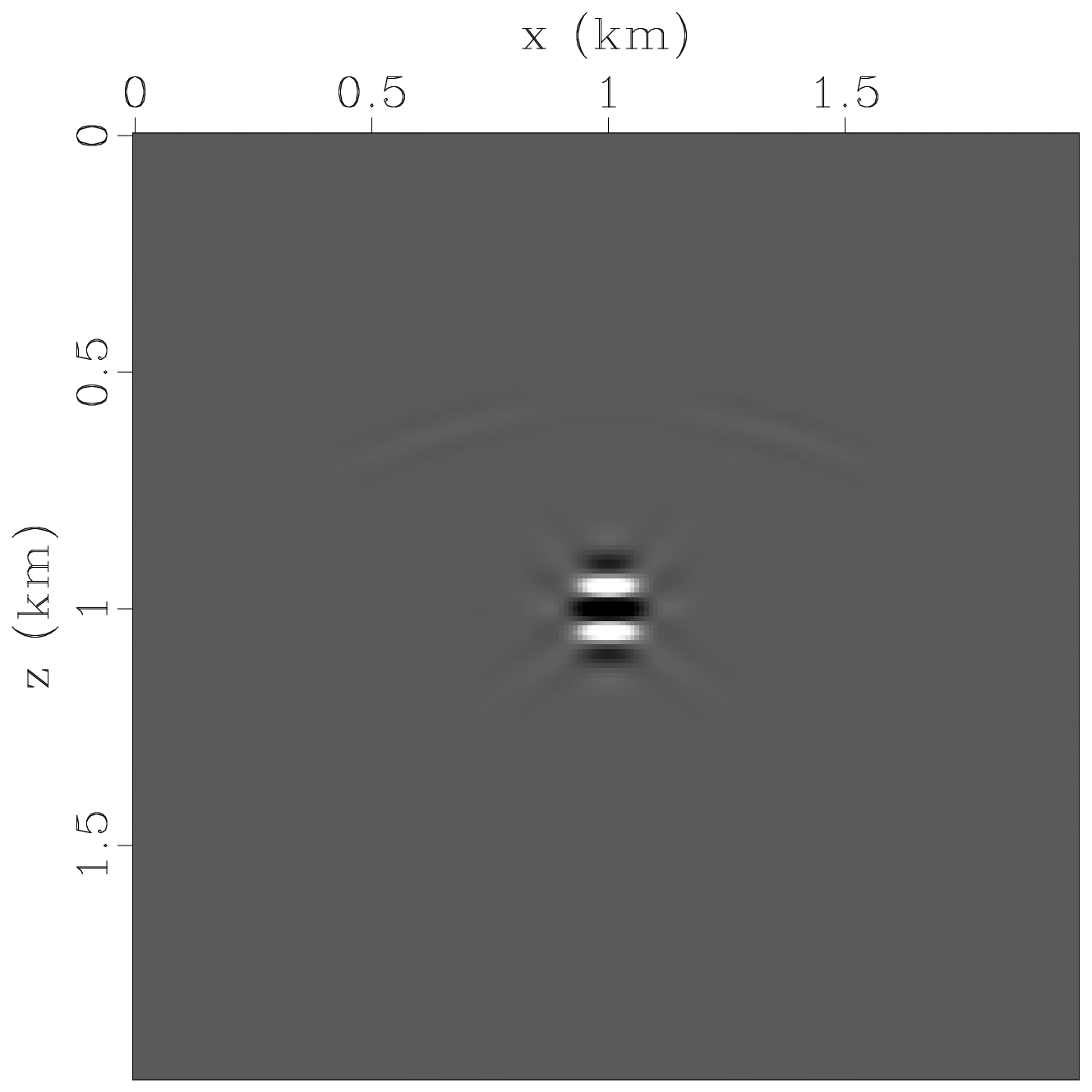}
  \caption{$\delta \beta, \left(0,0,1\right)$}
  \label{fig:bs3}  
\end{subfigure}
\begin{subfigure}{0.16\textwidth}
  \centering
  \includegraphics[width=1\linewidth]{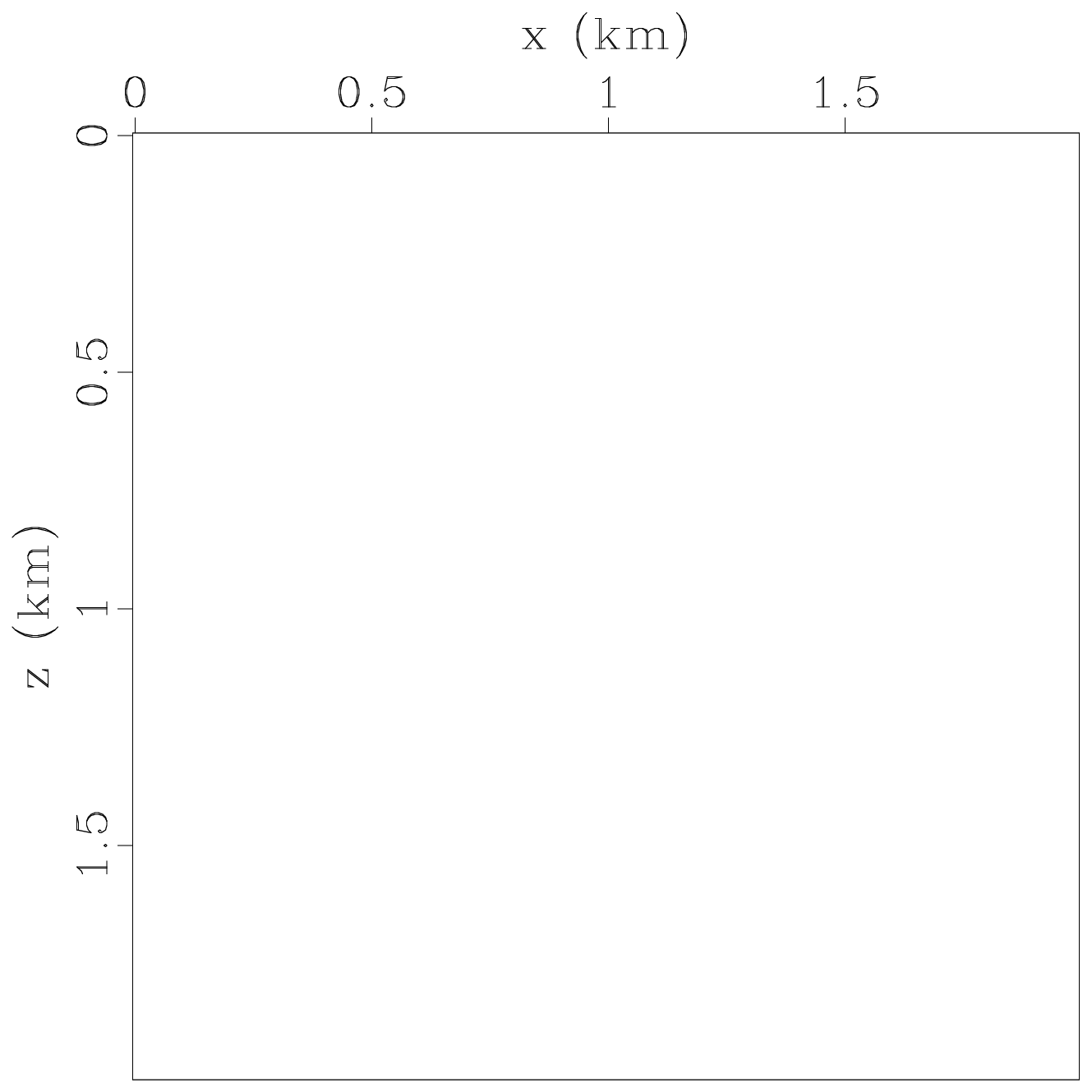}
  \caption{$\delta \alpha, \left(1,-1,0\right)$}
  \label{fig:as4}
\end{subfigure}
\begin{subfigure}{0.16\textwidth}
  \centering
  \includegraphics[width=1\linewidth]{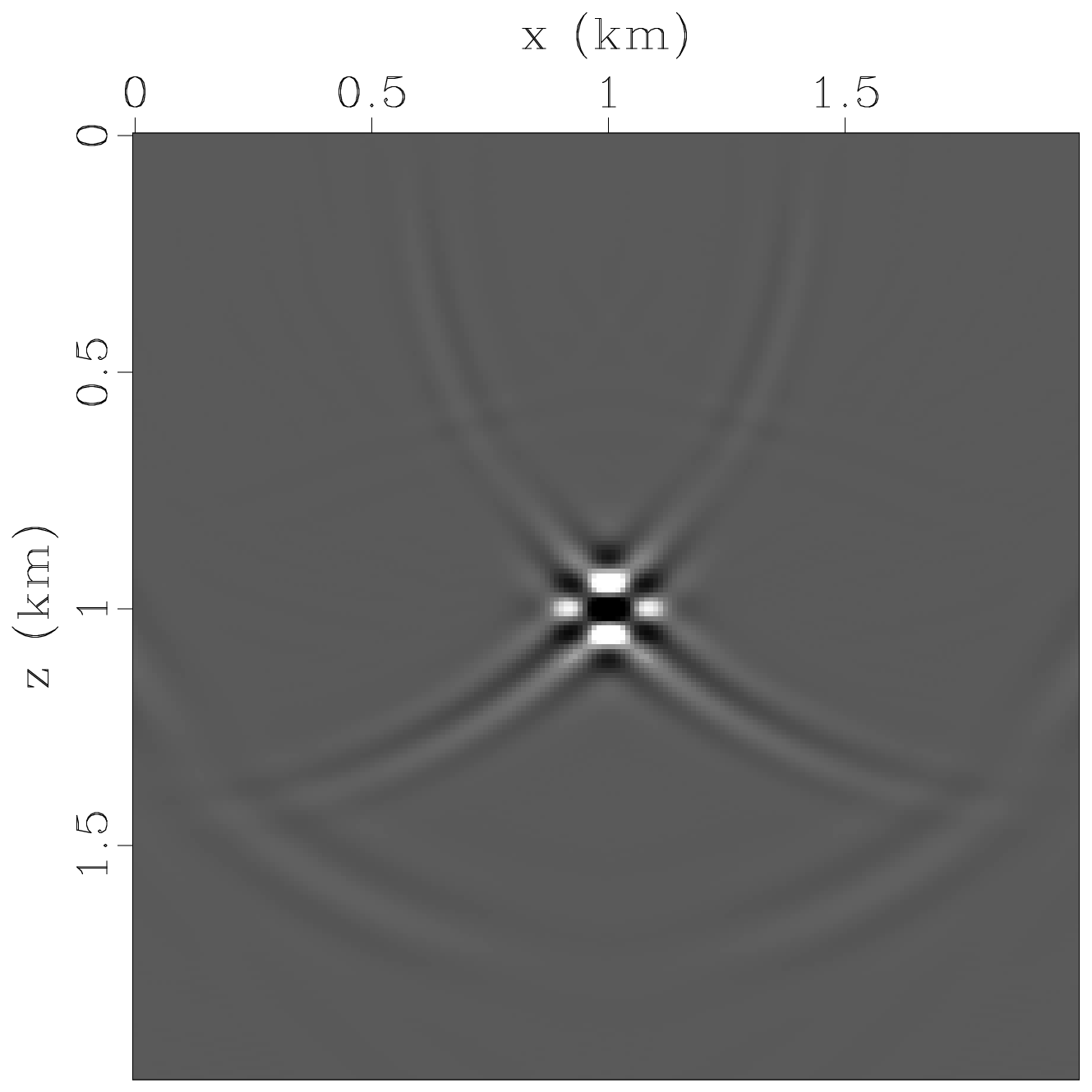}
  \caption{$\delta \beta, \left(1,-1,0\right)$}
  \label{fig:bs4}  
\end{subfigure}
\begin{subfigure}{0.16\textwidth}
  \centering
  \includegraphics[width=1\linewidth]{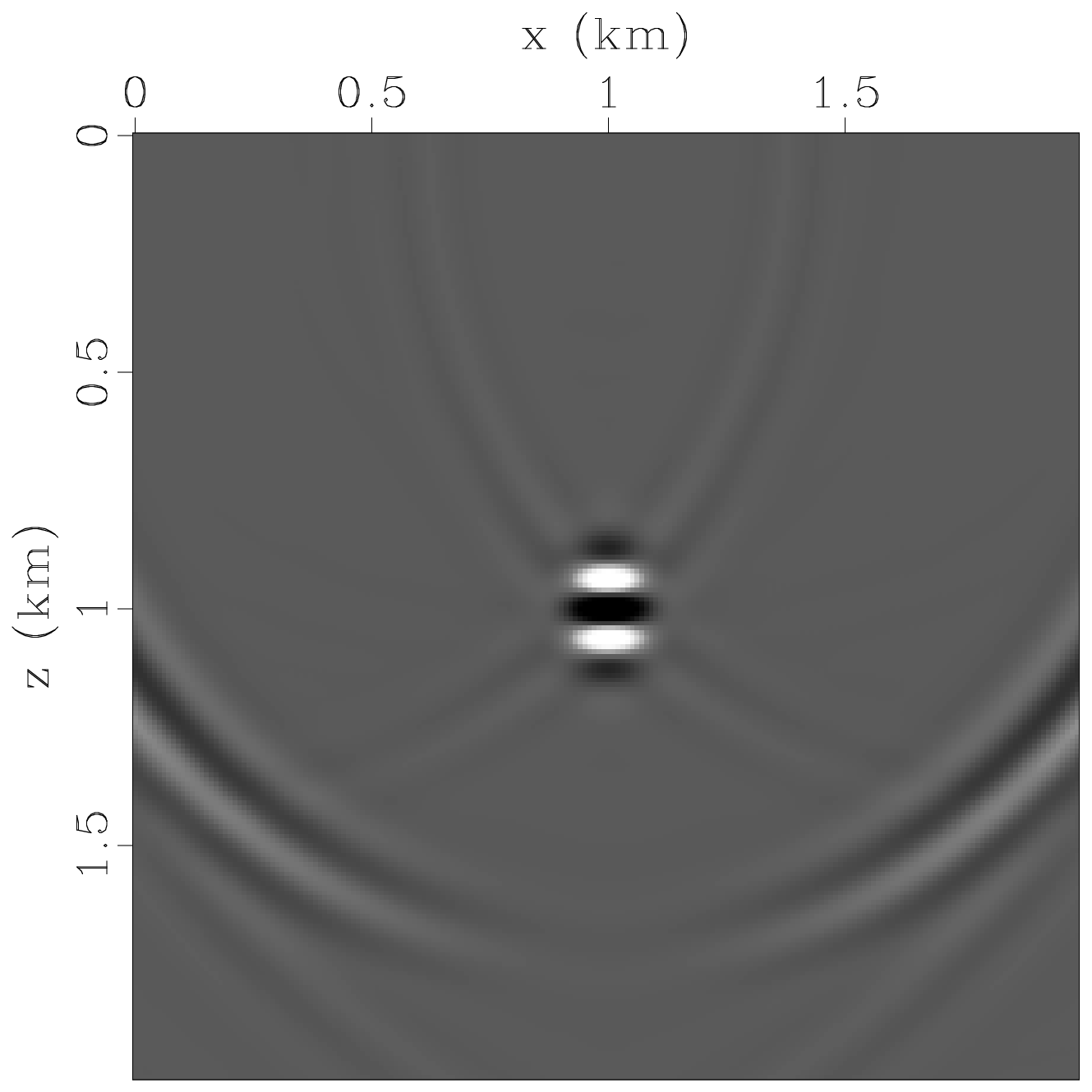}
  \caption{$\delta \alpha, \left(1,0,0\right)$}
  \label{fig:as5}
\end{subfigure}
\begin{subfigure}{0.16\textwidth}
  \centering
  \includegraphics[width=1\linewidth]{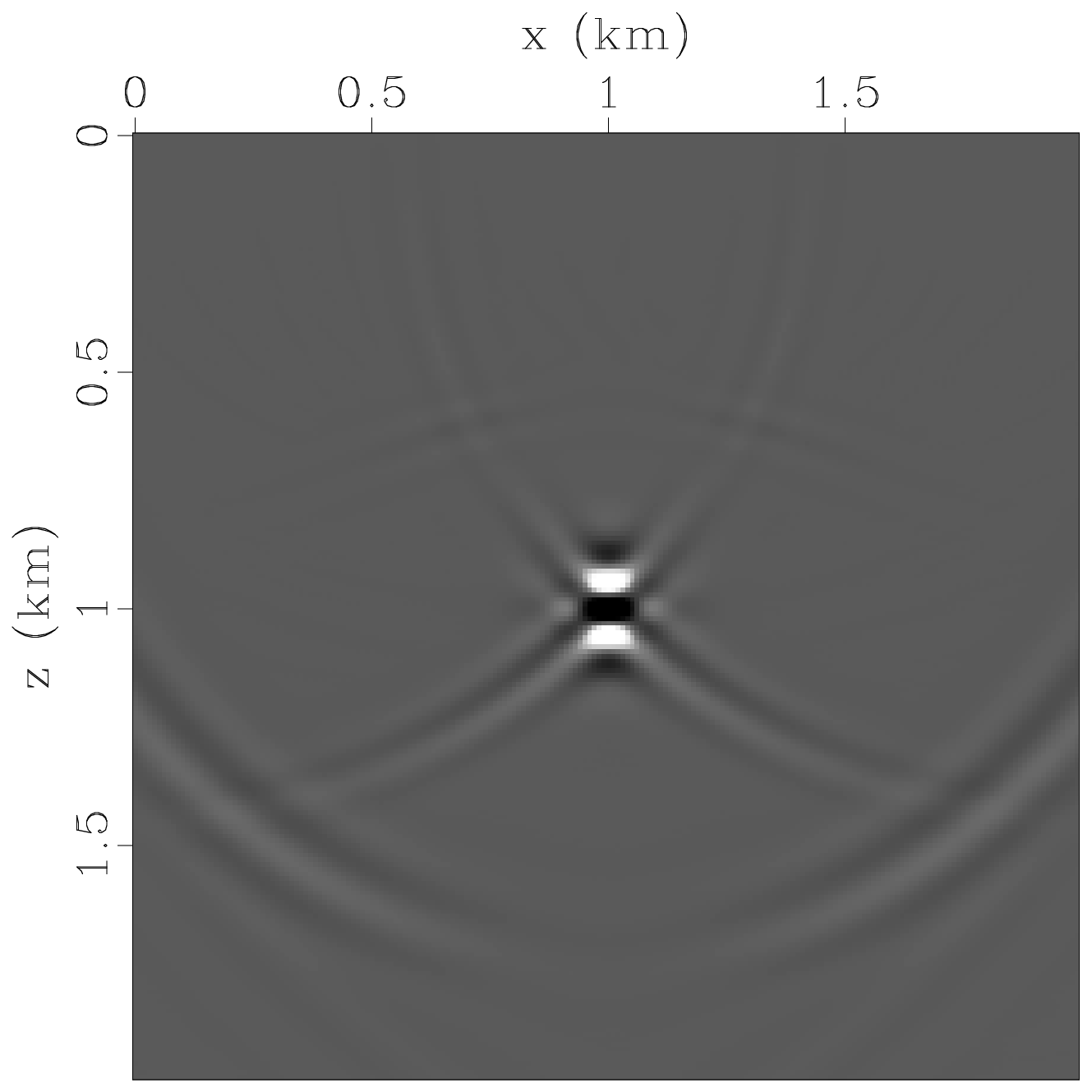}
  \caption{$\delta \beta, \left(1,0,0\right)$}
  \label{fig:bs5}  
\end{subfigure}
\begin{subfigure}{0.16\textwidth}
  \centering
  \includegraphics[width=1\linewidth]{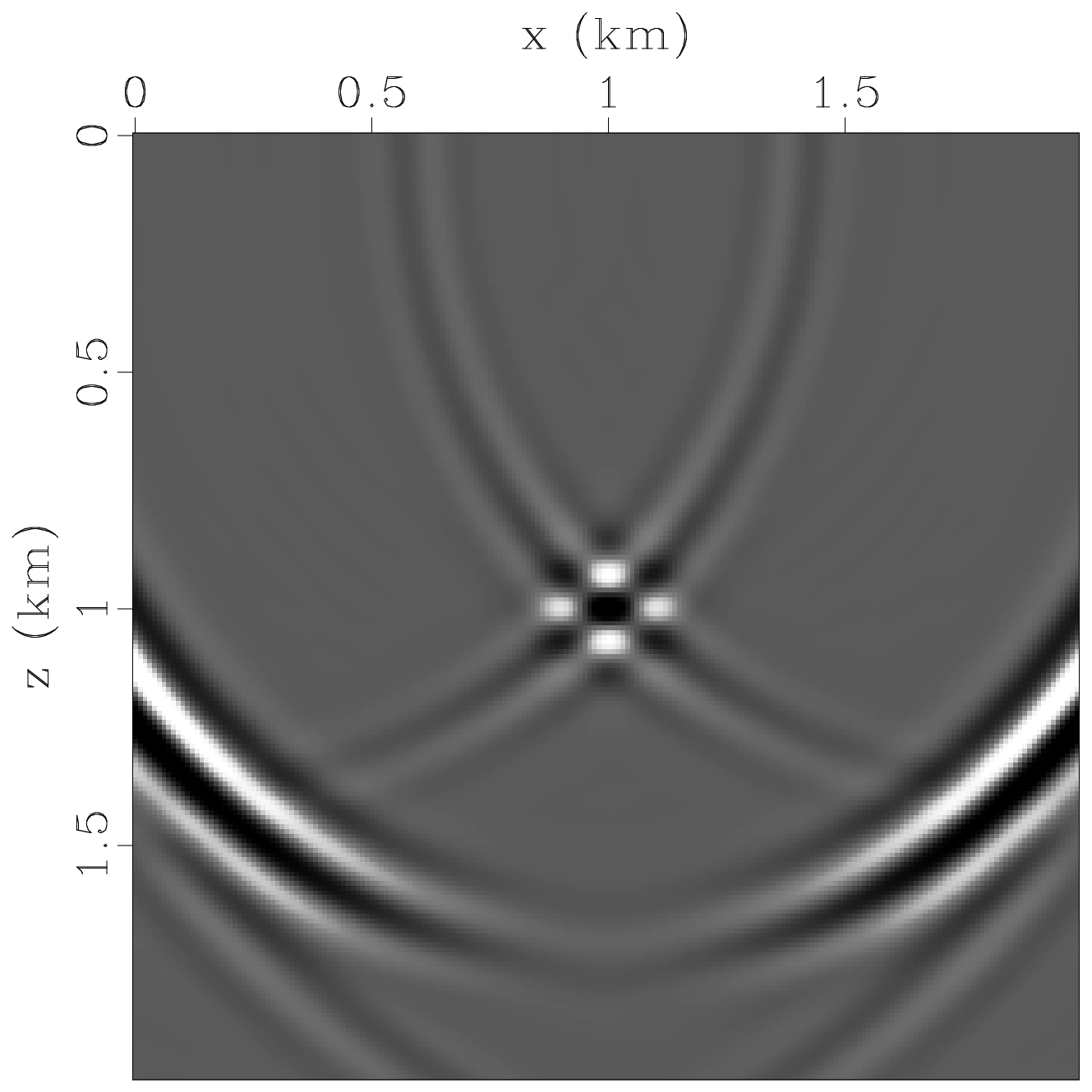}
  \caption{$\delta \alpha, \left(0,1,0\right)$}
  \label{fig:as6}
\end{subfigure}
\begin{subfigure}{0.16\textwidth}
  \centering
  \includegraphics[width=1\linewidth]{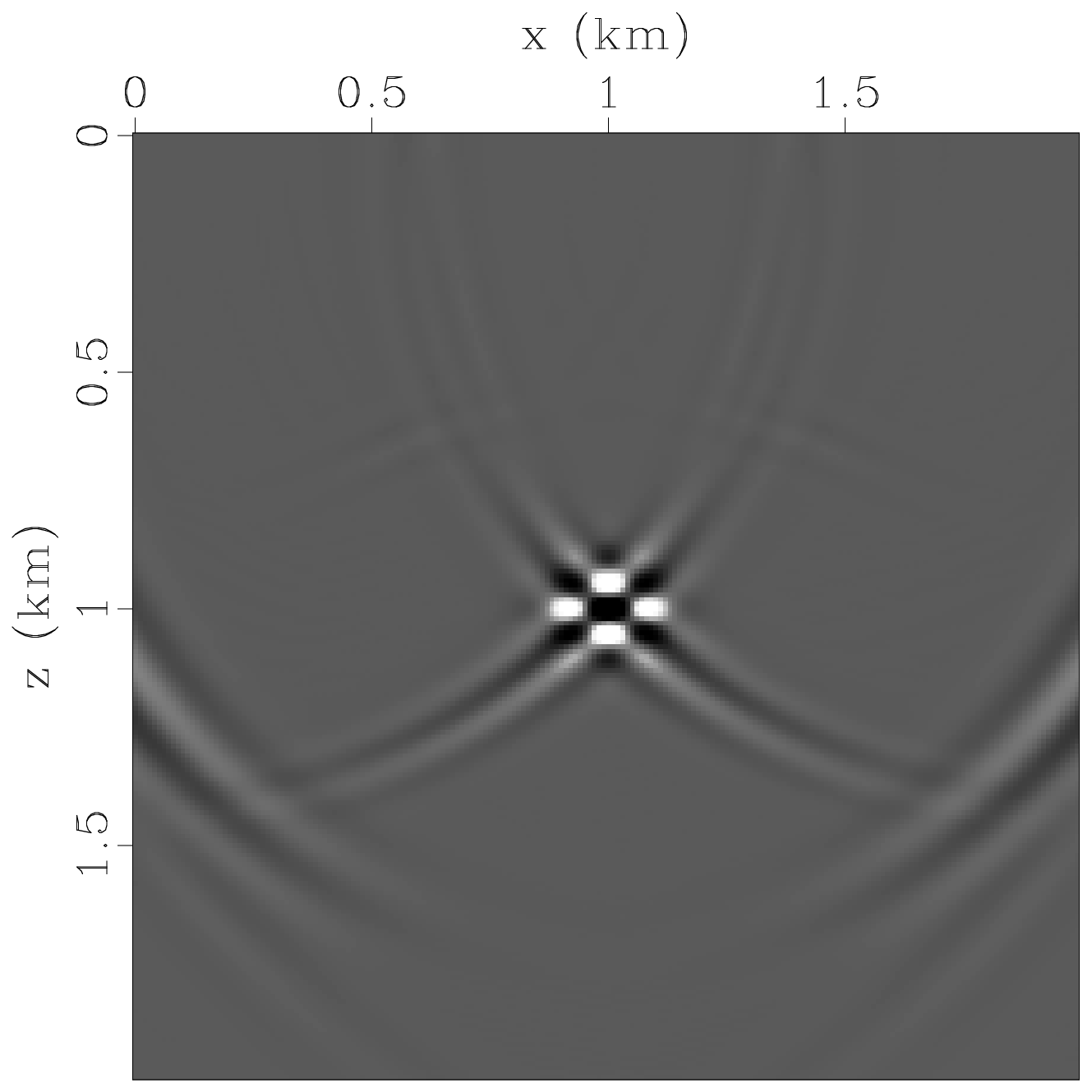}
  \caption{$\delta \beta, \left(0,1,0\right)$}
  \label{fig:bs6}  
\end{subfigure}
\begin{subfigure}{0.16\textwidth}
  \centering
  \includegraphics[width=1\linewidth]{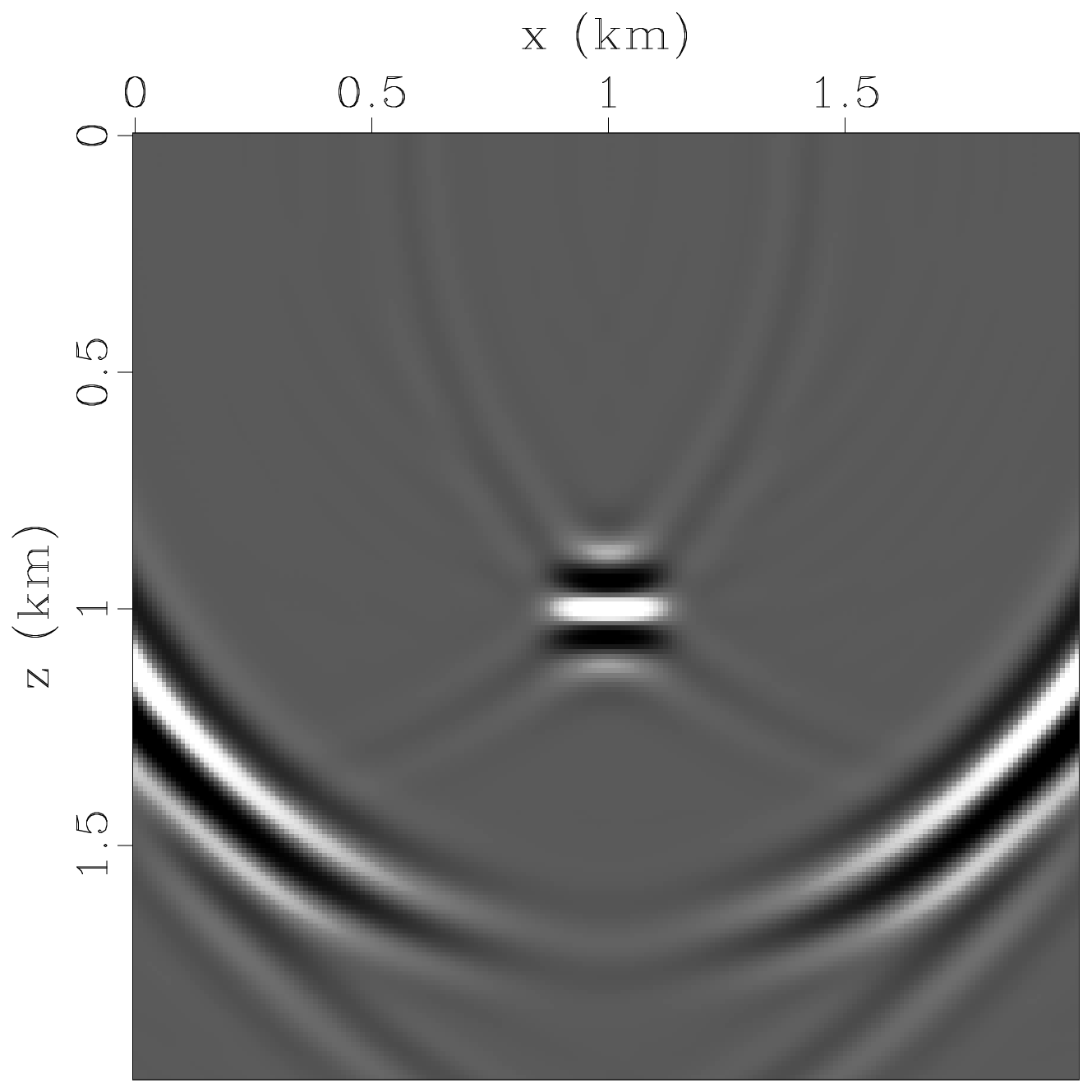}
  \caption{$\delta \alpha, \left(1,-2,0\right)$}
  \label{fig:as7}
\end{subfigure}
\begin{subfigure}{0.16\textwidth}
  \centering
  \includegraphics[width=1\linewidth]{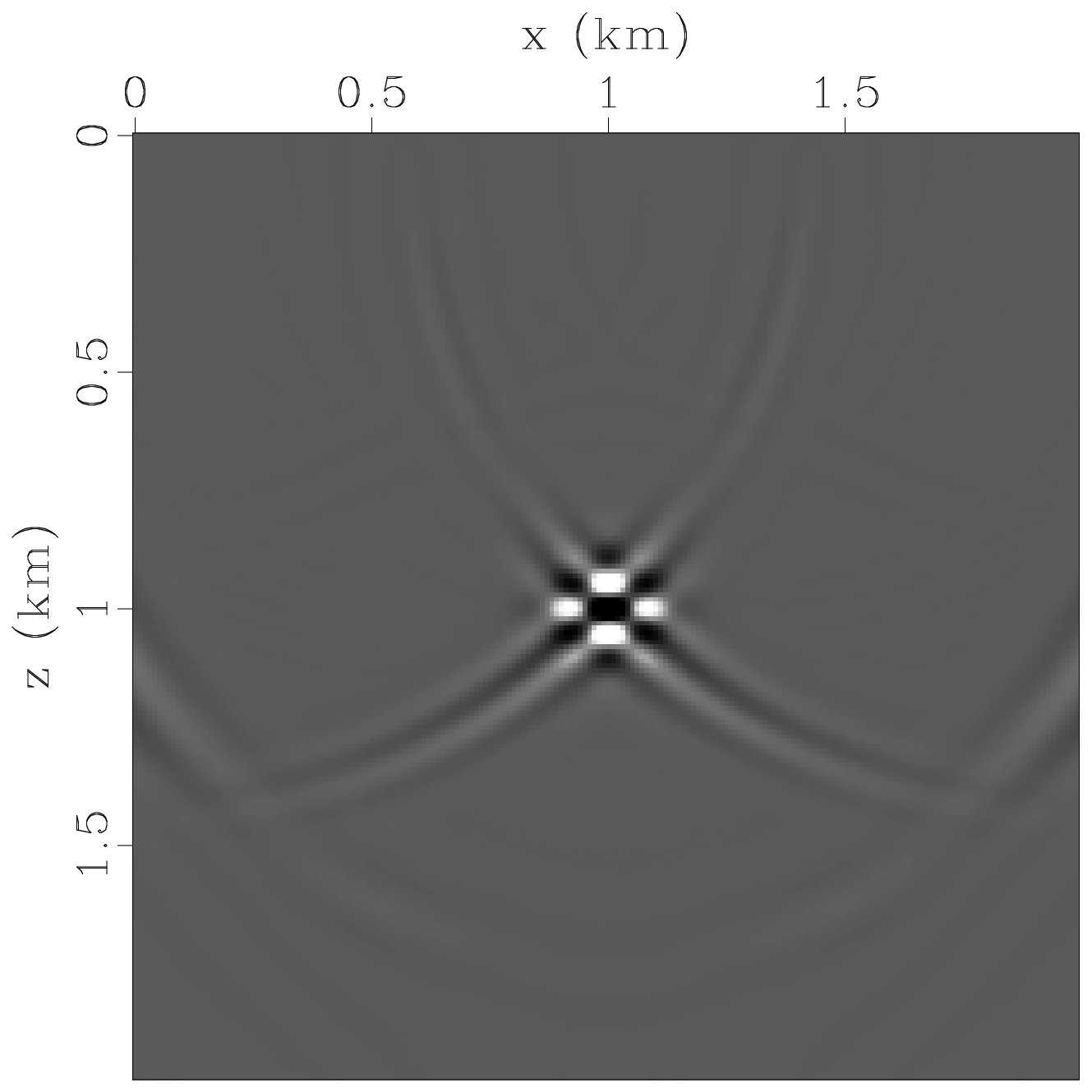}
  \caption{$\delta \beta, \left(1,-2,0\right)$}
  \label{fig:bs7}  
\end{subfigure}
\begin{subfigure}{0.16\textwidth}
  \centering
  \includegraphics[width=1\linewidth]{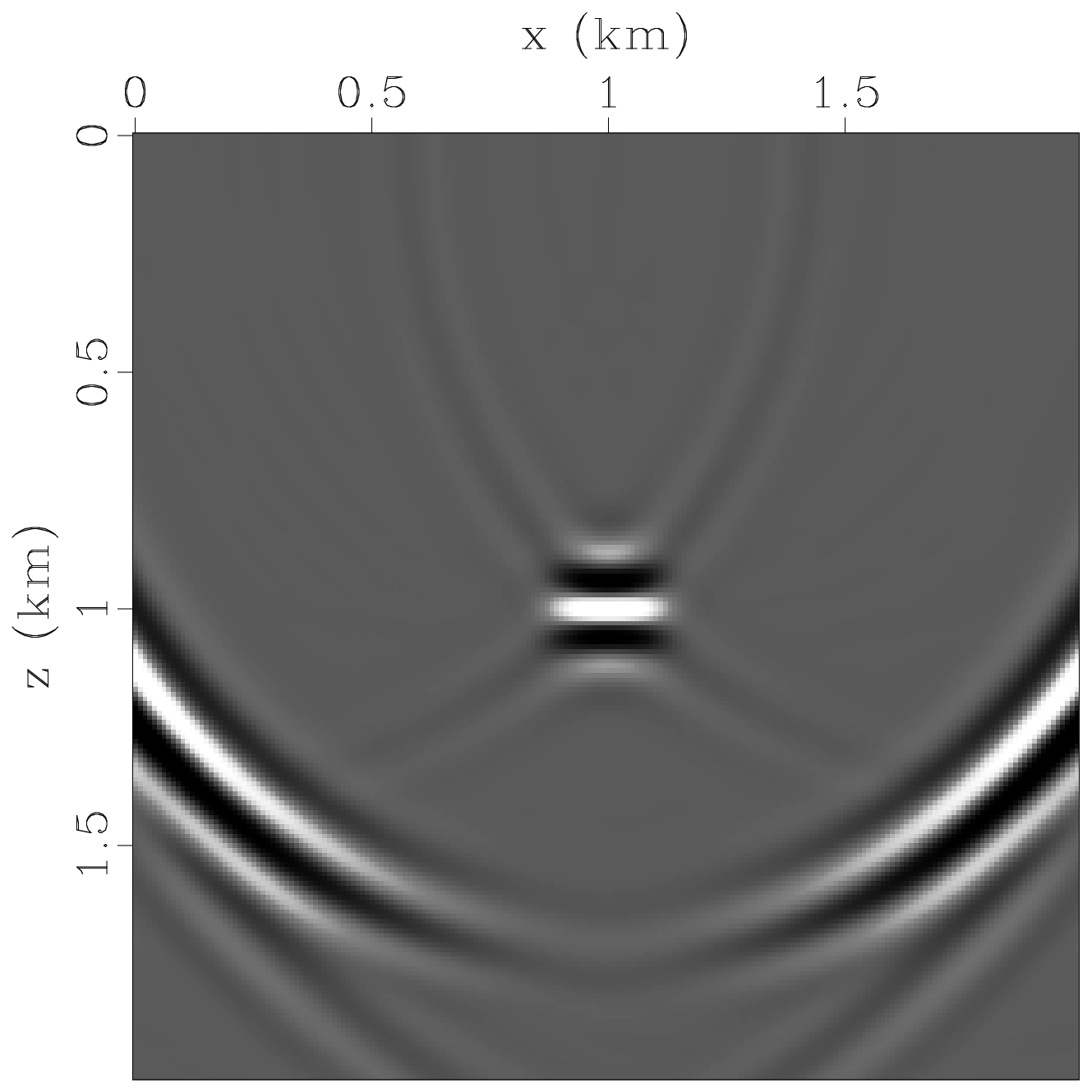}
  \caption{$\delta \alpha, \left(-1,2,0\right)$}
  \label{fig:as8}
\end{subfigure}
\begin{subfigure}{0.16\textwidth}
  \centering
  \includegraphics[width=1\linewidth]{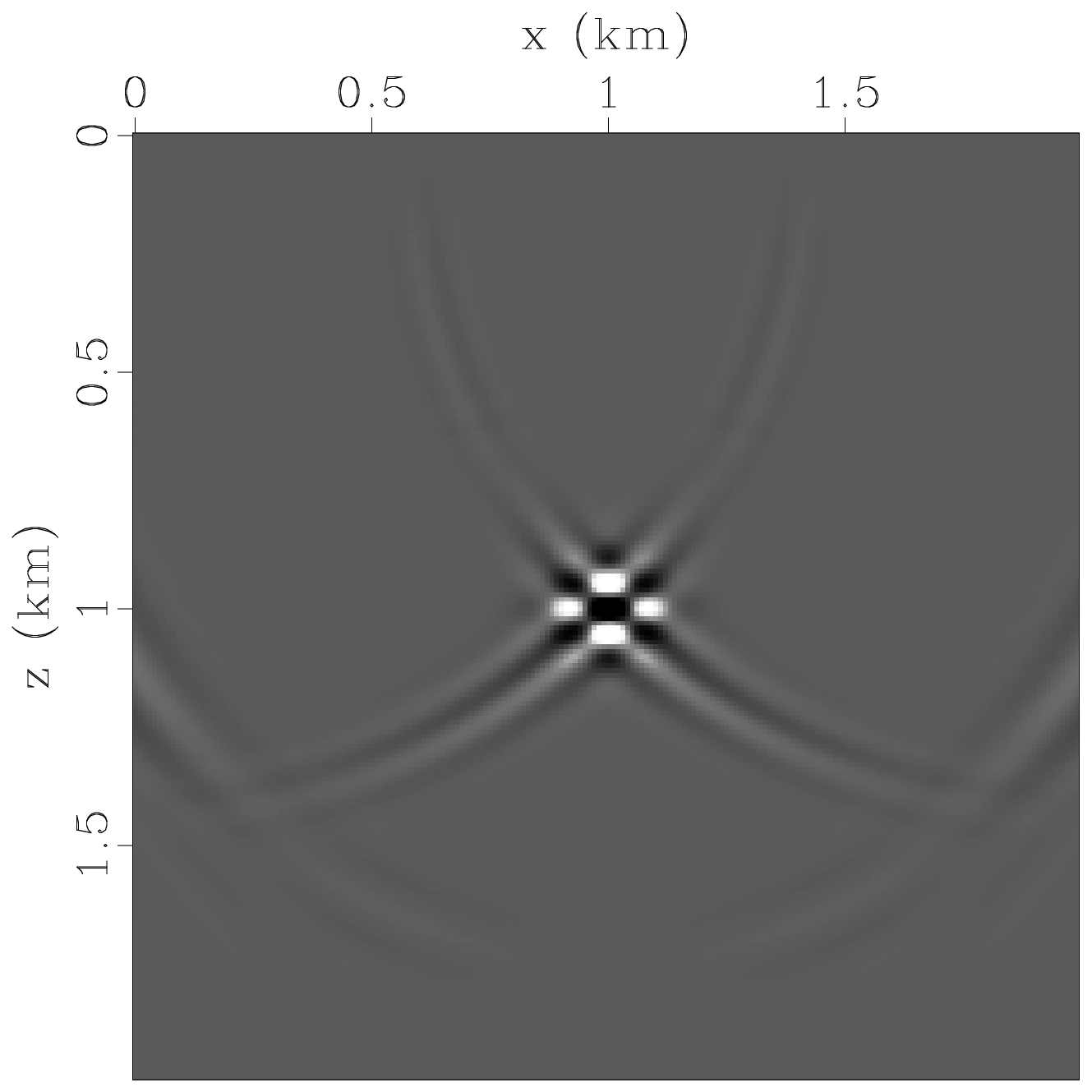}
  \caption{$\delta \beta, \left(-1,2,0\right)$}
  \label{fig:bs8}  
\end{subfigure}
\begin{subfigure}{0.16\textwidth}
  \centering
  \includegraphics[width=1\linewidth]{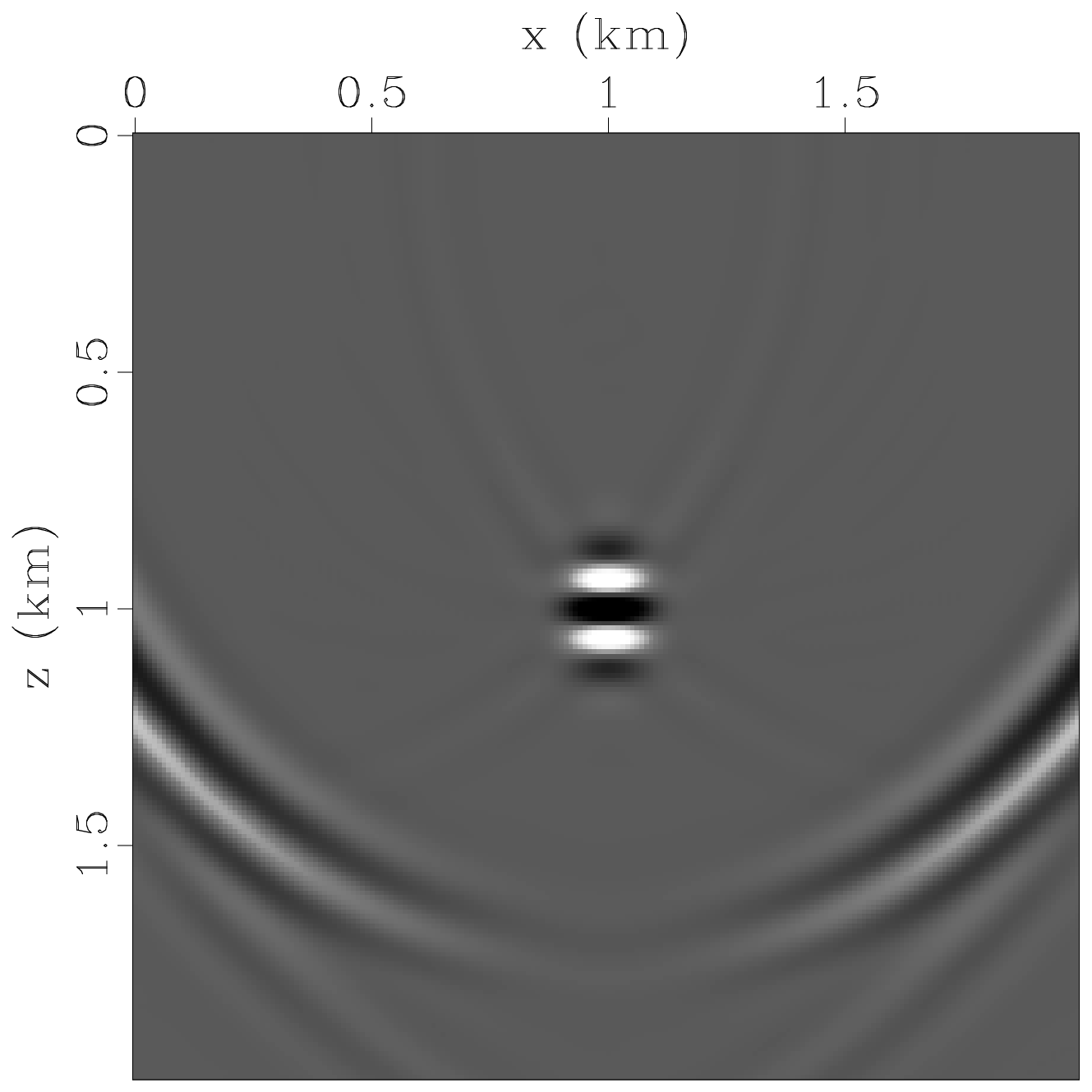}
  \caption{$\delta \alpha, \left(-2,1,0\right)$}
  \label{fig:as9}
\end{subfigure}
\begin{subfigure}{.16\textwidth}
  \centering
  \includegraphics[width=1\linewidth]{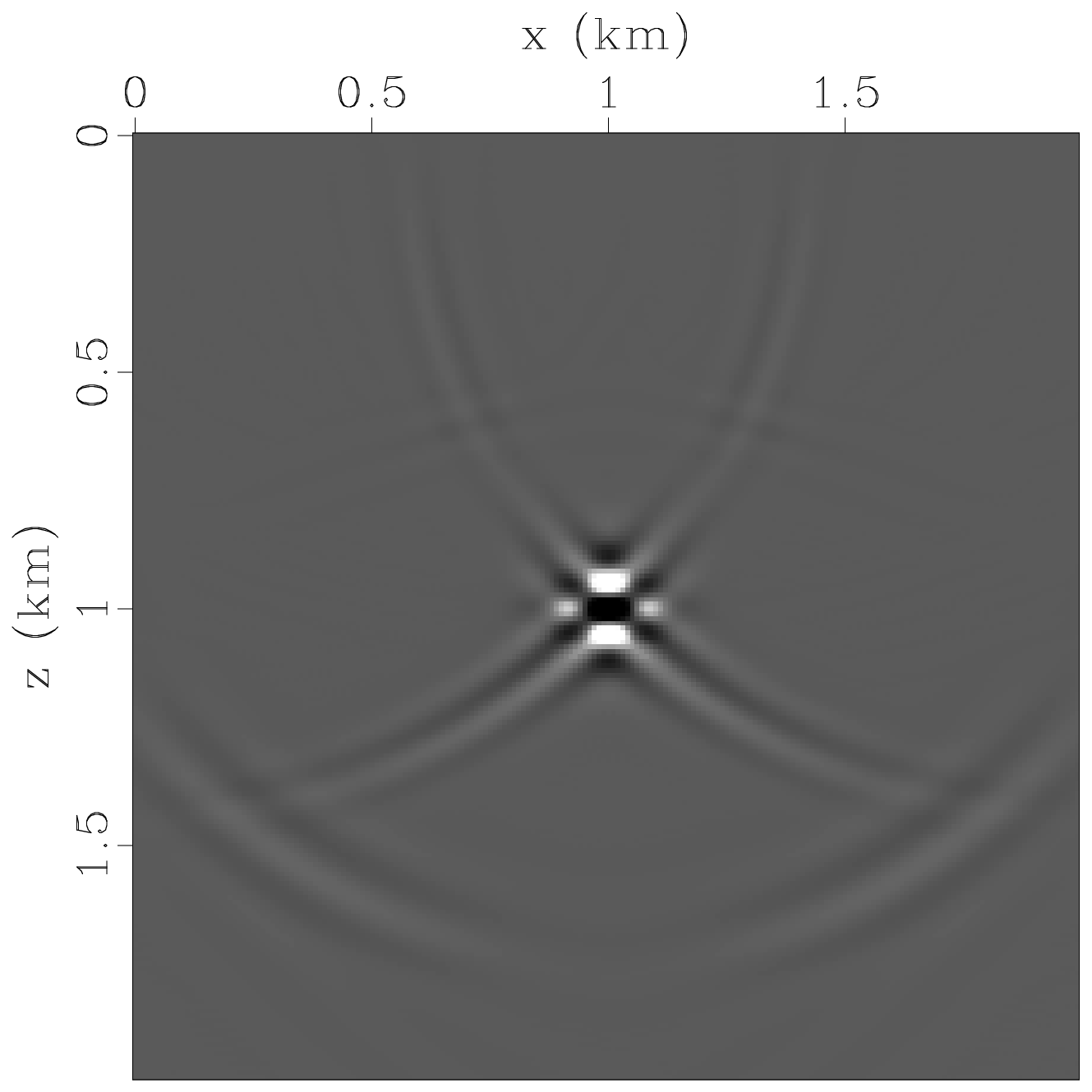}
  \caption{$\delta \beta, \left(-2,1,0\right)$}
  \label{fig:bs9}  
\end{subfigure}
\caption{Inverted source images with only surface receivers. Moment tensors $M_i={M_{11},M_{22},M_{12}}$ used for forward modeling: (a)(b) $M_1={1,1,0}$; (c)(d) $M_2={-1,-1,0}$; (e)(f) $M_3={0,0,1}$; (g)(h) $M_4={0,-1,0}$; (i)(j) $M_5={1,0,0}$; (k)(l) $M_6={0,1,0}$; (m)(n) $M_7={1,-2,0}$; (o)(p) $M_8={-1,2,0}$; (q)(r) $M_9={-2,1,0}$.}
\label{fig:images_sur}
\end{figure}

\begin{figure}
\begin{subfigure}{0.3\textwidth}
  \centering
  \includegraphics[width=1\linewidth]{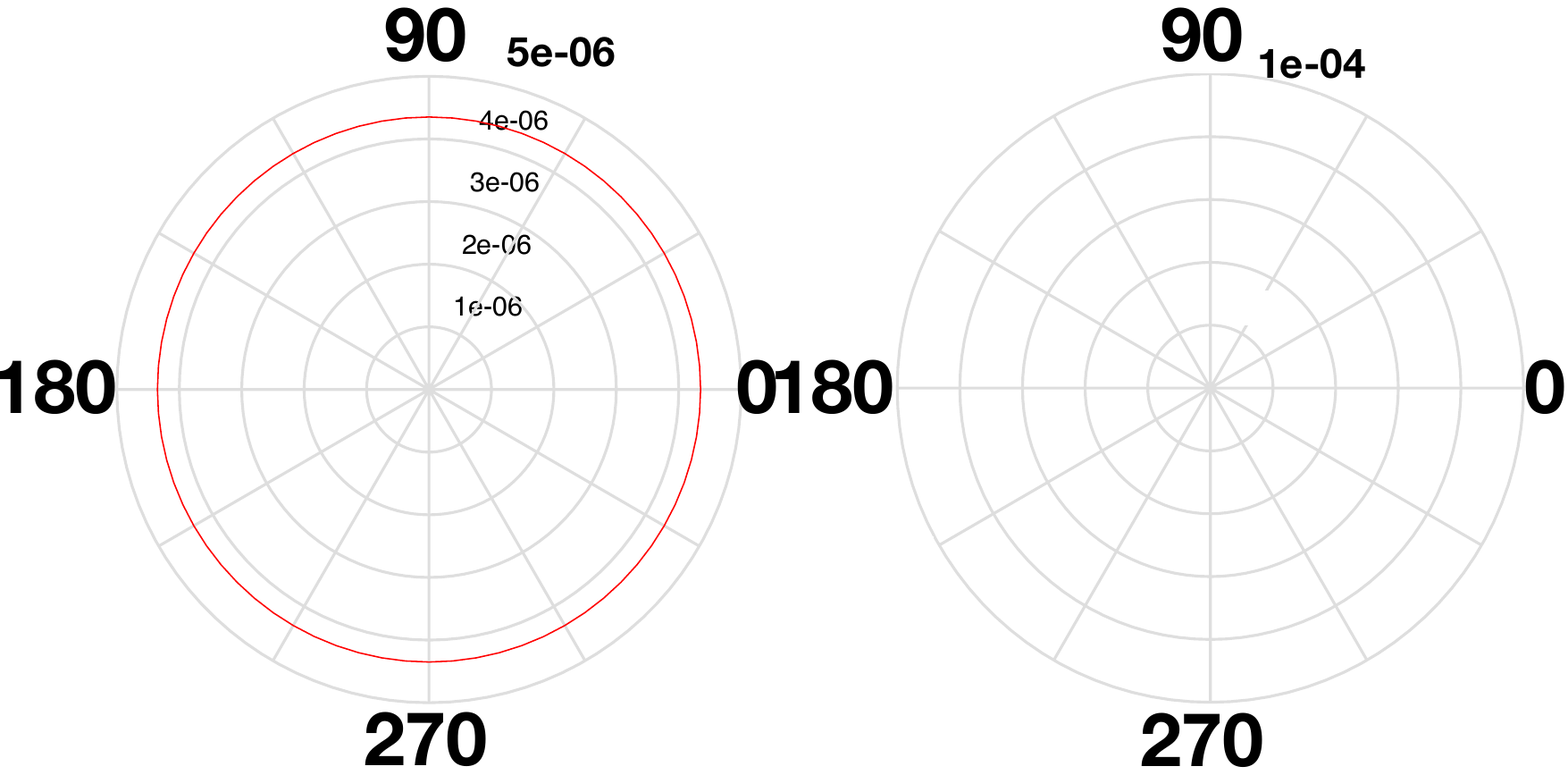}
  \caption{}
  \label{fig:pwsw1}
\end{subfigure}
\begin{subfigure}{0.3\textwidth}
  \centering
  \includegraphics[width=1\linewidth]{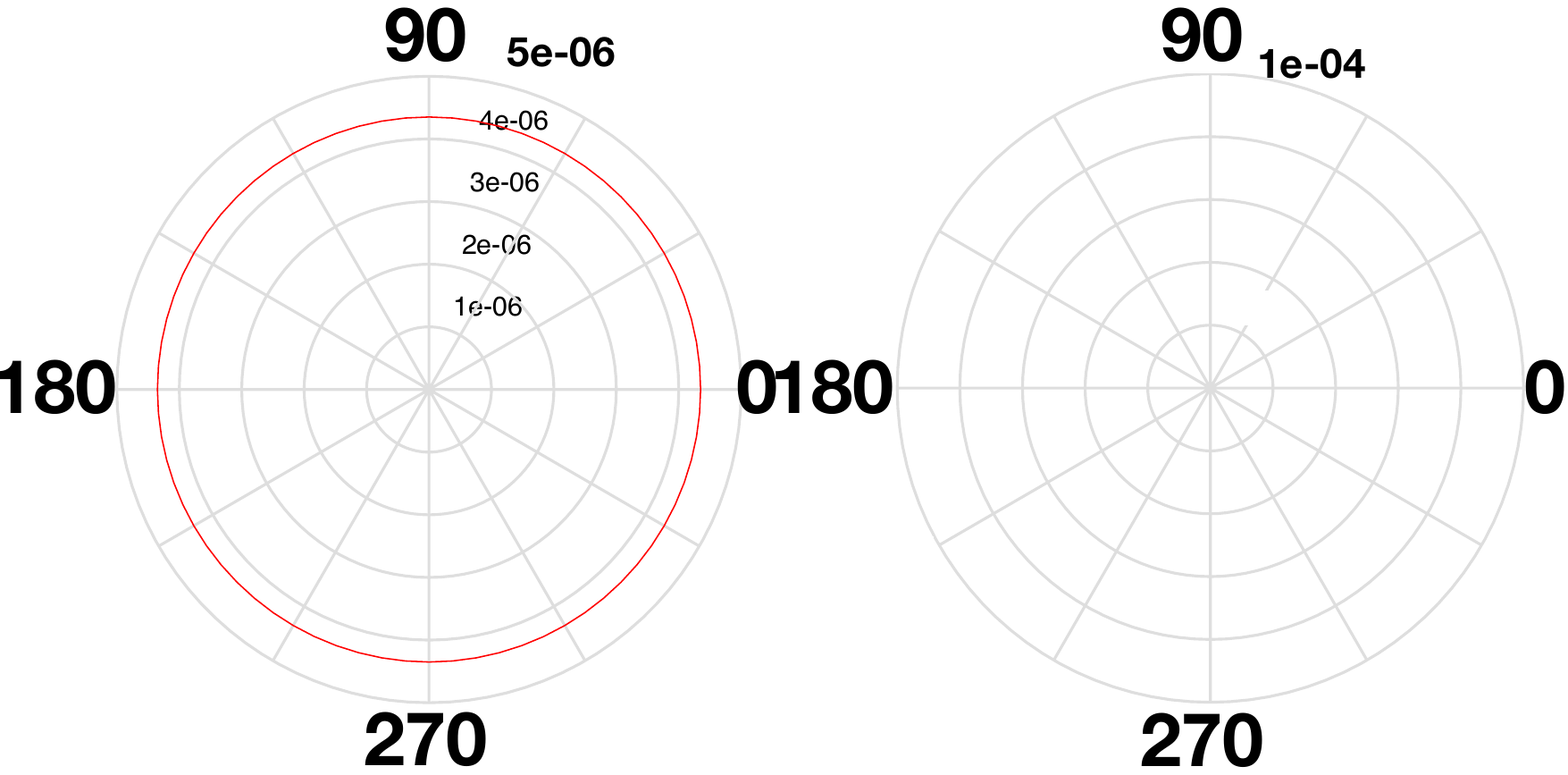}
  \caption{}
  \label{fig:pwsw2}
\end{subfigure}
\begin{subfigure}{0.3\textwidth}
  \centering
  \includegraphics[width=1\linewidth]{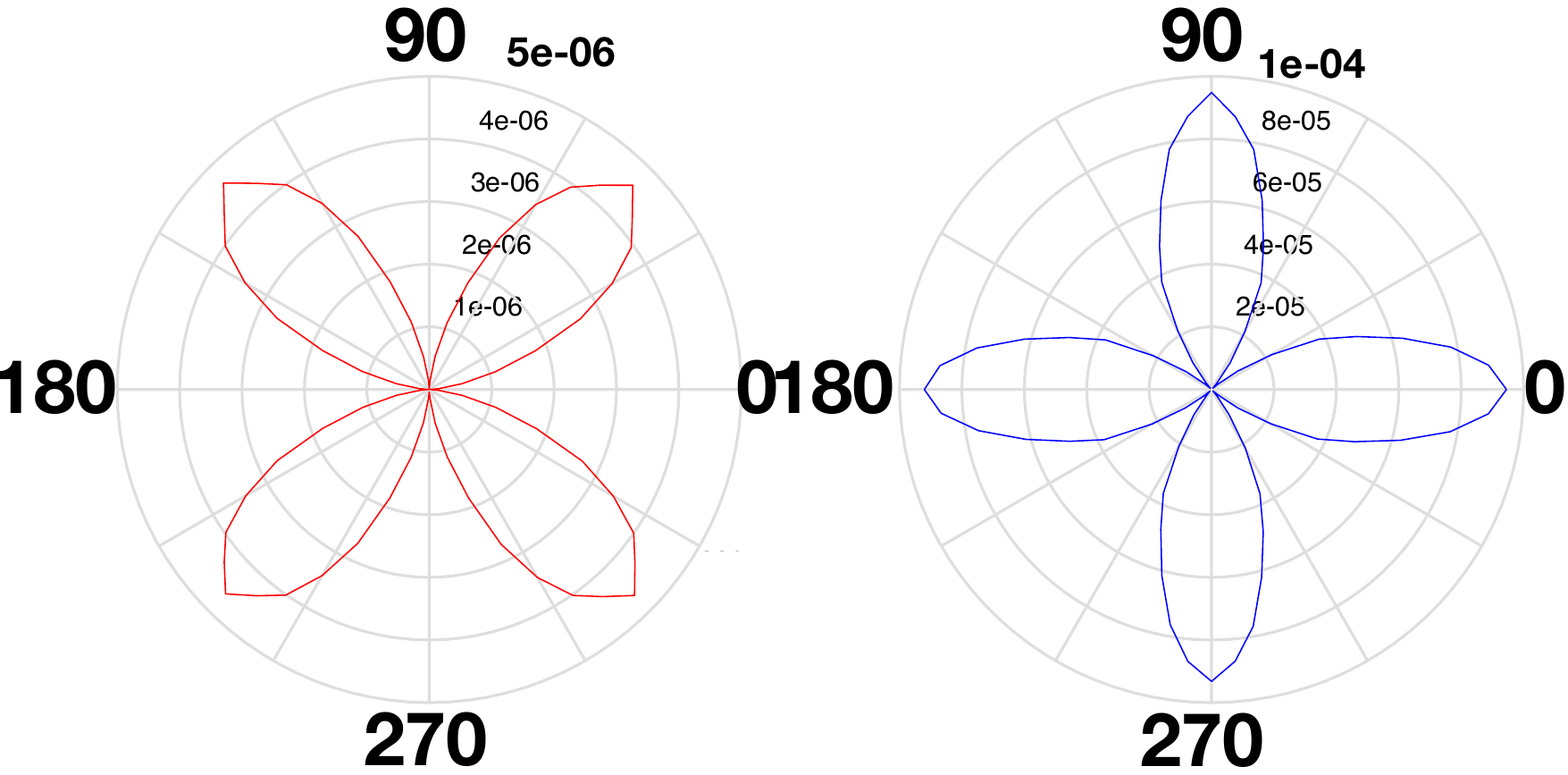}
  \caption{}
  \label{fig:pwsw3}
\end{subfigure}
\begin{subfigure}{0.3\textwidth}
  \centering
  \includegraphics[width=1\linewidth]{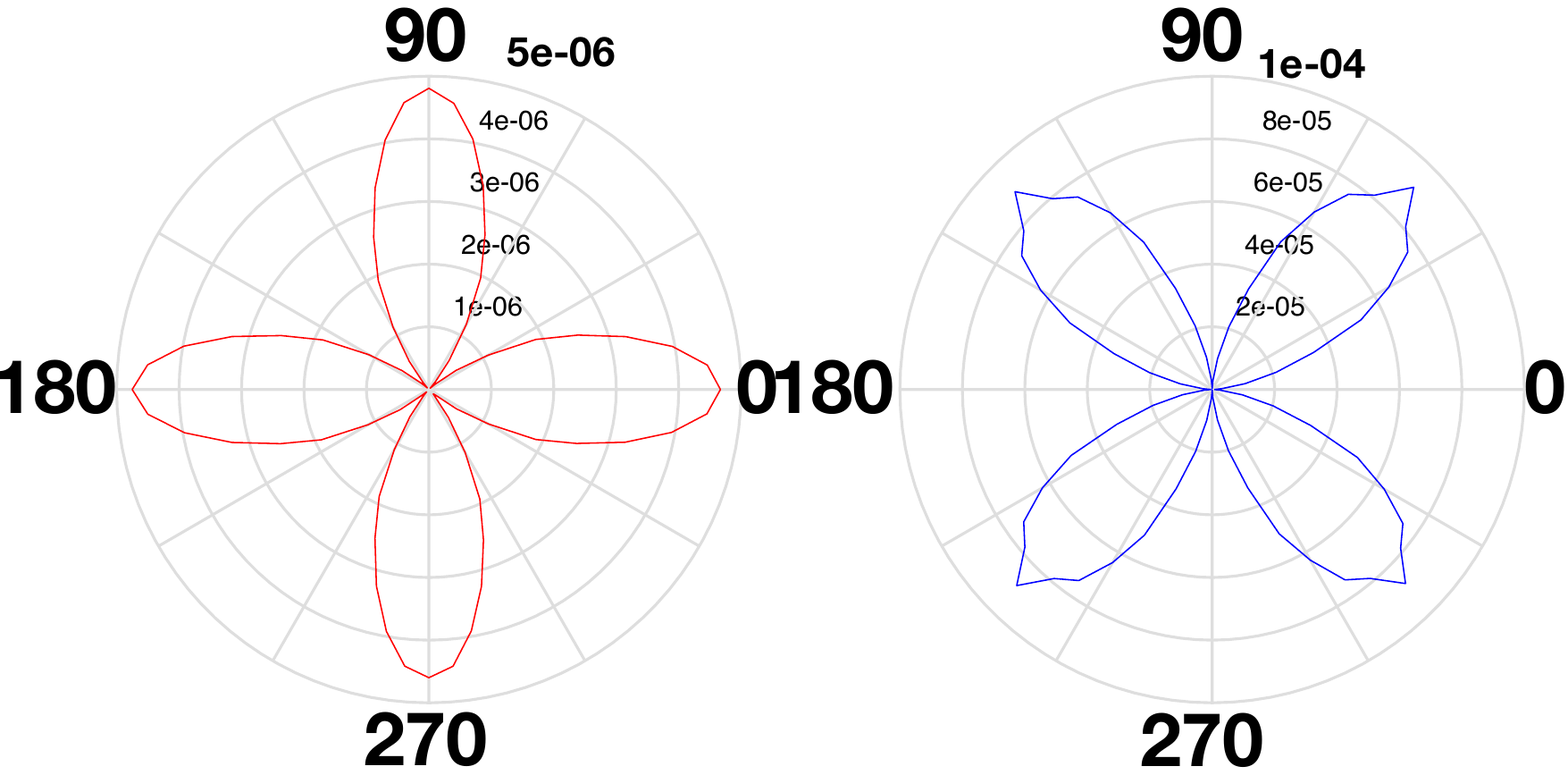}
  \caption{}
  \label{fig:pwsw4}
\end{subfigure}
\begin{subfigure}{0.3\textwidth}
  \centering
  \includegraphics[width=1\linewidth]{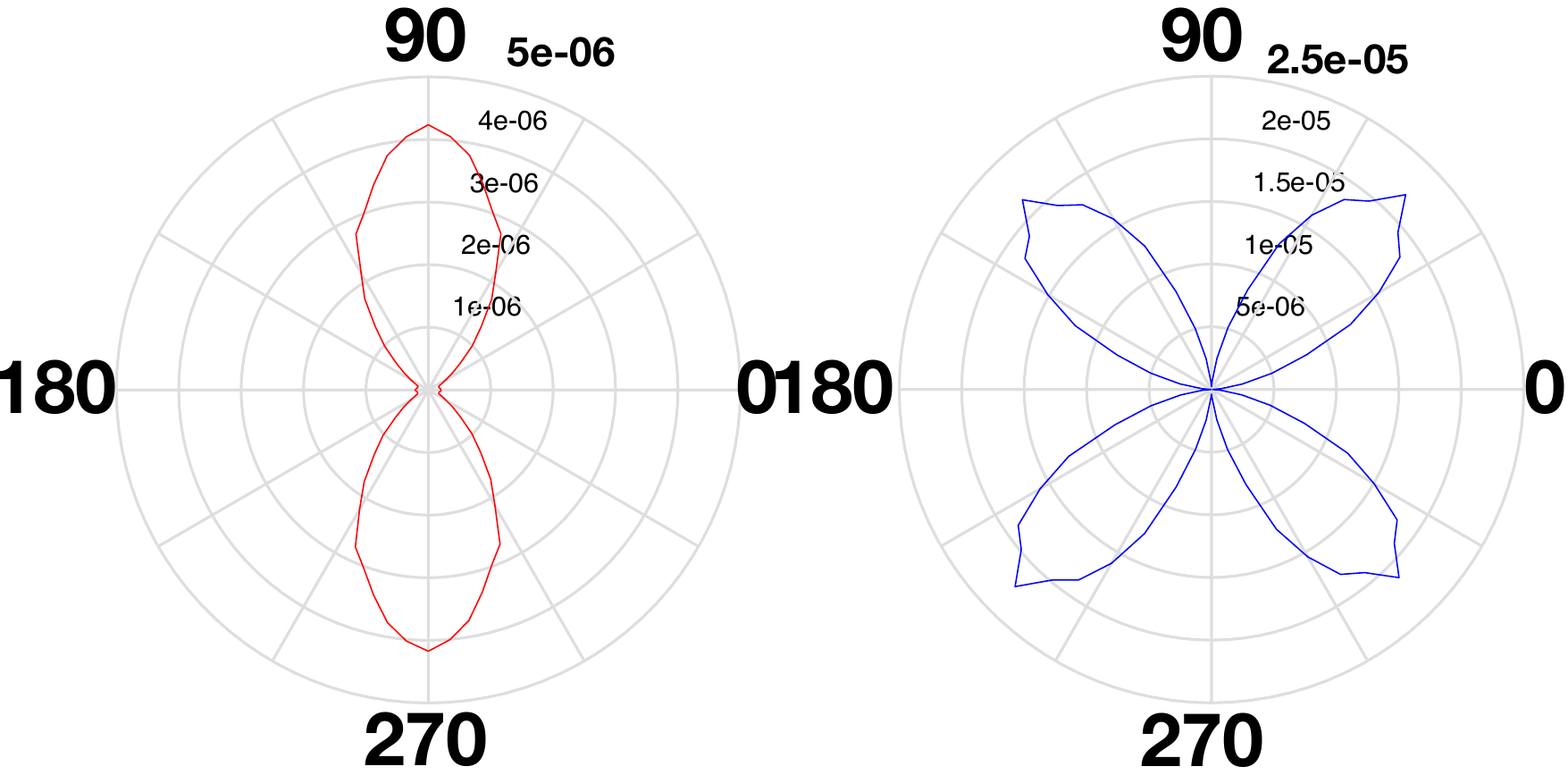}
  \caption{}
  \label{fig:pwsw5}
\end{subfigure}
\begin{subfigure}{0.3\textwidth}
  \centering
  \includegraphics[width=1\linewidth]{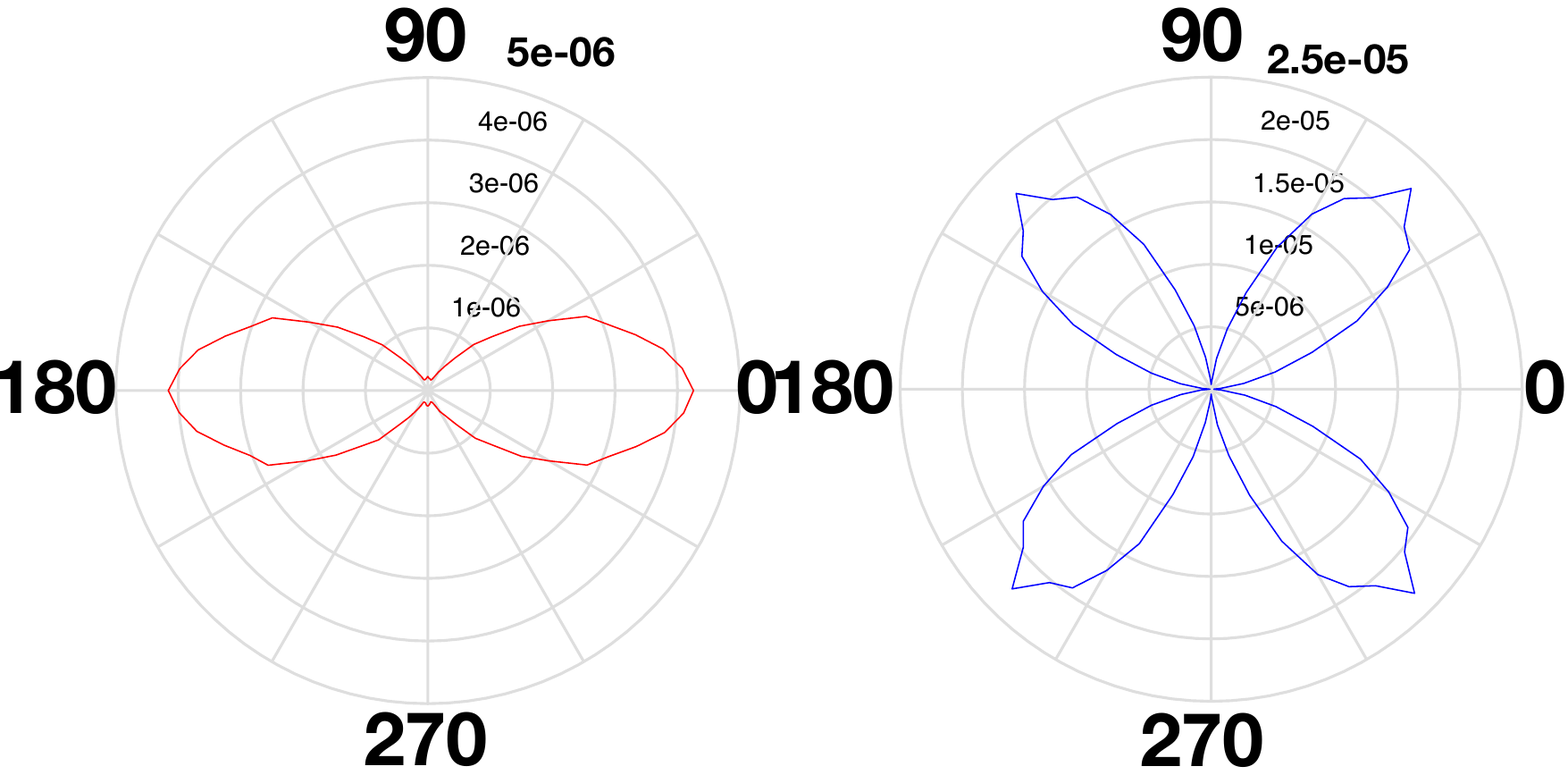}
  \caption{}
  \label{fig:pwsw6}
\end{subfigure}
\begin{subfigure}{0.3\textwidth}
  \centering
  \includegraphics[width=1\linewidth]{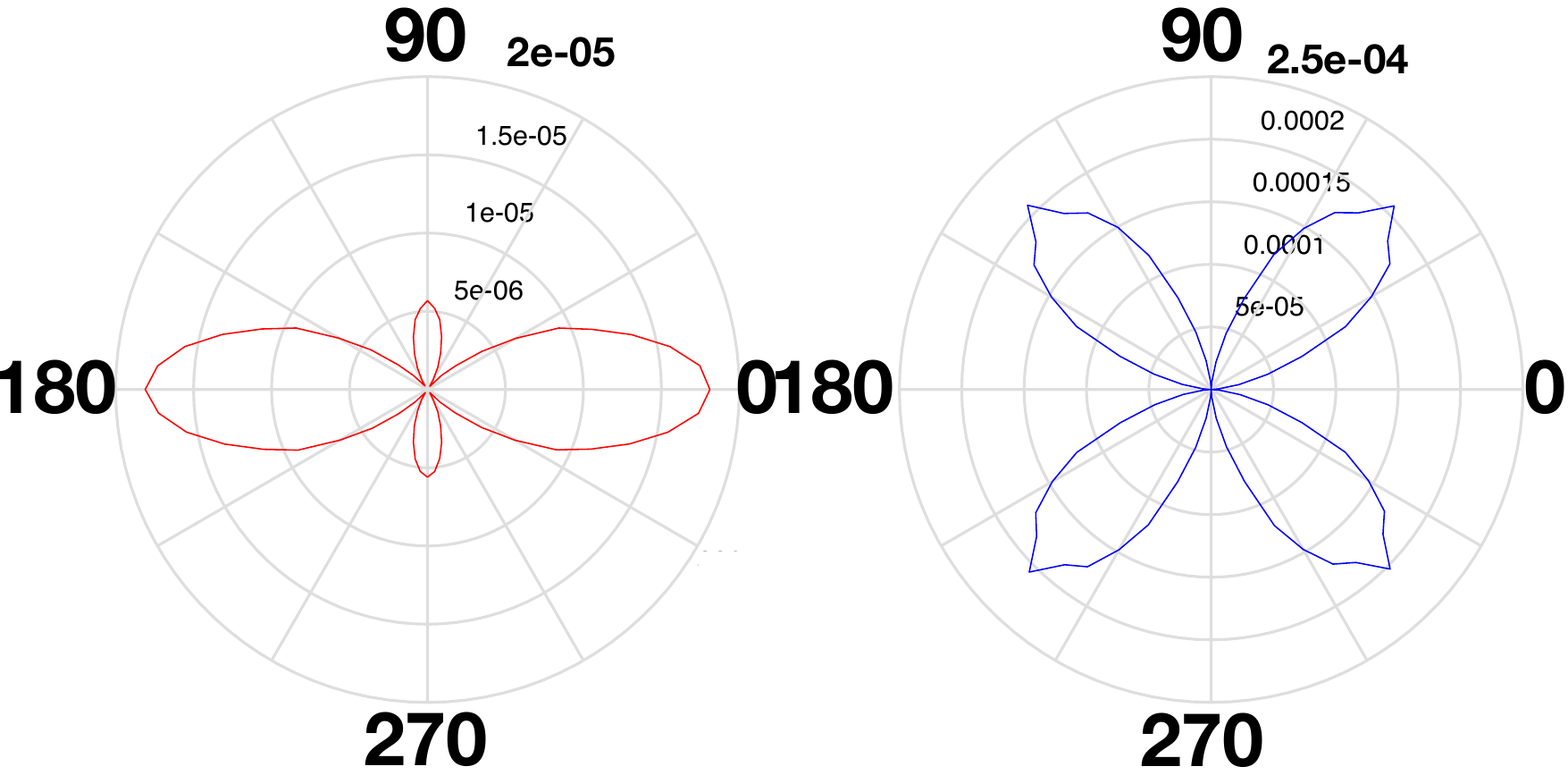}
  \caption{}
  \label{fig:pwsw7}
\end{subfigure}
\begin{subfigure}{0.3\textwidth}
  \centering
  \includegraphics[width=1\linewidth]{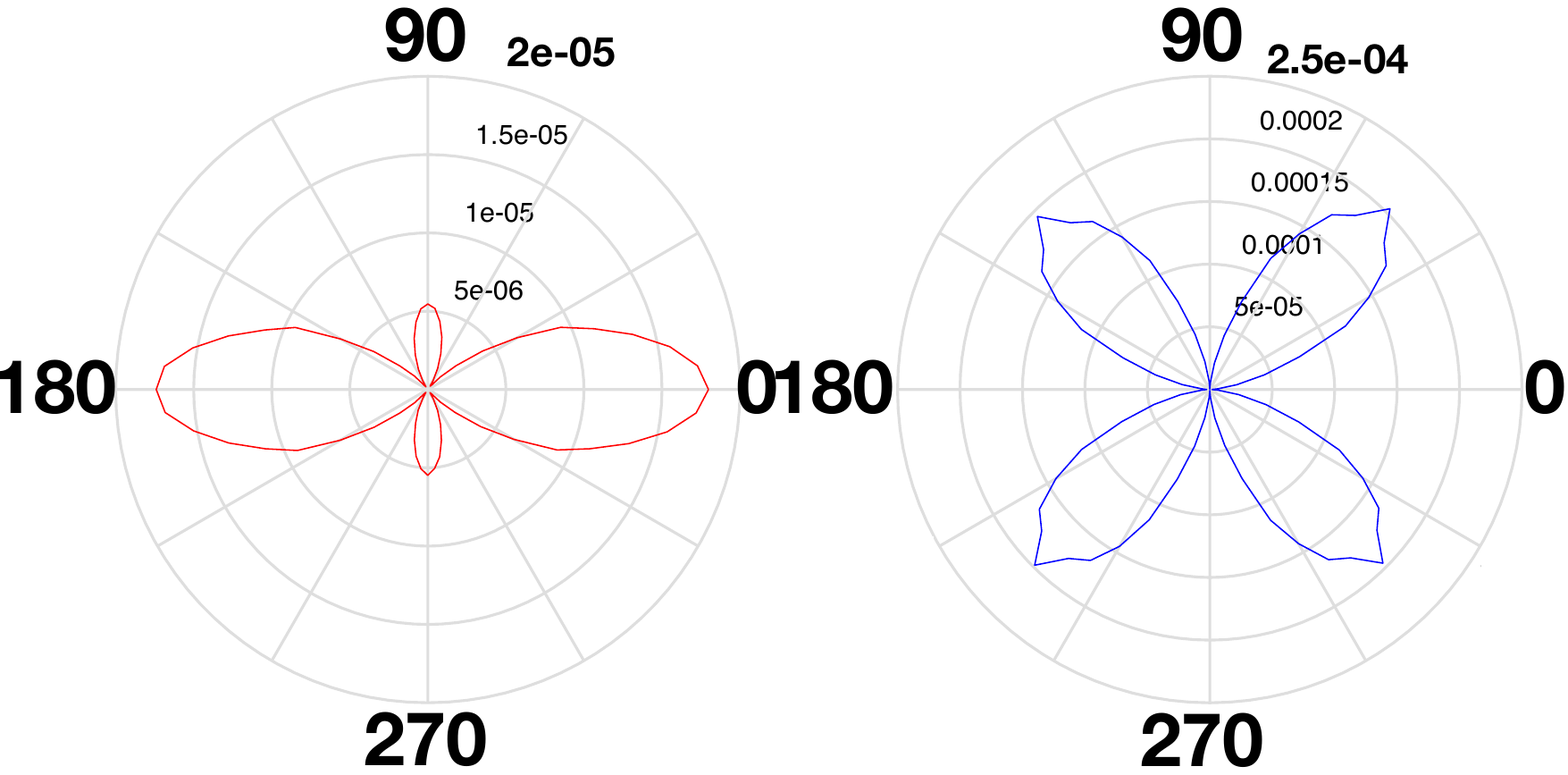}
  \caption{}
  \label{fig:pwsw8}
\end{subfigure}
\begin{subfigure}{0.3\textwidth}
  \centering
  \includegraphics[width=1\linewidth]{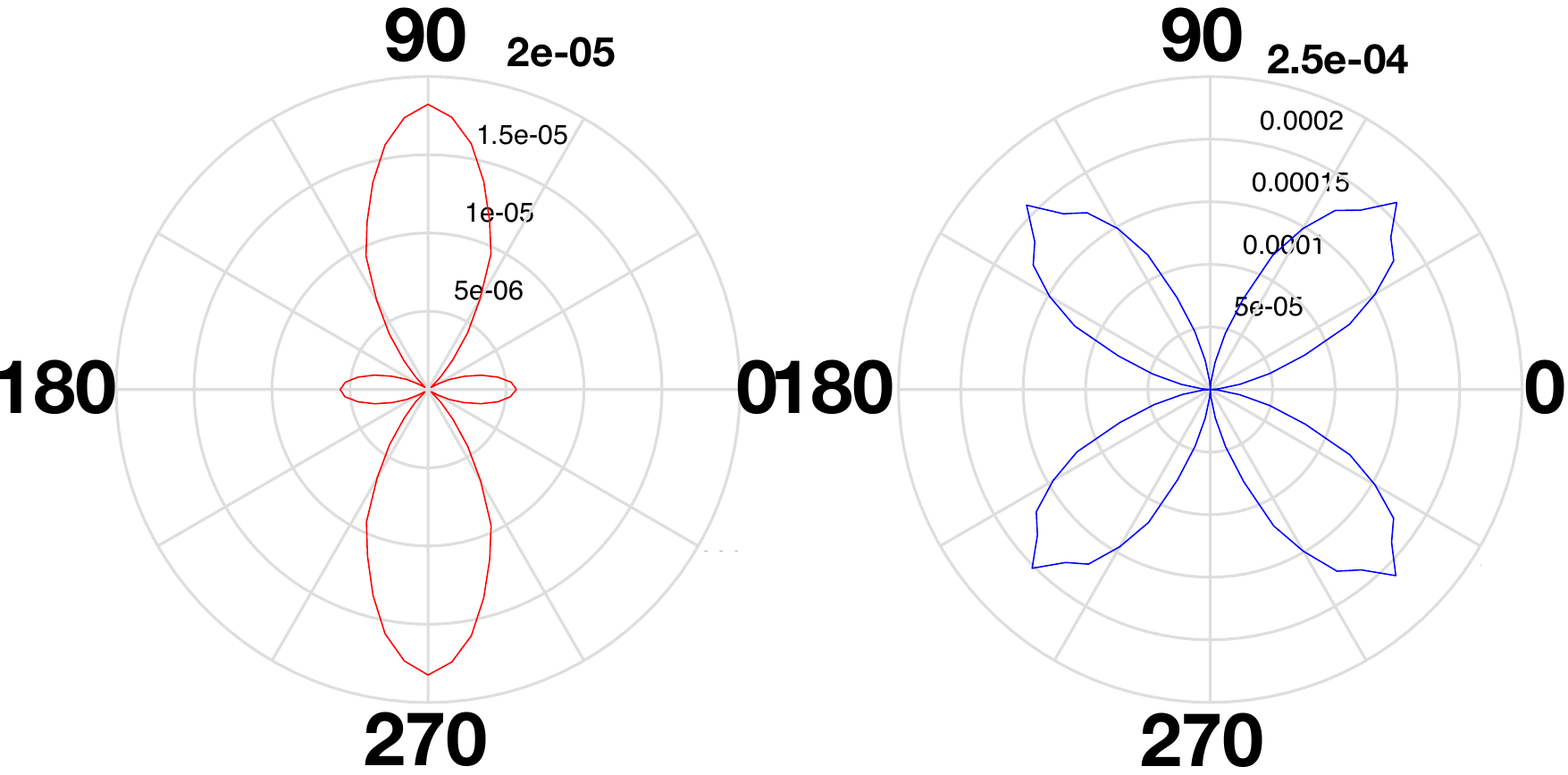}
  \caption{}
  \label{fig:pwsw9}
\end{subfigure}
\caption{Radiation pattern analysis of P- and S-wave with respect to aperture (acquisition angles). Red lines: P-wave sensitivity; blue lines: S-wave sensitivity. Moment tensors $M_i={M_{11},M_{22},M_{12}}$ used for forward modeling: (a) $M_1={1,1,0}$; (b) $M_2={-1,-1,0}$; (c) $M_3={0,0,1}$; (d) $M_4={1,-1,0}$; (e) $M_5={1,0,0}$; (f) $M_6={0,1,0}$; (g) $M_7={1,-2,0}$; (h) $M_8={-1,2,0}$; (i) $M_9={-2,1,0}$.}
\label{fig:radiation}
\end{figure}

In this section, we show the radiation pattern analysis of the source images, $\delta \alpha$ and $\delta \beta$, as well as the relationship between the source images and the seismic moment tensors, so that we can better understand the physical meaning of the source image term we introduced.
\newline
The most complete form used to describe the source mechanism is the seismic moment tensor. In 3D, nine force couples are used to fully represent every possible source \cite{aki2002quantitative}.
The moment tensor in 2D can be written in the following symmetric matrix form
\begin{equation}
\bf{ M} = {M_0}  \left( \begin{matrix} M_{ 11 } & M_{ 12 }  \\ M_{ 21 } & M_{ 22 }   \end{matrix} \right),
\label{eq:moment2d}
\end{equation}
where $M_{12}$ equals to $M_{21}$, as the system would otherwise be unstable and would rotate. $M_0$ refers to the moment magnitude. Thus we have three independent components in 2D to mathematically represent a seismic source mechanism.
\newline
In our proposed theory, $\delta \alpha$ and $\delta \beta$ can also be used to describe seismic source mechanisms because they are simply another formulation of equation~\ref{eq:equivalent_source}. Both equivalent source terms should have the same physical meaning. Since we merge the moment tensor and the space coordinates into a single term, the source images, we need to find out what exactly the source images represent. A good way to understand this representation is to find the connection between the source images and the well-known moment tensor. In order to understand the relationship between the two, a simple experiment is performed under ideal conditions. A source is ignited in the center of a 2D square homogeneous velocity model, with $V_p=2km/s$ and $V_s=\frac{2}{\sqrt{3}}km/s$. Every grid point on all four boundaries is set as a receiver, in order for the full Green's function to be measured. We use a moment tensor to describe the source in forward modeling, and then the event is recorded by the receivers. We subsequently perform an inversion of the source images using equations~\ref{eq:alpha_grd} and ~\ref{eq:beta_grd}, under the assumption that the original velocities and the original source time function are known. We can then compare the inverted results of source images with the moment tensor. The inverted source images are shown in Figure~\ref{fig:images_all}. There are other ways of representing the moment tensor with images. For example, a diffraction stack migration combines ray-theory Green's functions with a reverse time imaging method to simultaneous locate and characterize the seismic source properties \cite{chambers2014moment}. Although the idea seems quite similar to ours, the theory part is quite different. In this paper, we use our inverted source images to represent the source mechanism.
\newline
These inverted source images show different features, indicating different source types. The cases of $M_{1}$ and $M_{2}$ show an explosive and an implosive sources, which means the moment tensor is either a positive or a negative diagonal matrix. Both $\delta \alpha$ and $\delta \beta$, Figures~\ref{fig:a1} to~\ref{fig:b2}, show round dots at the source location, which means that the source is symmetric in all directions in space. In other words, Figures~\ref{fig:a1} to~\ref{fig:b2} show the cases of isotropic sources, just like what we used for forward modeling. For cases of $M_{3}$ and $M_{4}$ (Figures~\ref{fig:a3} to~\ref{fig:b4}), $\delta \alpha$ values equal to zero because according to equation~\ref{eq:alpha_grd}, $(w_{11}+w_{22})$ equals to $0$, and therefore, the whole gradient is zero. In the source images of double coupled (DC) and compensated-linear-vector-dipole (CLVD) sources $(M_3$ to $M_9)$, positive (white color) and negative (black color) values can be directly connected to the source type by looking at the sign distribution at the source location. This is because positive values mean that the wave is propagating outward, and negative values mean that the wave is propagating inward. On the other hand, the source signatures in these source images exhibit the same features with the beach ball plots of the same moment tensors. These observations lead us to the conclusion that source images in Figure~\ref{fig:images_all} represents components of a moment tensor. 
\newline
For $M_{1}$ and $M_2$ (Figures~\ref{fig:a1} to~\ref{fig:b2}), source images exhibit the same features, no matter whether the source is an implosive or explosive one, which means that we cannot tell the difference between the expansive and compressive sources. The same issue also shows in Figure~\ref{fig:a7} to~\ref{fig:b8}. Both cases have the similar features in $\delta \alpha$ and $\delta \beta$, but the moment tensors for the two sources have opposite signs. This is reasonable as the moment tensor in 2D case has three independent parameters $(M_{11}, M_{22}$ and $M_{12})$, whereas here, we only have two parameters $(\delta \alpha,\delta \beta)$. As a result, some information is lost.
\newline
In field experiments, we rarely get a perfectly illuminated acquisition system, unlike in the experimental setup described above. Therefore, here, we present a more realistic set of tests with surface receivers only. The other conditions, including the moment tensors, remain the same as in the previous tests.
\newline
In this set of tests, the source images share the same features with the source signatures and in almost all cases, the resolution along the horizontal direction is poor and makes it difficult to recognize what source types each case refers to. This is expected, since we have only surface receivers. We lose most of the source signature information from left, right and bottom. This issue could be even more severe in 3D cases with 2D plane receivers, because we should lose more information and six independent moment tensor components need to be represented by two source images. There is a limitation in identifying source types, when using these source images due to the fact they highly rely on with acquisition systems (illumination). 
\newline
To understand the sensitivity of source images with respect to illumination, a radiation pattern analysis in $2D$ case is presented here, with radiation pattern plots shown in Figure~\ref{fig:radiation}.
\newline
With these radiation pattern analysis results, we may recognize whether the source image difference is identifiable or not under certain acquisition configuration. For example, in cases $3$ and $4$, both P- and S-waves show different sensitivities from angle $-45(315)$ to $45$ degrees, which reflect the presence of surface receivers only. Relatively large differences in source images are observed, which allow us to identify the different source types (Figures~\ref{fig:as3} to ~\ref{fig:bs4}). However, we cannot detect any difference between cases 5 and 9 (surface receivers only) or between cases 6, 7 and 8 (vertical well receivers only), since they share similar sensitivities at certain angles.
\newline
Although the source images indicate the source types, we still can hardly rely on these parameters to identify which types they are, as the source images highly depend on the acquisition aperture and inversion quality of the other model parameters than source images, which is the trade-off commonly experienced in FWI. With these tests and analysis, we now understand the physical meaning of the source images. The most reliable information provided by the source images, which is also the most important one, is the source location in space. Besides the spatial locations, $\delta \alpha$ and $\delta \beta$ also represent the source mechanism to a certain extent (when the model parameters are good enough and the acquisition is dense). If additional information is needed, such as the moment tensor, one can first locate the source by the proposed method according to the inverted source images, and then invert the moment tensor with a conventional seismology method \cite{droujinine2011elastic}. 
\section{Velocity Update With Source Function Independent Objective Function}
The velocity models can be updated by minimizing the data misfit in equation~\ref{eq:obj_func}. The gradients with respect to $\alpha$ and $\beta$ can be written as:
\begin{equation}
\nabla E_{\alpha}=-\int_t dt (\frac{\partial u_x}{\partial x}+\frac{\partial u_z}{\partial z})(\frac{\partial \hat u_x}{\partial x}+\frac{\partial \hat u_z}{\partial z});
\label{eq:E_alpha}
\end{equation}
\begin{equation}
\nabla E_{\beta}=-\int_t dt (2\frac{\partial u_x}{\partial x} \frac{\partial \hat u_x}{\partial x})+(2\frac{\partial u_z}{\partial z} \frac{\partial \hat u_z}{\partial z})+(\frac{\partial u_x}{\partial z}+\frac{\partial u_z}{\partial x})(\frac{\partial \hat u_x}{\partial z}+\frac{\partial \hat u_z}{\partial x}).
\label{eq:E_beta}
\end{equation}
In case of a poor initial guess of the velocity model, the initial source terms will be far from the true ones in both space and time, especially the source origin time. Using such an inaccurate source function as the initial input in the inversion is not acceptable because a slight time shift in the source origin time leads to severe cycle skipping problems in the data. To avoid such cycle skipping problems, we can use some advanced objective functions such as the adaptive waveform inversion introduced by \cite{sun2019adaptive}. We propose to use a convolution based objective function \cite{choi2011source,wang2018microseismic}, which has source time function independency, as we mitigate the role of the source origin time in the objective function. We also add a model total variation (TV) term as a regularization in the objective function. Thus, the whole objective function can be written as:
\begin{equation}
{ E_{indp} }\left( m \right) = \frac { 1 }{ 2 } \sum _{ r }{ \int { \left\| \textbf u \ast { \textbf {d} }_{ ref }-\textbf {d} \ast { \textbf u }_{ ref }  \right\|^2_2 dt }  } + \lambda  \sqrt{\frac{\partial m}{\partial x}^2 + \frac{\partial m}{\partial z}^2 + \epsilon ^2} ,
\label{eq:indp_obj}
\end{equation}
where the subscript ${\ast}_{ ref }$ represents the reference traces in both the true and the predicted data. $\lambda$ is a factor that balances the weight of the regularization term. $\epsilon$ is a factor that controls the smoothness of the gradient. The larger $\epsilon$ is, the smoother the gradient will be. The symbol $*$ is the convolution operator. In practice, our experimentation suggests that a near-offset trace is a better reference than a far-offset trace. The TV term makes the gradients blocky. However, in our experiment, the main goal of adding this TV regularization is to obtain smooth velocity updates in every iteration with a large $\epsilon$. In equation~\ref{eq:indp_obj}, a new source function that contains both the predicted and the true data source function information is convolved in both terms equally, which means that the effect of the delay in the source origin time is muted. Thus, the true source origin time is no longer necessary. The velocity gradients of the objective function are calculated by minimizing equation~\ref{eq:indp_obj} with respect to $\alpha$ and $\beta$. The gradients can then be given as (see Appendix B):
\begin{equation}
\frac { \partial E_{indp} }{ \partial V } = \sum _{ i }^{ nr }{ \left[ \left( \frac { \partial { u _ i } }{ \partial V } *{ \textbf {d} }_{ ref } \right) \cdot \textbf {r}_{ i }-\left( { \textbf {d}_{ i } }*\frac { \partial { \textbf u }_{ ref } }{ \partial V }  \right) \cdot \textbf {r}_{ i } \right]} + \lambda \nabla \cdot \left(\frac{\nabla V}{ \sqrt{\frac{\partial V}{\partial x}^2 + \frac{\partial V}{\partial z}^2 + \epsilon ^2} }\right).
\label{eq:vel_grd_cnv}
\end{equation}
The adjoint wavefields for the gradient calculation are thus simulated by injecting the adjoint sources:
\begin{equation}
\textbf {r}^{ (1) }_{ i,j }=\textbf {d}_{ i,ref }\bigotimes \left( { \textbf u }_{ i,j }\left( { x }_{ r },t \right) *{ \textbf {d} }_{ i,ref }-{ \textbf {d} }_{ i,j }*{ \textbf u }_{ i,ref }\left( { x }_{ ref },t \right)  \right) ,
\label{eq:new_resi1}
\end{equation}
\begin{equation}
\textbf {r}^{ (2) }_{ i,j }=-\textbf {d}_{ i,j }\bigotimes \left( { \textbf u }_{ i,j }\left( { x }_{ r },t \right) *{ \textbf {d} }_{ i,ref }-{ \textbf {d} }_{ i,j }*{ \textbf u }_{ i,ref }\left( { x }_{ ref },t \right)  \right) ,
\label{eq:new_resi2}
\end{equation}
at the $i^{th}$ receiver locations and the reference trace location, respectively, where $\bigotimes$ refers to the cross-correlation operation in time. The two seismograms in equations~\ref{eq:new_resi1} and~\ref{eq:new_resi2} are back-propagated at meanwhile to keep the computational cost equivalent to FWI.
\newline
A nested approach is used as the inversion strategy, i.e. $(1)$ updating the source images with equations~\ref{eq:alpha_grd} and~\ref{eq:beta_grd} (for five iterations and this was decided by experimentation); $(2)$ updating the velocities using equation~\ref{eq:vel_grd_cnv} with the adjoint sources equations~\ref{eq:new_resi1} and~\ref{eq:new_resi2} (for ten iterations); $(3)$ updating $\textbf{w}$ using equation~\ref{eq:E_wt} (for five iterations). We repeat the above steps until convergence is achieved, which, in our case, means the objective function is no longer decreasing. Details are discussed in the implementation section below with the workflow.

\section{Numerical Examples}
\begin{figure}
\centering
\begin{subfigure}{0.48\textwidth}
  \centering
  \includegraphics[width=1\linewidth]{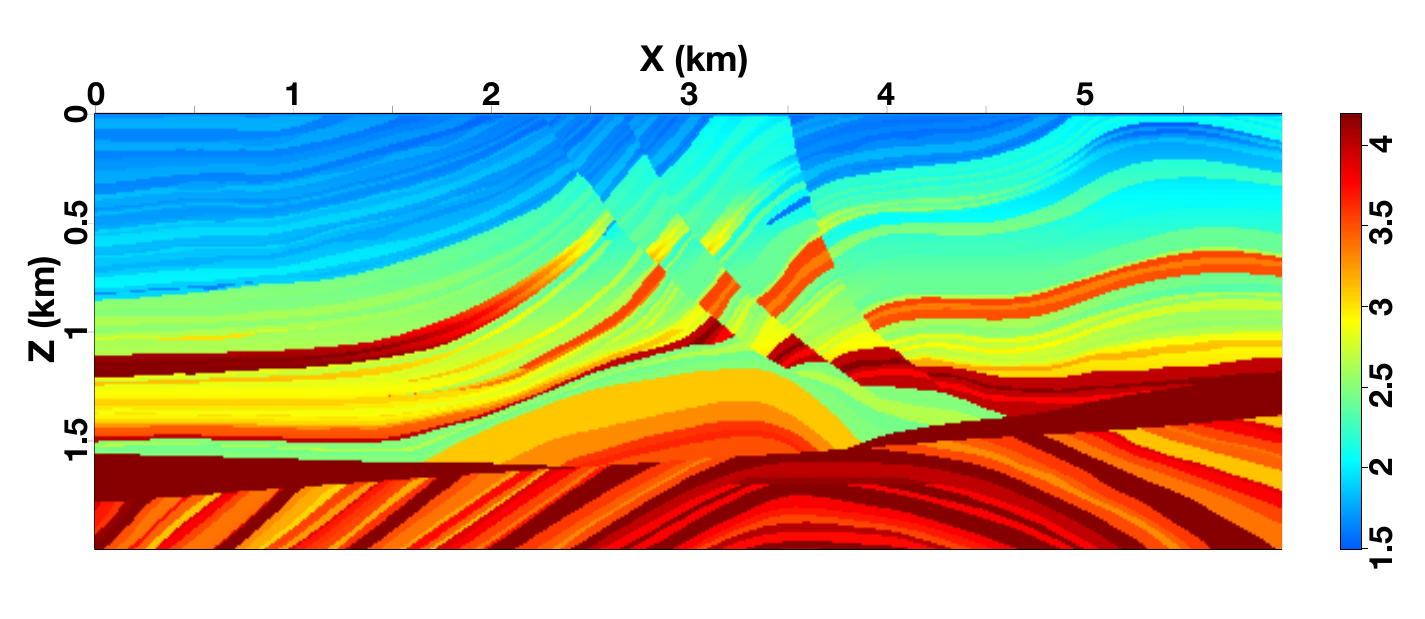}
  \caption{}
  \label{fig:vptrue}
\end{subfigure}
\begin{subfigure}{0.48\textwidth}
  \centering
  \includegraphics[width=1\linewidth]{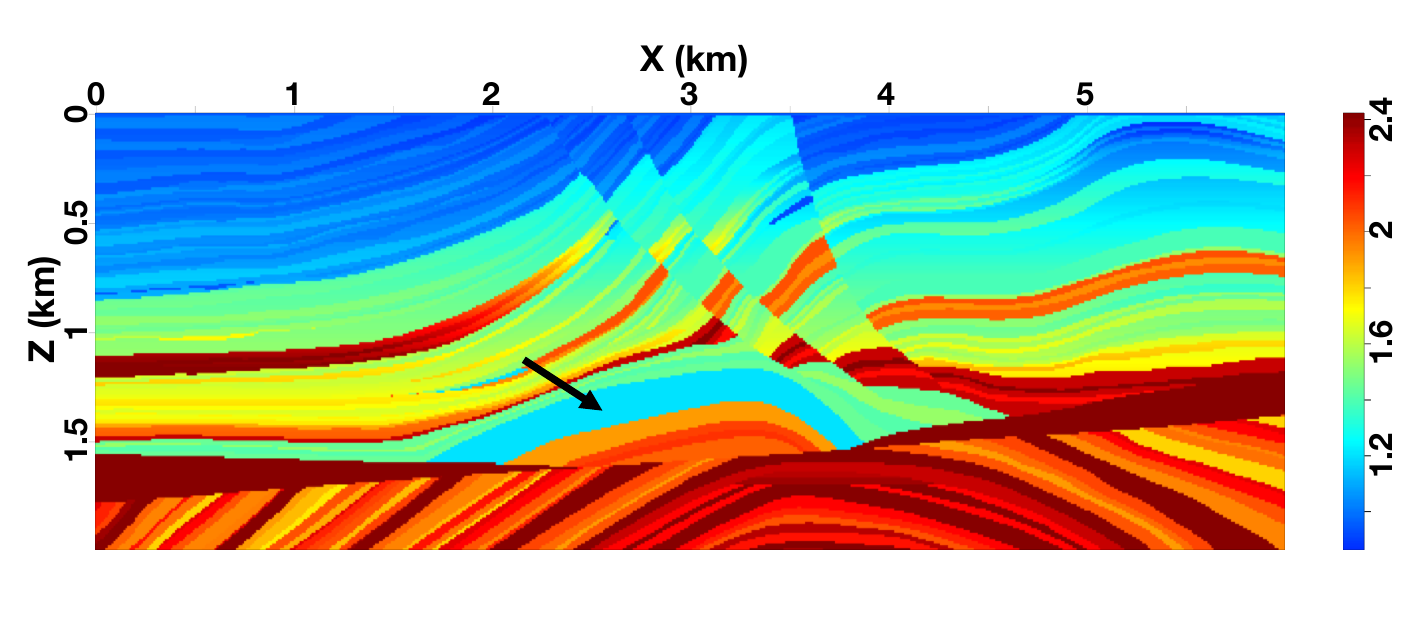}
  \caption{}
  \label{fig:vstrue}
\end{subfigure}
\begin{subfigure}{0.48\textwidth}
  \centering
  \includegraphics[width=1\linewidth]{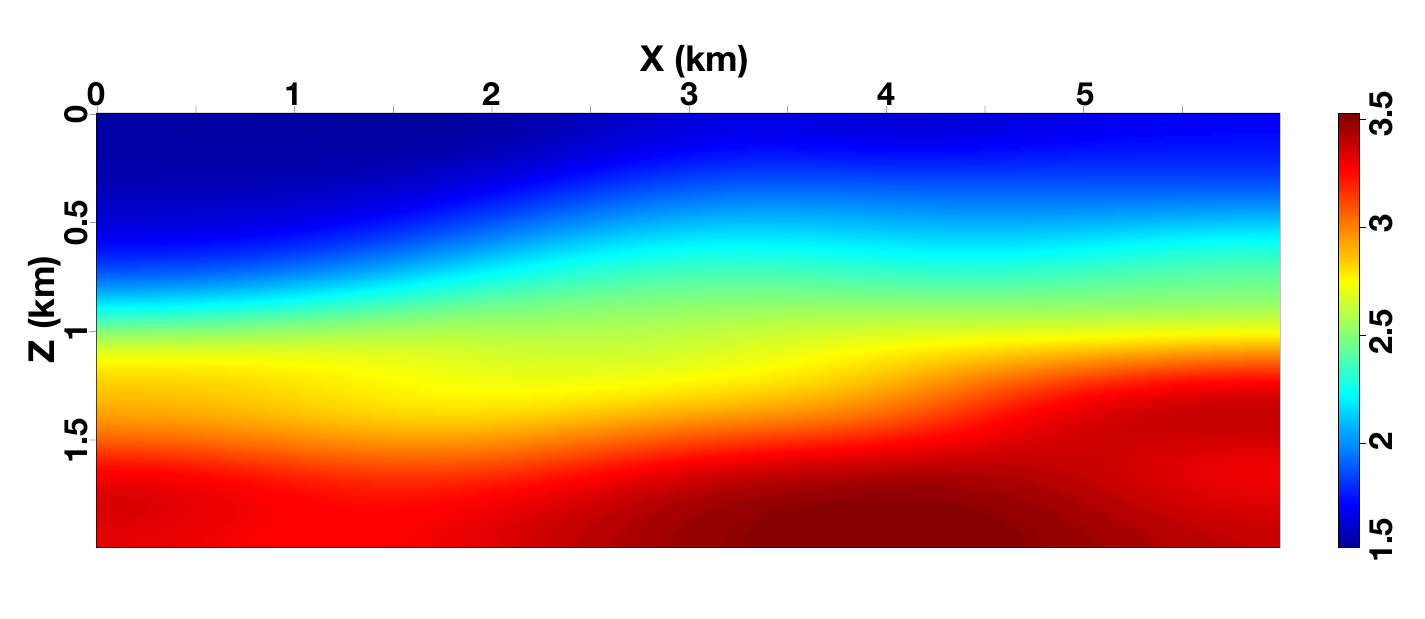}
  \caption{}
  \label{fig:vpini}
\end{subfigure}
\begin{subfigure}{0.48\textwidth}
  \centering
  \includegraphics[width=1\linewidth]{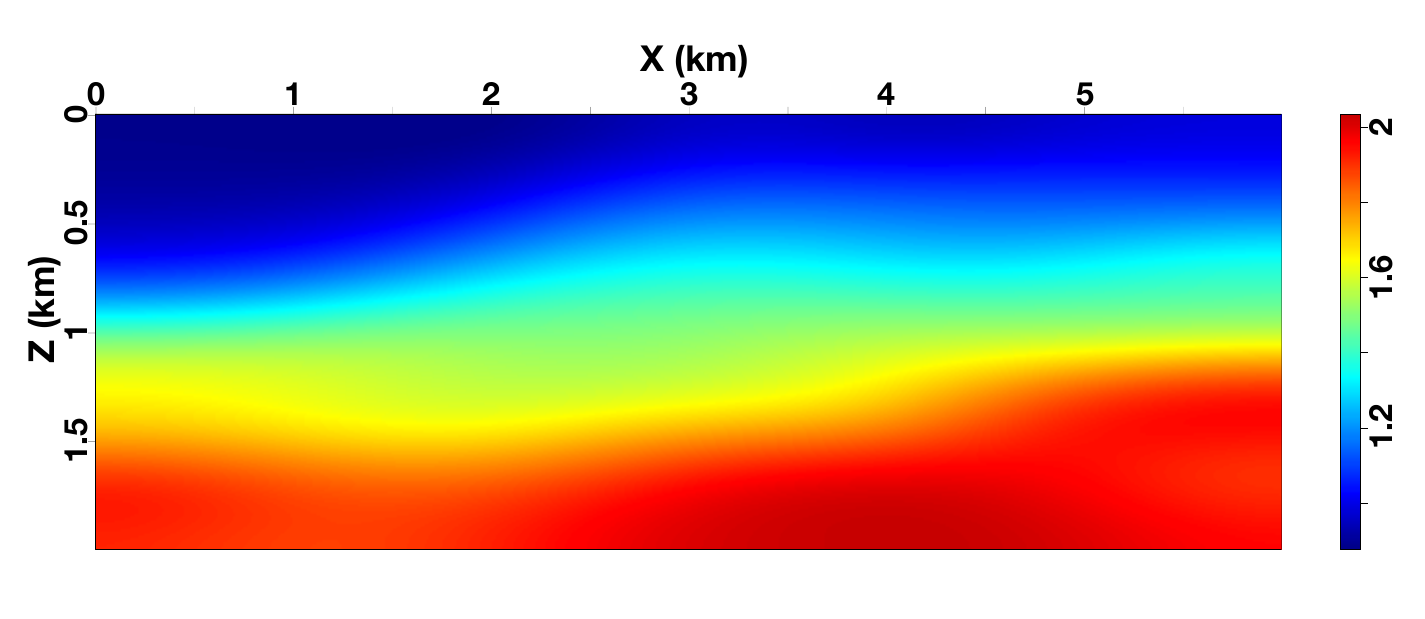}
  \caption{}
  \label{fig:vsini}
\end{subfigure}
\begin{subfigure}{0.48\textwidth}
  \centering
  \includegraphics[width=1\linewidth]{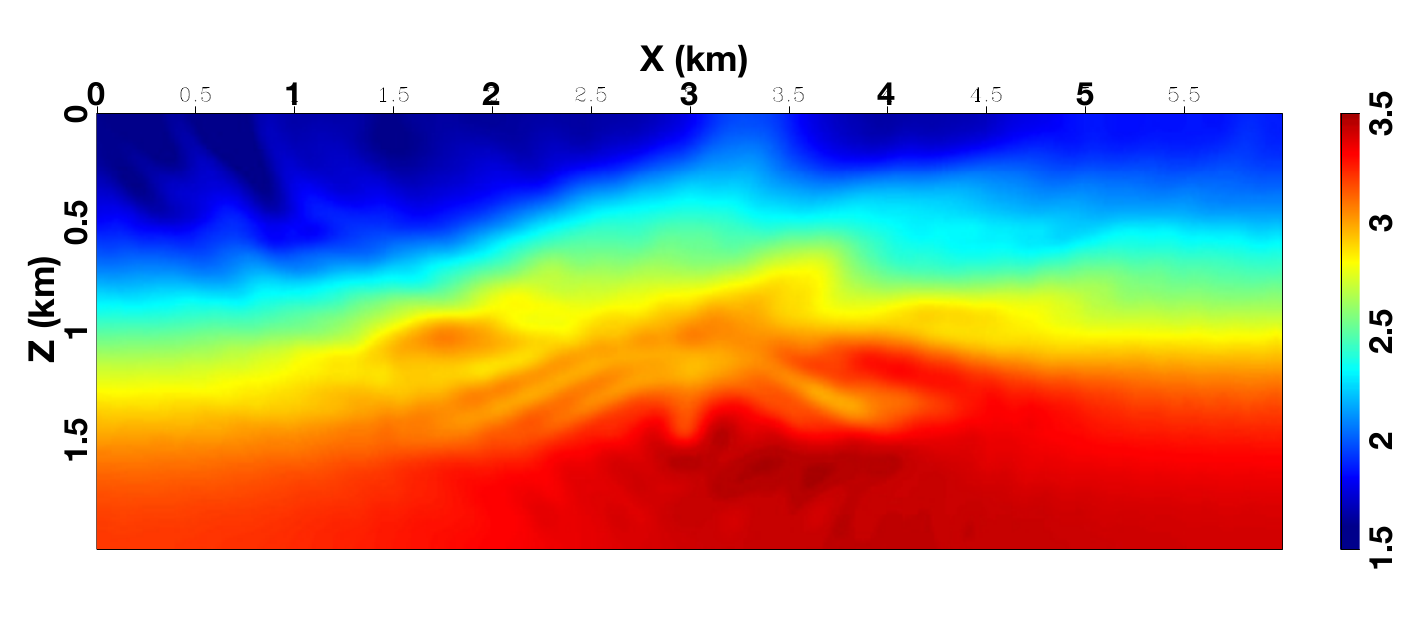}
  \caption{}
  \label{fig:vpinv}
\end{subfigure}
\begin{subfigure}{0.48\textwidth}
  \centering
  \includegraphics[width=1\linewidth]{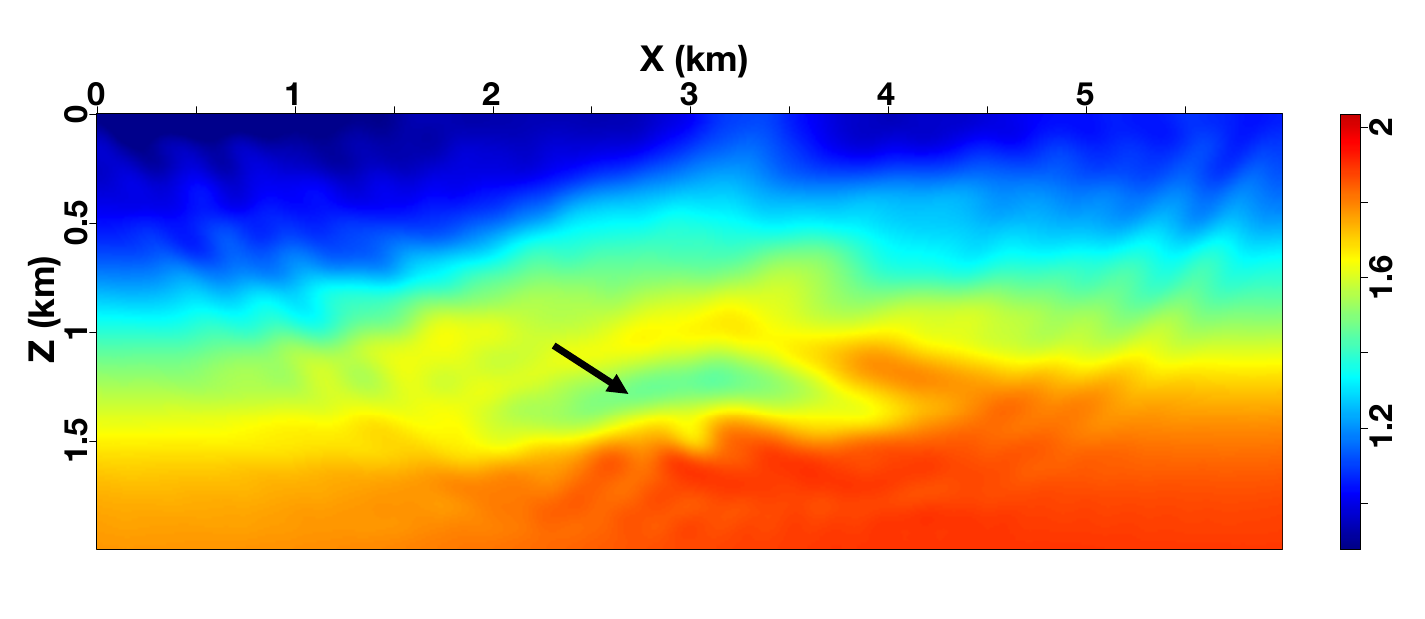}
  \caption{}
  \label{fig:vsinv}
\end{subfigure}
\caption{Original (a) Vp and (b) Vs 2D velocity profiles; initial (c) Vp and (d) Vs; inverted (e) Vp and (f) Vs 2D velocity profiles.}
\label{fig:velocity}
\end{figure}

In this section, we establish our work on a modified elastic Marmousi model whose S-wave velocity is modified by adding a half-circle-shaped low velocity layer (representing the hydraulic injected area) pointed out by an arrow in Figures~\ref{fig:vptrue} and~\ref{fig:vstrue}. Five sources are separately triggered at locations $x(i)=2.4+0.2(i-1) $ km, $i=1,...,5, z=1.55$ km.
\newline
The model size is $6$km by $2$km with a $0.01$km spatial interval in the $X$ and $Z$ directions, respectively. Receivers are set on all grid points on the top surface and in a vertical well, which locates from $(x=1,z=0.5)$ km to $(x=1,z=1.5)$ km in space with the same $0.01$km receiver interval. Totally $600$ receivers on the top surface and $100$ receivers in the well work as our monitoring system. We use a $10$Hz dominant frequency Ricker wavelet as the true source function. In this case, five grid points represent the minimum wavelength. The record has been convolved with a high-pass filter with the minimum frequency of $3$Hz to make the experiment more realistic. The initial $\alpha$ and $\beta$ are generated by applying strong smoothing to the true velocities and then making $\alpha$ and $\beta$ $10\%$ slower. As a result, shown in Figures~\ref{fig:vpini} and~\ref{fig:vsini}, the initial models are almost linearly increasing with depth. The smoothing operator here is a Gaussian smoothing window with a length of 300m in the X and 100m in the Z directions and we repeat the smoothing for 3 times. The $10\%$ deviation makes the starting model here even harder to invert than the linear increasing velocity model, which is widely used for Marmousi inversion experiments. We intended to show the effectiveness of our method using this initial model. When we generate the initial $V_s$ model, we do not include the modified hydraulic injected area information in the smoothing because it is our target zone, which is the low velocity zone pointed by an arrow in Figure~\ref{fig:vstrue}. The record length is 5 seconds with a $1$ms temporal sampling interval. The top surface is considered as a free surface boundary and the other three boundaries have absorbing boundary conditions applied. The modeling is implemented using a finite difference algorithm.
\newline
It is assumed that the five sources are temporally separated, which allows us to separate the event from each other. This is commonly satisfied in the field monitoring. Each source has a unique moment tensor and origin time. Moment tensors $M_1$ to $M_5$ are set to be the source types of the five sources, where the first source on the left uses $M_3$, the second source uses $M_4$, the third source uses $M_1$, the fourth source uses $M_2$ and the last source uses $M_5$. 
\newline
\subsection{Implementation}
\begin{figure}
  \centering
  \includegraphics[width=1\textwidth]{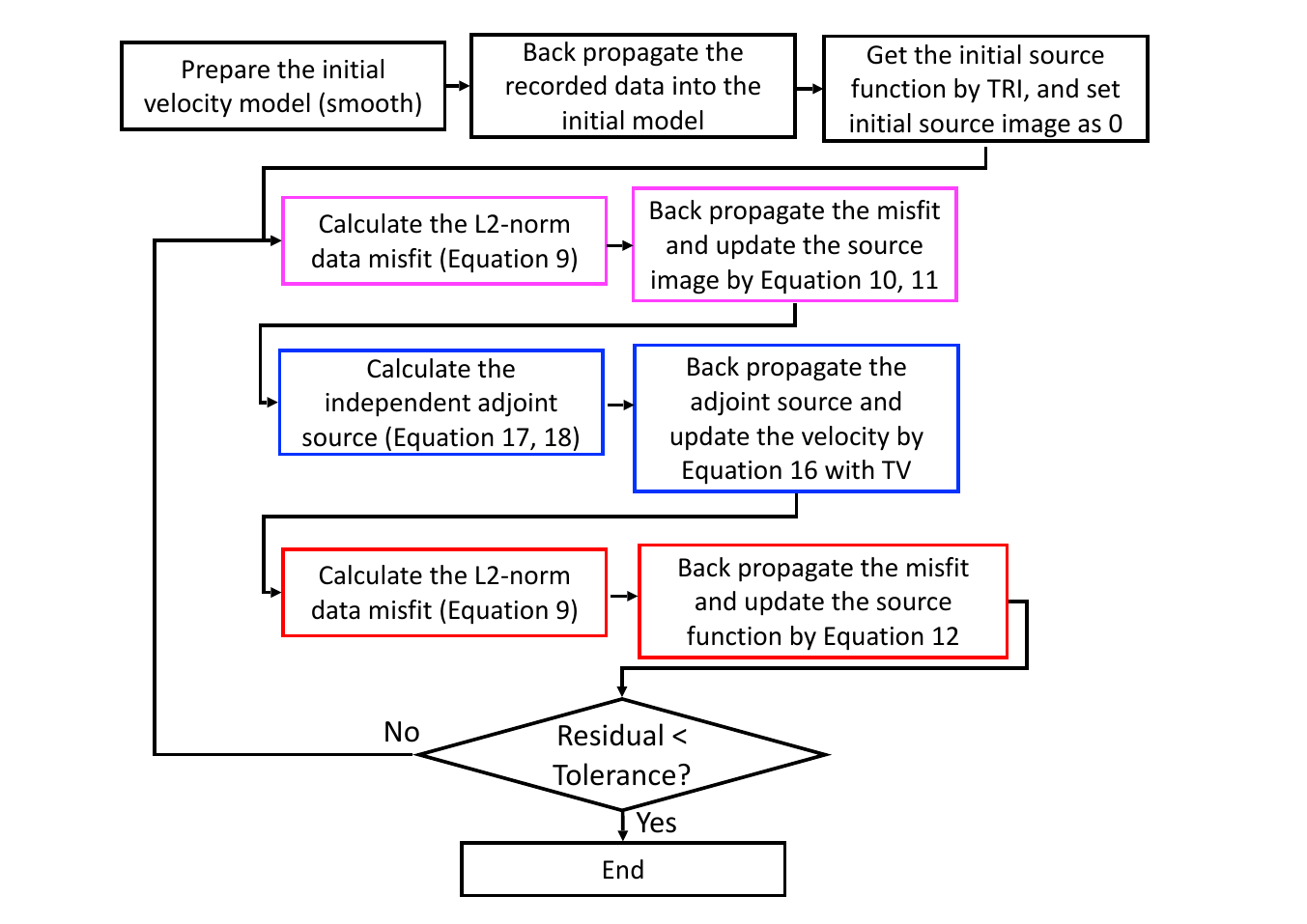}
  \caption{Nested approach workflow with equations for each step. Magenta: source image update, Equations \ref{eq:alpha_grd} and \ref{eq:beta_grd}; blue: velocity model update, Equation \ref{eq:vel_grd_cnv}; red: source wavelet update, Equations \ref{eq:E_wt}.}
\label{fig:flow}
\end{figure}

We implement our method on this numerical experiment using a nested approach, as shown in Figure~\ref{fig:flow}. We need an initial velocity model and source term to start the inversion. The initial velocities have been discussed already. The initial source images are completely zero $(\delta \alpha=\delta \beta=0)$, which means that the source spatial distribution is not available. To obtain the initial source function, the true record is back-propagated using the initial velocity model and then the wavefield at the maximum energy point is extracted from the wavefield, which is similar to the time-reversal imaging method and the maximum energy imaging condition. Once the initial parameters are ready, we can then start the inversion and update the three parameters iteratively until we get a convergent result. To suppress the source signatures at the five passive source locations, we use the pseudo Hessian to condition the gradient, which is calculated from the squared values of the adjoint wavefields to balance the energy contribution. To mitigate the cross-talk between different parameters, we need to carefully choose the optimization scheme.
\newline
The data residual (Equation~\ref{eq:obj_func}) reaches its minimum after ten outer-loop iterations. In each outer loop, five inner loops of source image and source time function updates and ten inner loops of velocity model updates are executed. The weighting factor $\lambda$ corresponding to the TV term is set to $0.5$ in the first iteration and linearly decreases to $0.05$ in the last iteration as a function of the outer-loop iteration number. Because of the unreliable initial velocities, we start with higher weights for the regularization in the beginning. As the velocities improve, we use smaller weights for the TV term to allow the velocity to have higher resolution information. We use a constant five percent of the current iteration TV maximum value as the smoothing factor $\epsilon$. Our objective is to have a smooth velocity model at the end by using a small $\epsilon$, as we are focusing on transmission waves. Usually $\epsilon$ is set to be $1\%$ to keep the model update blocky and approximately $5\%$ to keep it smooth. If the smoothing parameter is too large, the model update will lose most of the details, which may have negative effect on the inversion. The weighting factor $\lambda$ is dependent on to what extent you believe in the initial model. Since our initial model in the experiment is quite far from the original one, we need a relatively larger $\lambda$ in the first few iterations. The velocity updating step lengths are fixed for the whole inversion, that is the step length for $\alpha$ is $0.05$ and the step length for $\beta$ is $0.03$. The step length for $\beta$ is approximately $\frac{1}{\sqrt{3}}$ of that of that for $\alpha$. If the computational resources are sufficient, an even smaller step-length would be better. However from our experimentation, keep the maximum update no more than 50 meters per iteration is a good enough step-length for the modified Marmousi model.
\subsection{Results and Discussions}
\begin{figure}
\centering
\begin{subfigure}{0.48\textwidth}
  \centering
  \includegraphics[width=1\linewidth]{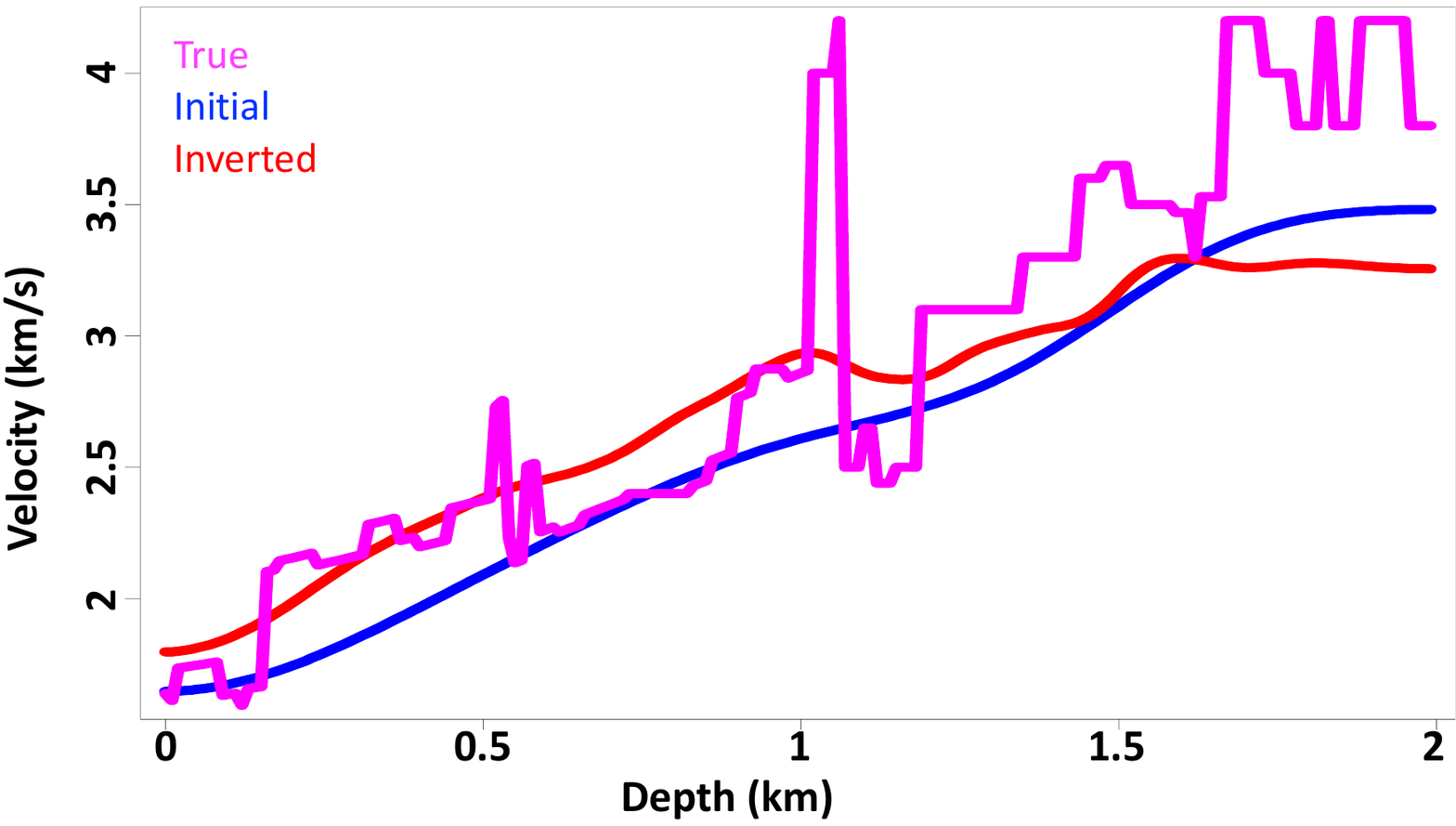}
  \caption{}
  \label{fig:vpprofile}
\end{subfigure}
\begin{subfigure}{0.48\textwidth}
  \centering
  \includegraphics[width=1\linewidth]{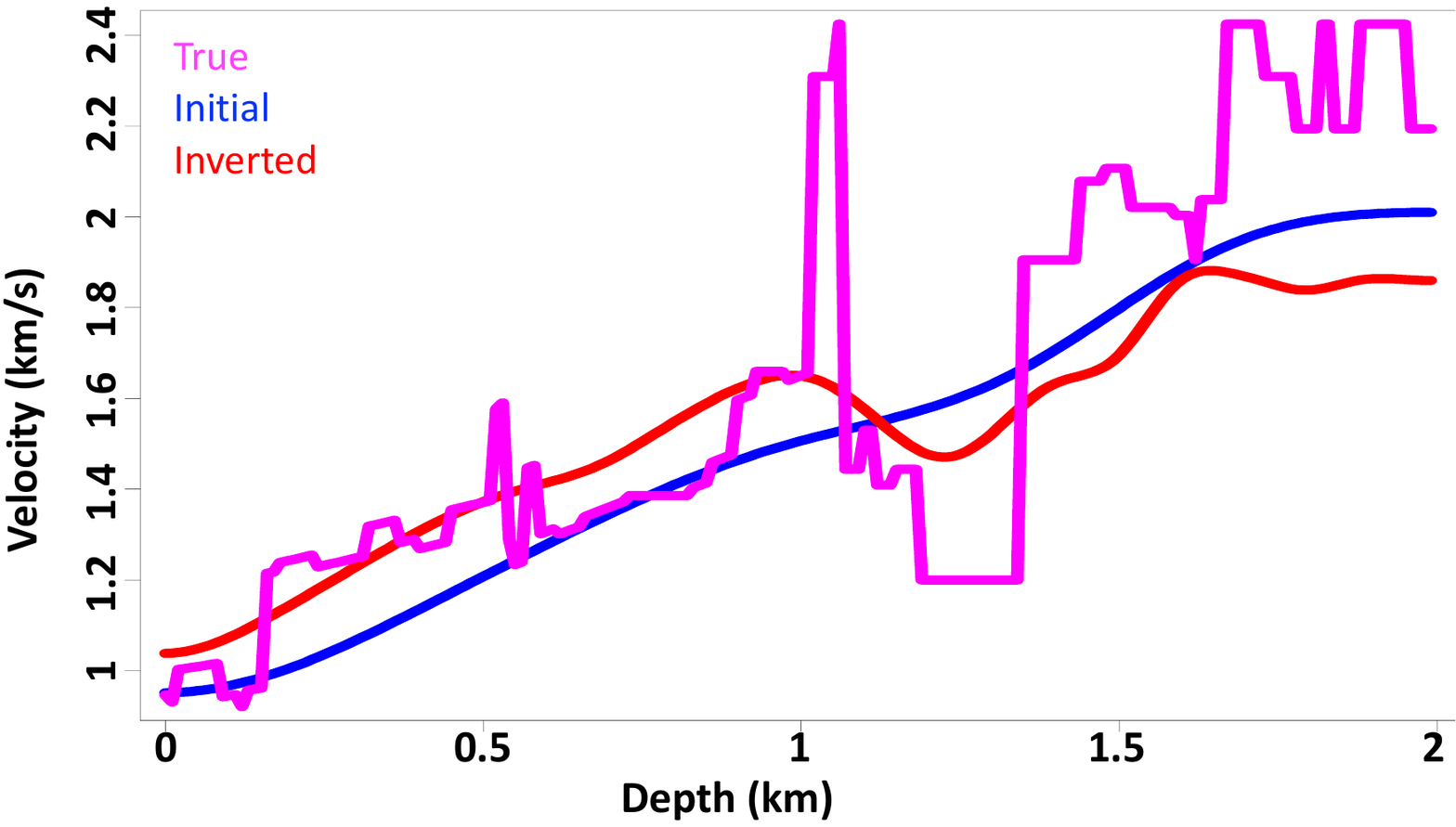}
  \caption{}
  \label{fig:vsprofile}
\end{subfigure}
\caption{Vertical profiles at $x=3km$ of (a) Vp and (b) Vs. Magenta, blue and red curves represent the original, initial and retrieved vertical velocity profiles, respectively.}
\label{fig:vprofile}
\end{figure}

\begin{figure}
\centering
\begin{subfigure}{0.4\textwidth}
  \centering
  \includegraphics[width=1\linewidth]{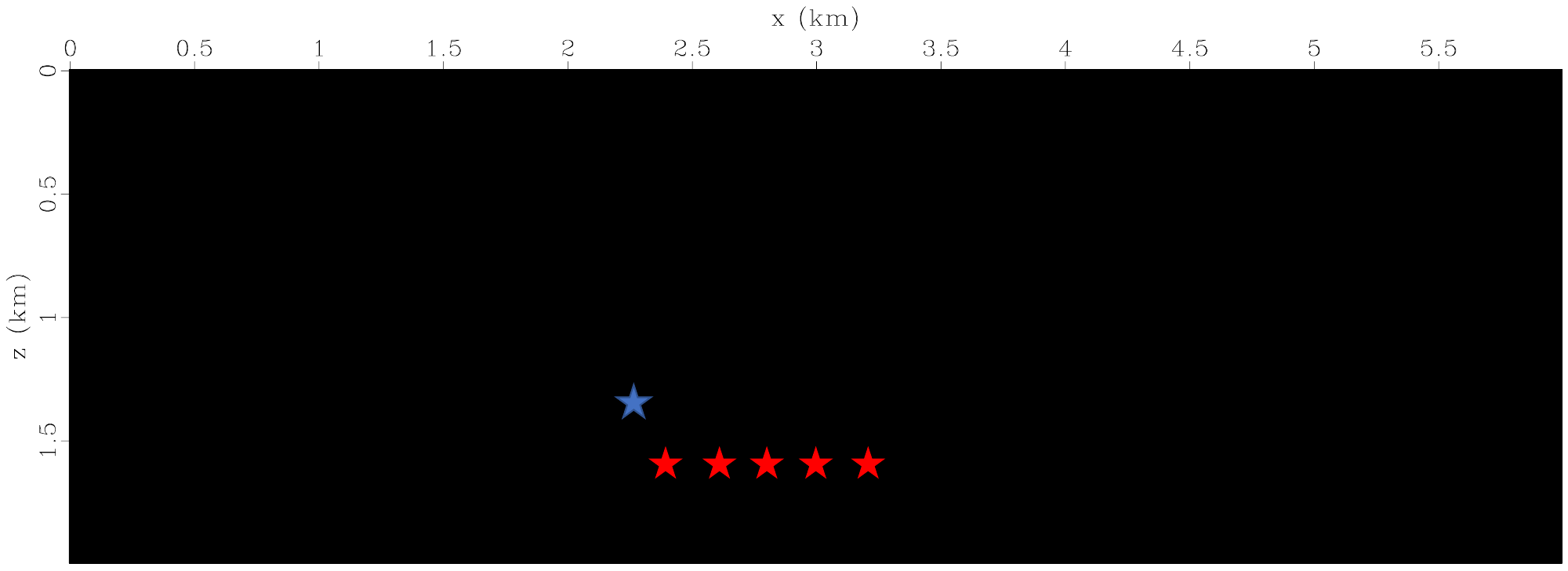}
  \caption{$\delta \alpha$ of source $No.1$, $M_3$}
  \label{fig:a1inv}
\end{subfigure}
\begin{subfigure}{0.4\textwidth}
  \centering
  \includegraphics[width=1\linewidth]{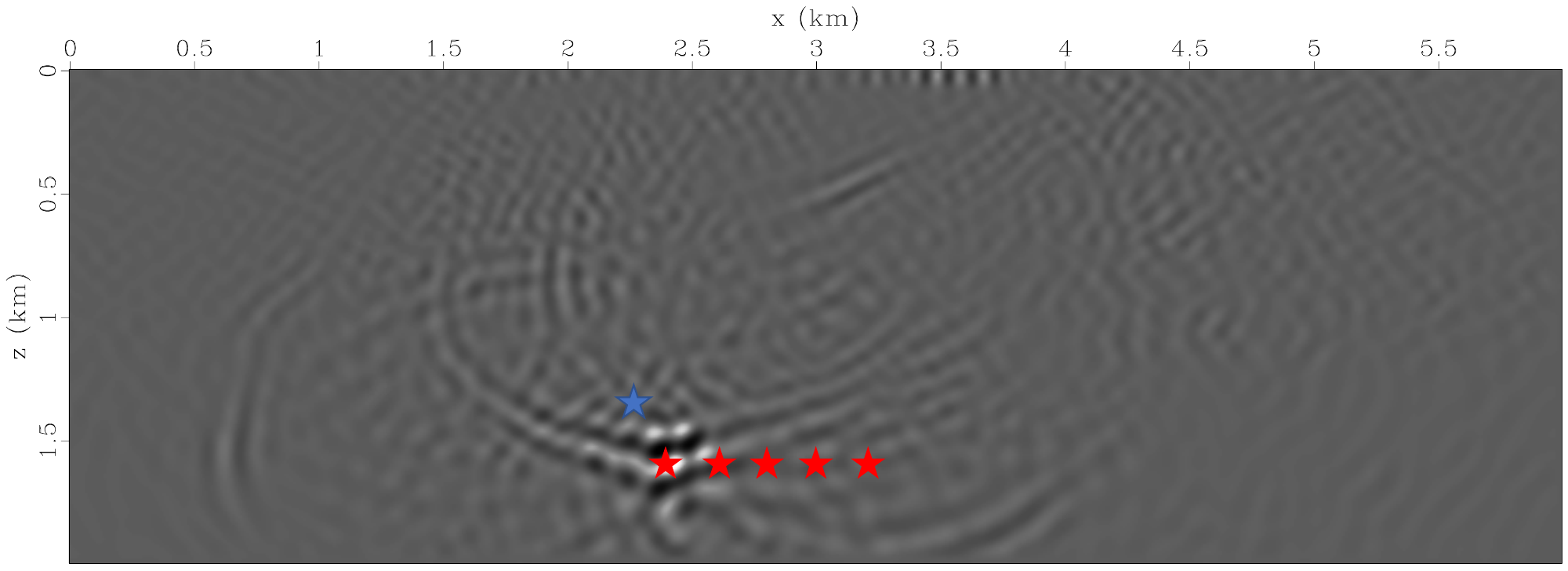}
  \caption{$\delta \beta$ of source $No.1$, $M_3$}
  \label{fig:b1inv}
\end{subfigure}
\begin{subfigure}{0.4\textwidth}
  \centering
  \includegraphics[width=1\linewidth]{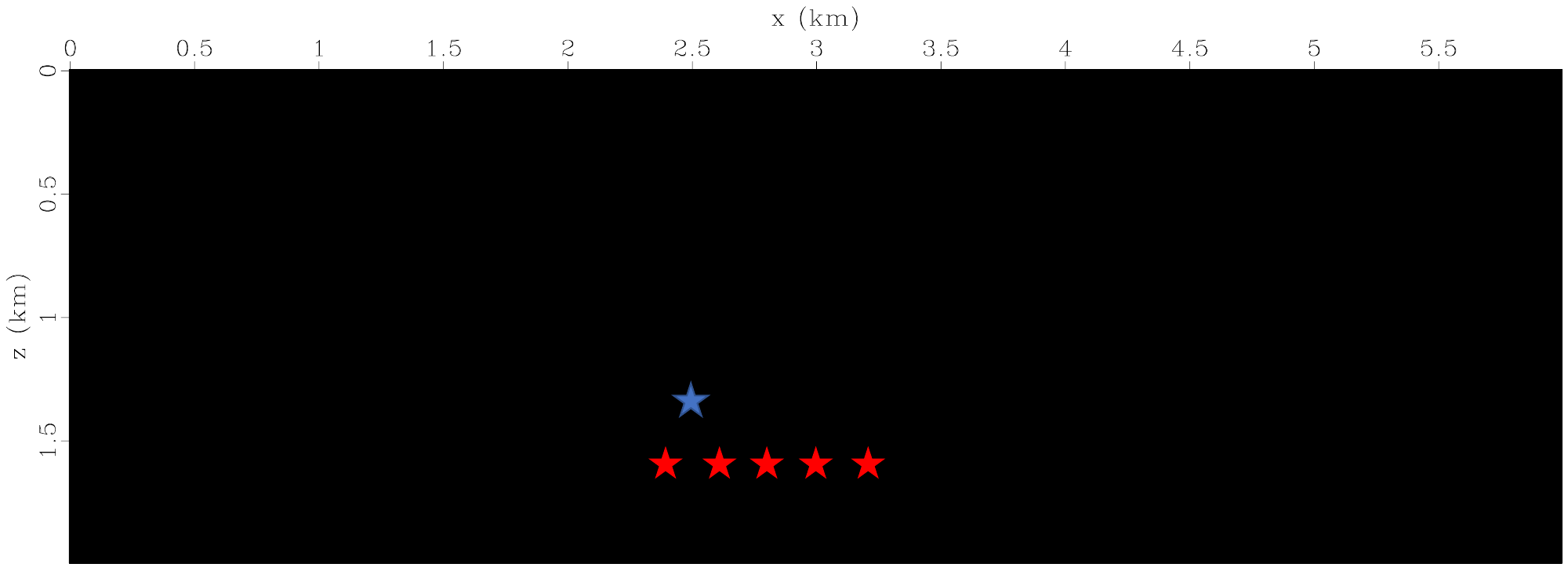}
  \caption{$\delta \alpha$ of source $No.2$, $M_4$}
  \label{fig:a2inv}
\end{subfigure}
\begin{subfigure}{0.4\textwidth}
  \centering
  \includegraphics[width=1\linewidth]{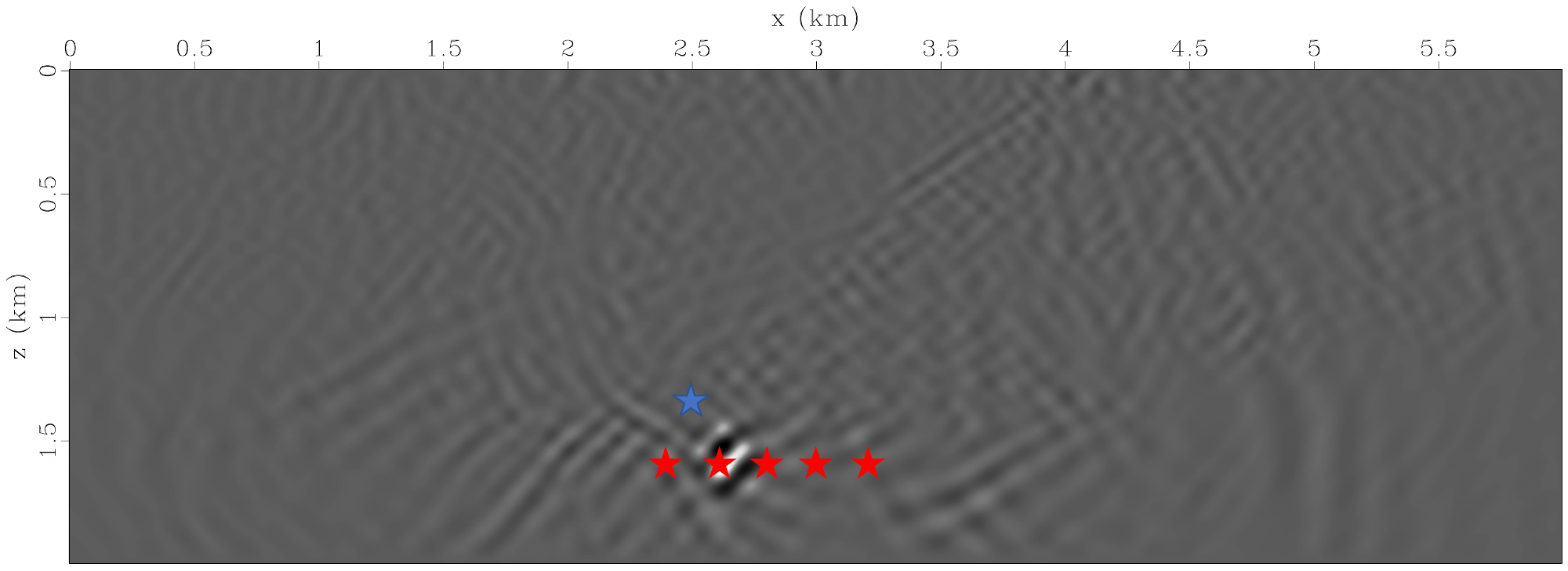}
  \caption{$\delta \beta$ of source $No.2$, $M_4$}
  \label{fig:b2inv}
\end{subfigure}
\begin{subfigure}{0.4\textwidth}
  \centering
  \includegraphics[width=1\linewidth]{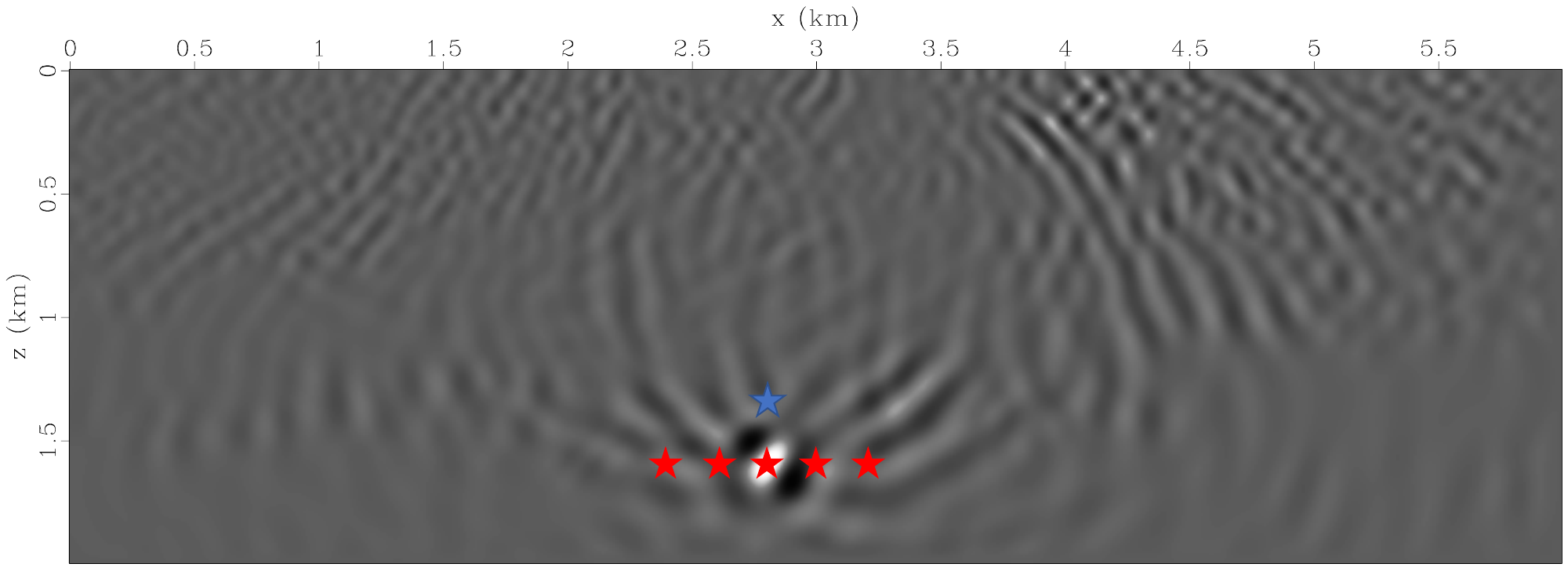}
  \caption{$\delta \alpha$ of source $No.3$, $M_1$}
  \label{fig:a3inv}
\end{subfigure}
\begin{subfigure}{0.4\textwidth}
  \centering
  \includegraphics[width=1\linewidth]{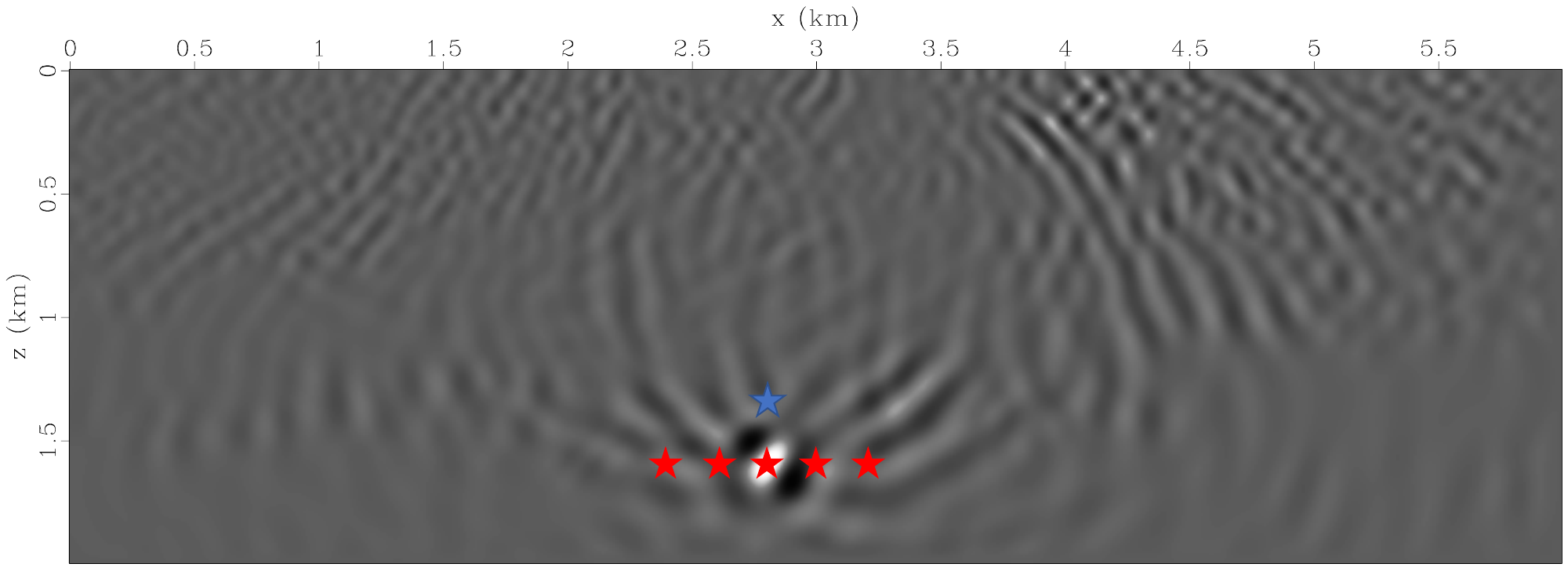}
  \caption{$\delta \beta$ of source $No.3$, $M_1$}
  \label{fig:b3inv}
\end{subfigure}
\begin{subfigure}{0.4\textwidth}
  \centering
  \includegraphics[width=1\linewidth]{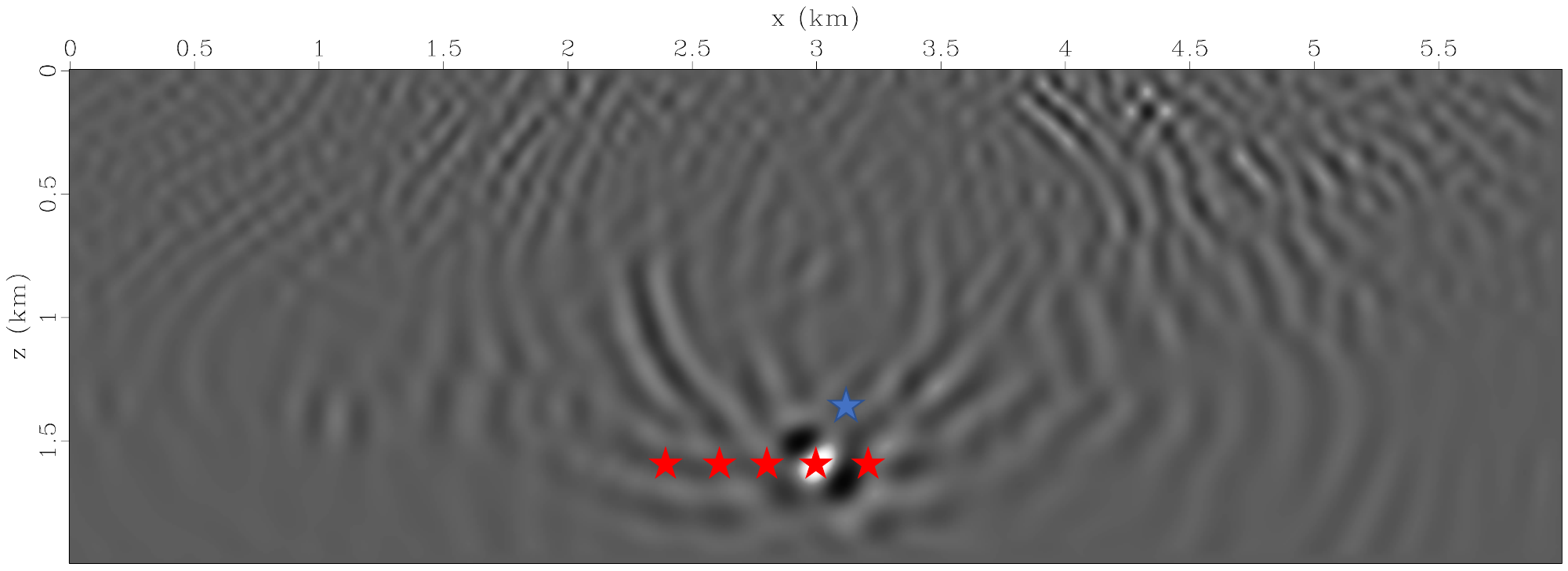}
  \caption{$\delta \alpha$ of source $No.4$, $M_2$}
  \label{fig:a4inv}
\end{subfigure}
\begin{subfigure}{0.4\textwidth}
  \centering
  \includegraphics[width=1\linewidth]{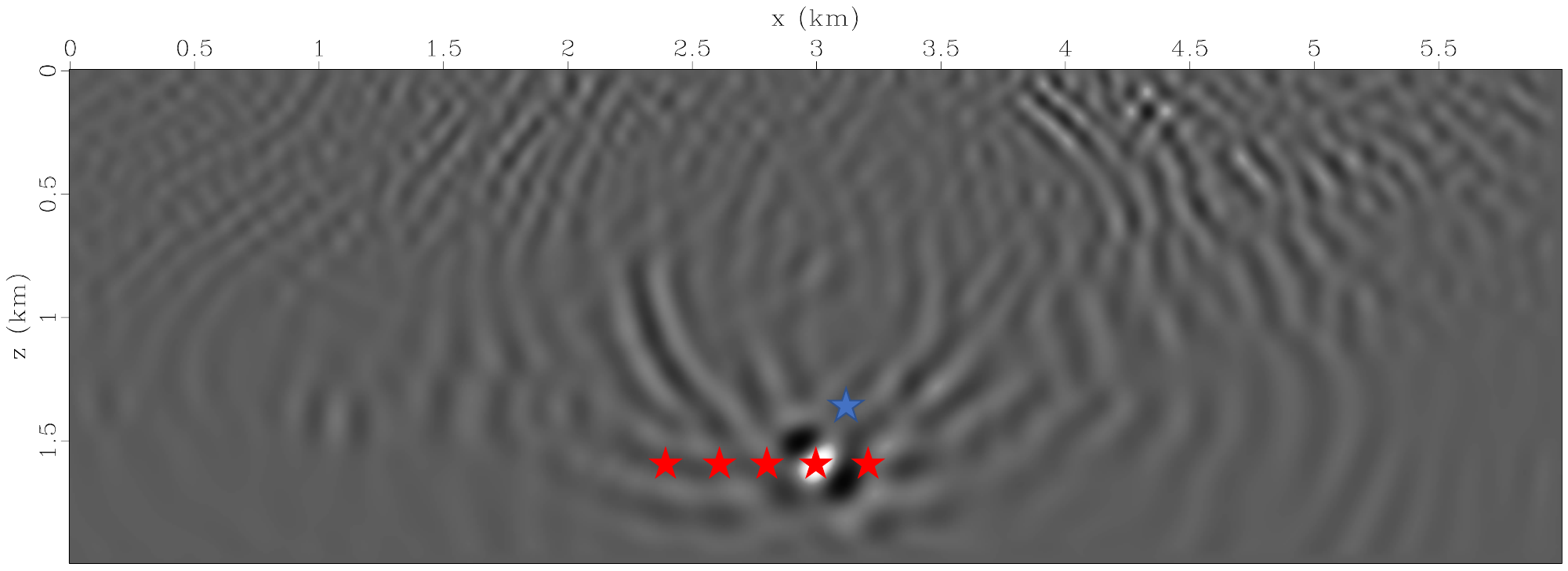}
  \caption{$\delta \beta$ of source $No.4$, $M_2$}
  \label{fig:b4inv}
\end{subfigure}
\begin{subfigure}{0.4\textwidth}
  \centering
  \includegraphics[width=1\linewidth]{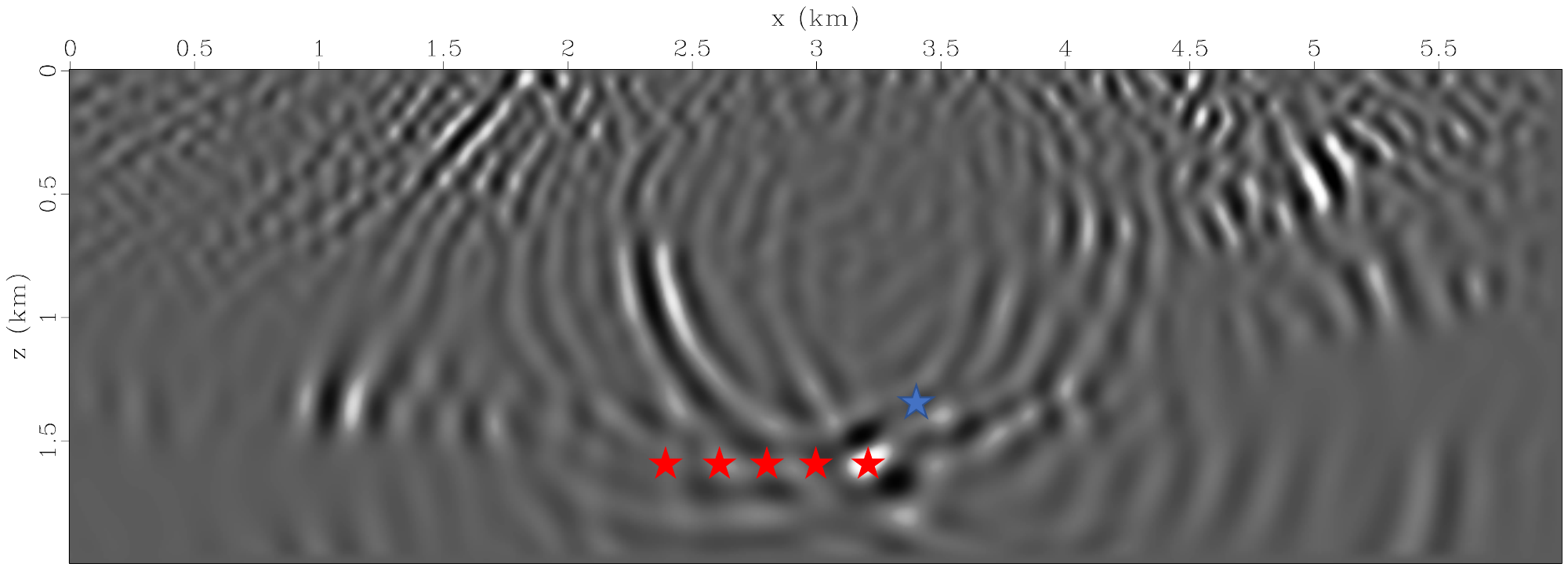}
  \caption{$\delta \alpha$ of source $No.5$, $M_5$}
  \label{fig:a5inv}
\end{subfigure}
\begin{subfigure}{0.4\textwidth}
  \centering
  \includegraphics[width=1\linewidth]{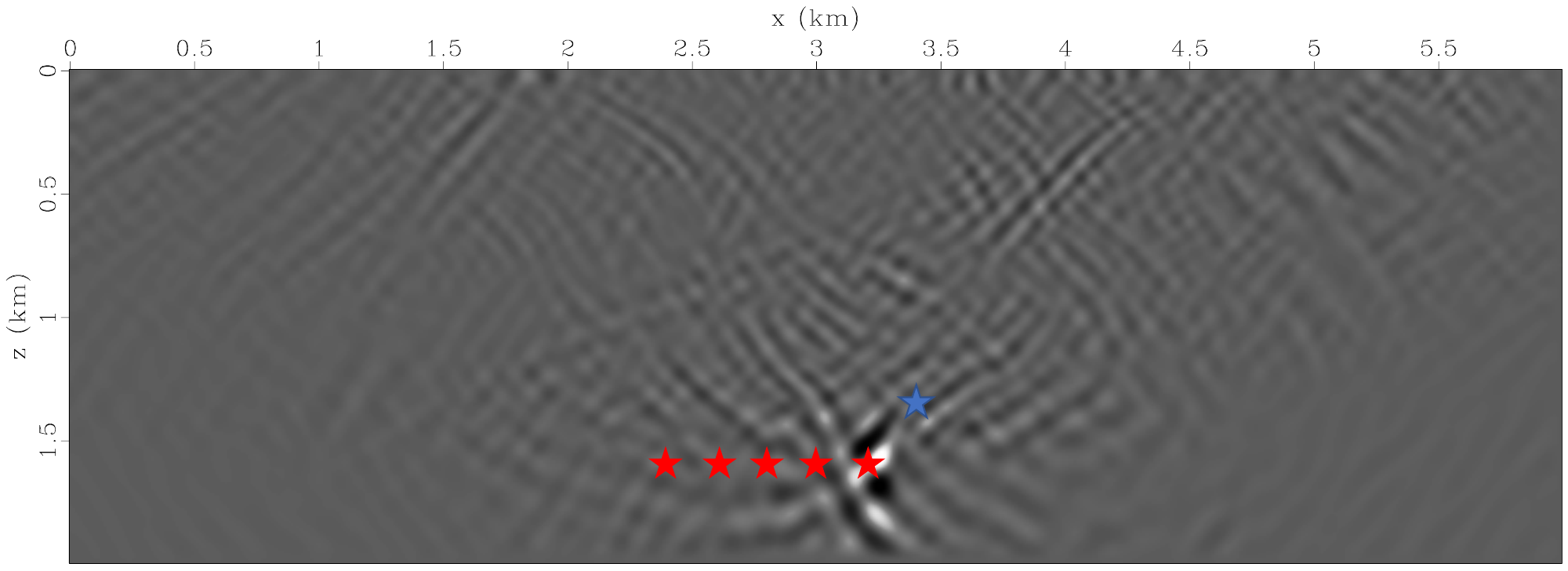}
  \caption{$\delta \beta$ of source $No.5$, $M_5$}
  \label{fig:b5inv}
\end{subfigure}
\caption{Inverted source images (left) $\delta \alpha$ (right) $\delta \beta$; True source locations $(x,z)$ are marked as red stars: (1) $(2.4km,1.55km)$, (2) $(2.6km,1.55km)$, (3) $(2.8km,1.55km)$, (4) $(3.0km,1.55km)$, (5) $(3.2km,1.55km)$. Blue stars refers to the initial source locations using TRI method.}
\label{fig:imagesinv}
\end{figure}

\begin{figure}
\centering
\begin{subfigure}{0.45\textwidth}
  \centering
  \includegraphics[width=1\linewidth]{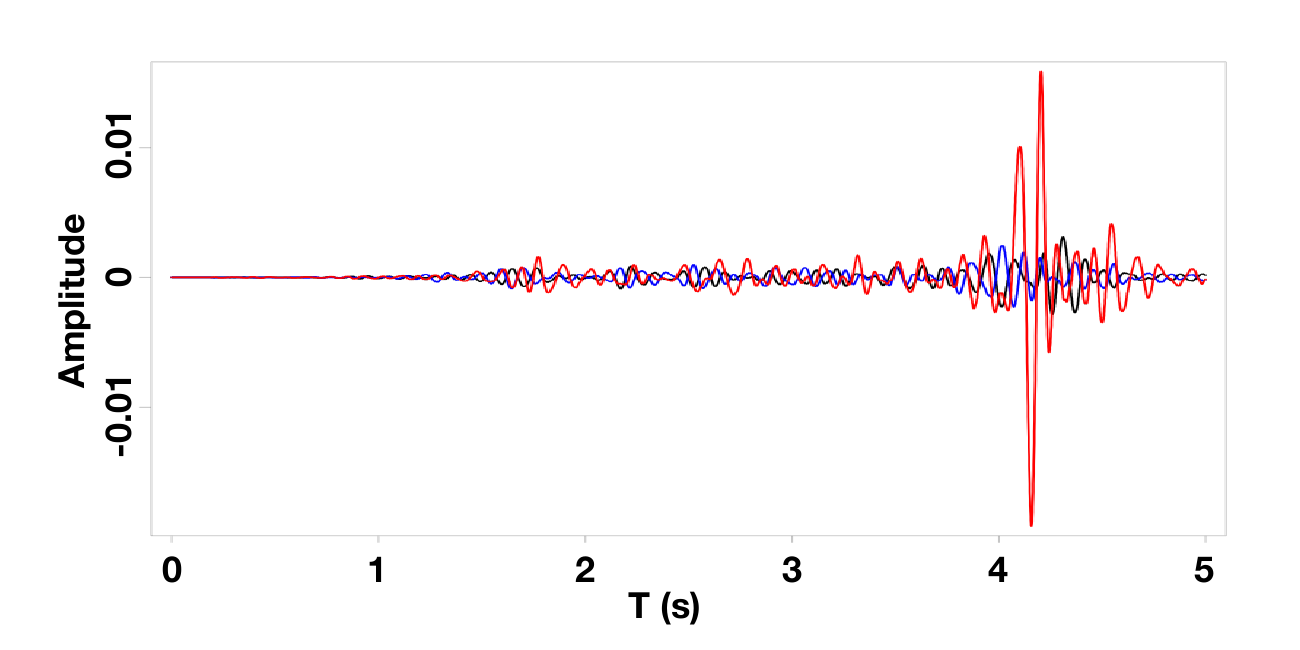}
  \caption{}
  \label{fig:w1}
\end{subfigure}
\begin{subfigure}{0.45\textwidth}
  \centering
  \includegraphics[width=1\linewidth]{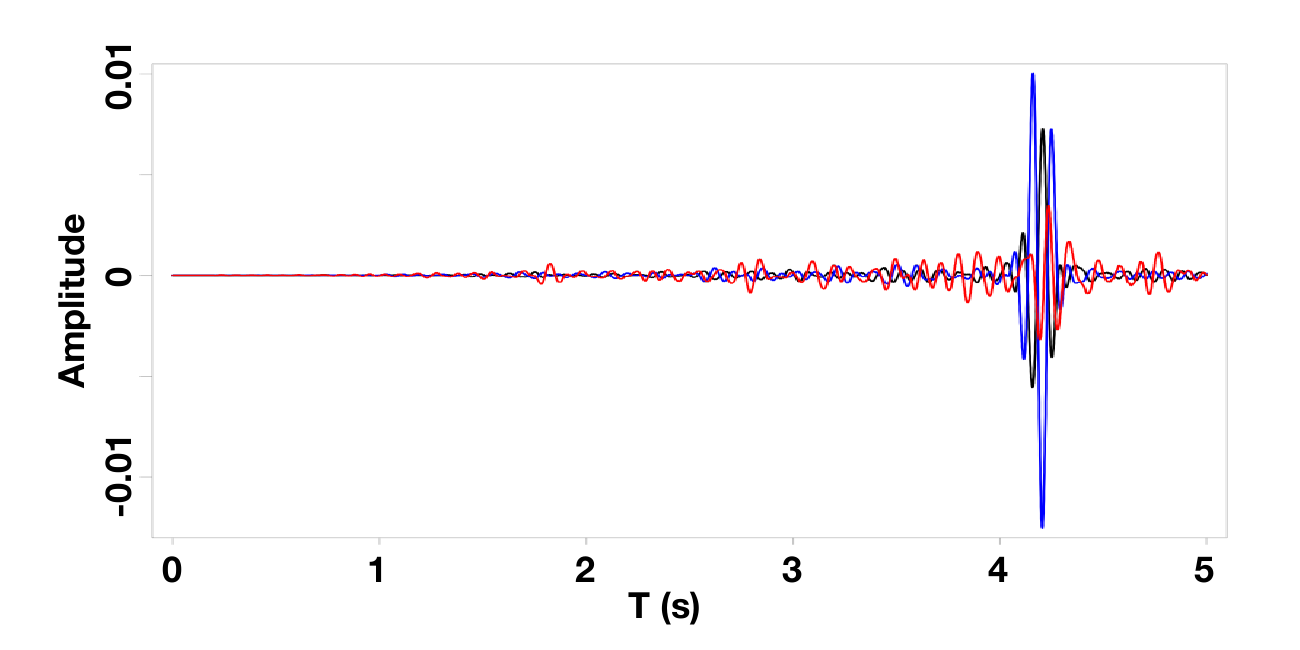}
  \caption{}
  \label{fig:w2}
\end{subfigure}
\begin{subfigure}{0.45\textwidth}
  \centering
  \includegraphics[width=1\linewidth]{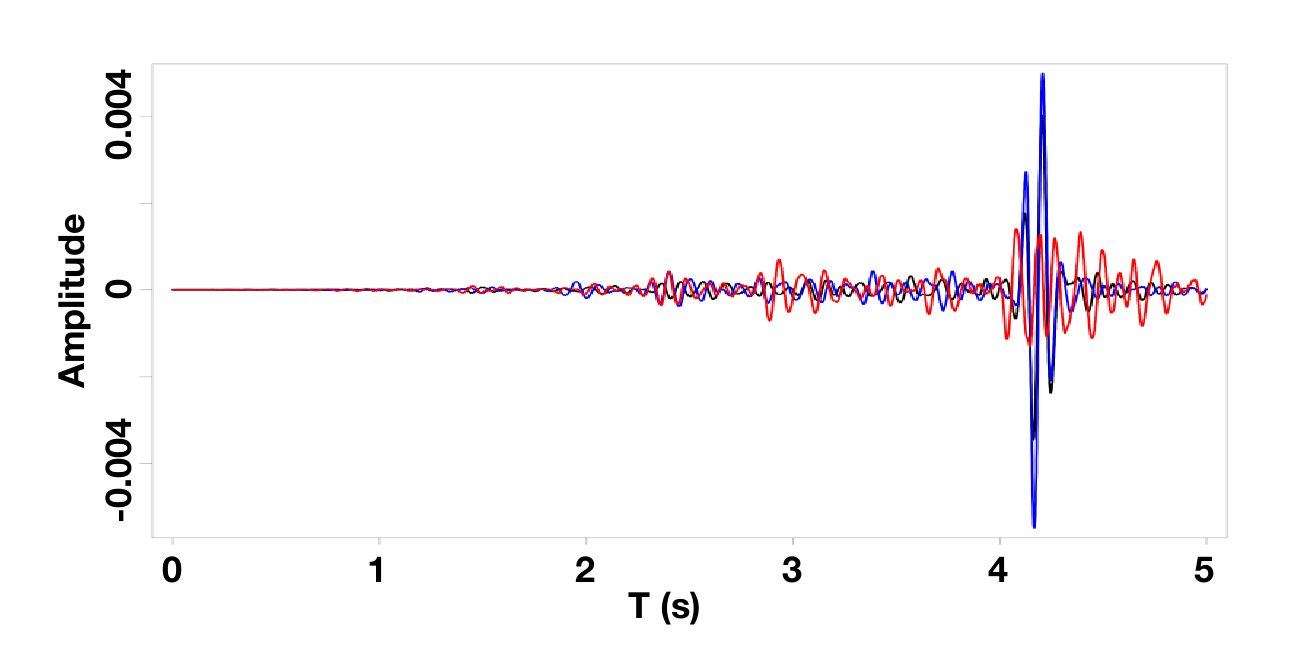}
  \caption{}
  \label{fig:w3}
\end{subfigure}
\begin{subfigure}{0.45\textwidth}
  \centering
  \includegraphics[width=1\linewidth]{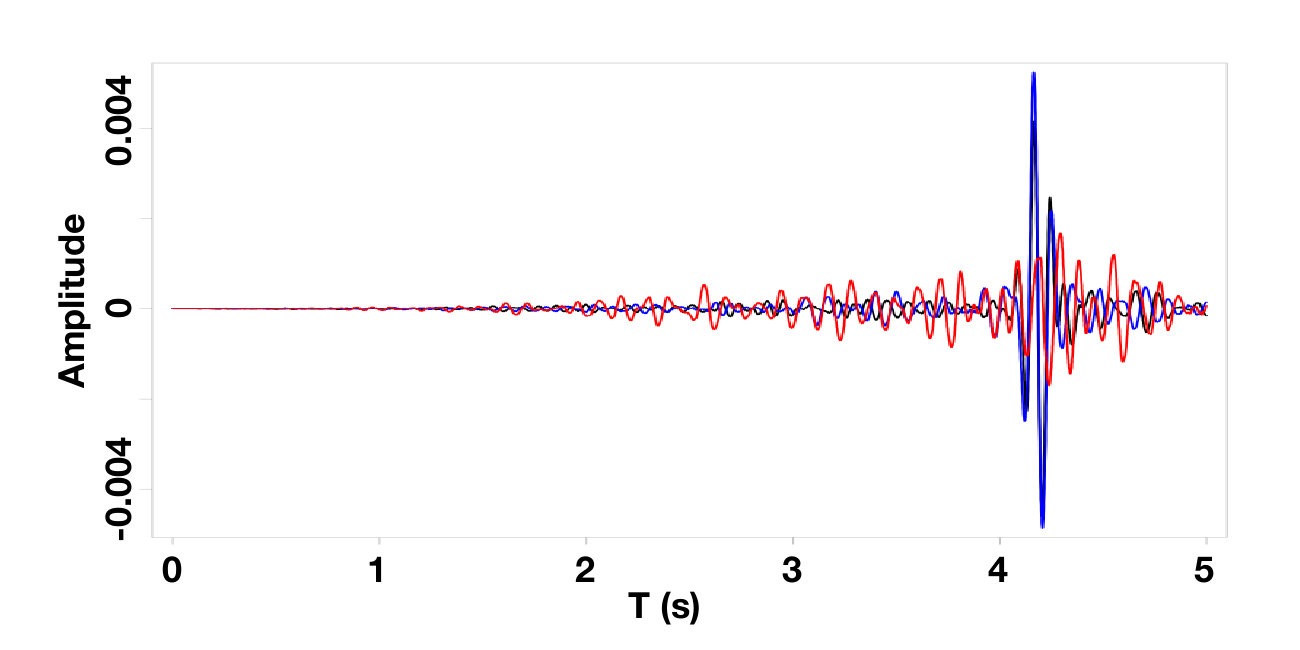}
  \caption{}
  \label{fig:w4}
\end{subfigure}
\begin{subfigure}{0.45\textwidth}
  \centering
  \includegraphics[width=1\linewidth]{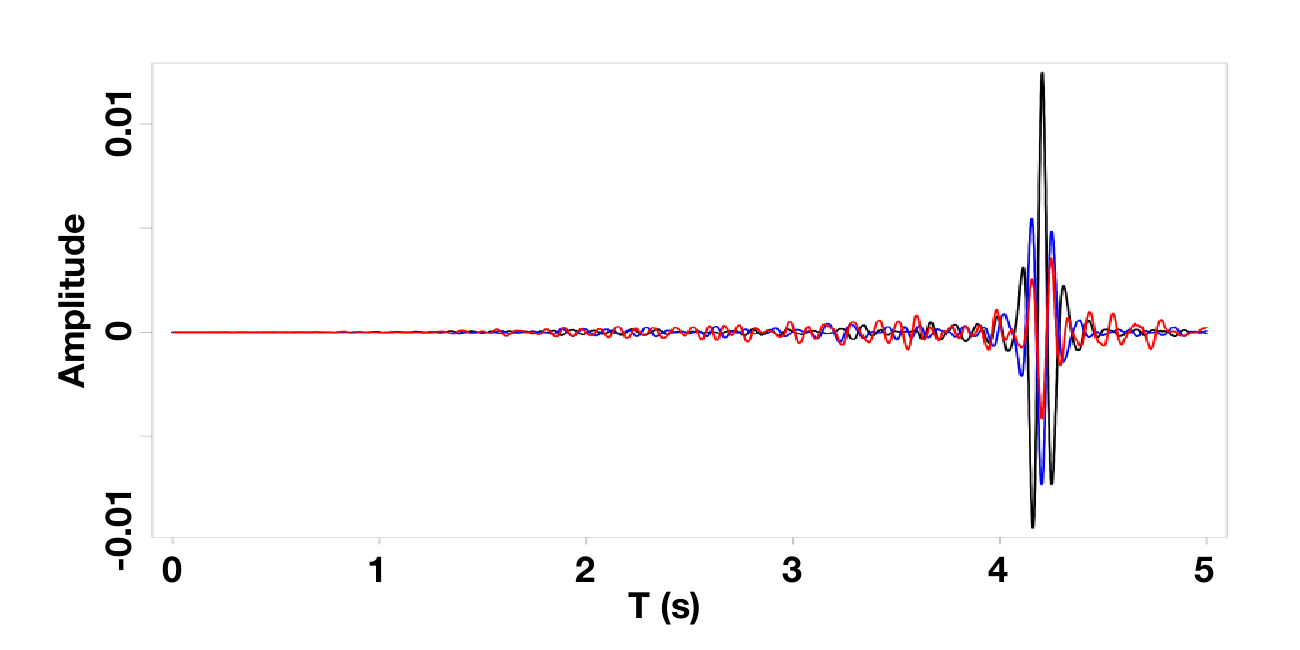}
  \caption{}
  \label{fig:w5}
\end{subfigure}
\caption{Inverted source functions for source (a) No.1 (b) No.2 (c) No.3 (d) No.4 (e) No.5. Black: $w_{11}$, blue: $w_{22}$, red: $w_{12}$.}
\label{fig:wavelet}
\end{figure}

The final results of $\alpha$ and $\beta$ are shown in Figures~\ref{fig:vpinv} and~\ref{fig:vsinv}. The inverted velocity models generally exhibit low resolution, due to the frequency band of the data and the limited illumination with only five nearby sources. Due to the distribution of the micro-seismic events and the receivers, transmission waves are the dominant recorded energy. Under the FWI scheme, transmission waves mainly update the low wave-number components of the model, whereas the reflection waves update the high wave-number parts. Thus, the inverted velocities in the experiment are mostly controlled by the smooth background updates. We may also find that the lower part of the model seems to have a higher resolution than the upper part. This is expected since the parts of the model below the sources are single scattered waves, whereas the reflection energy above the source can only be recorded after multi-scattering. In other words, the recorded reflection energy from the bottom reflections is higher than that from the upper reflections, which results in the upper part of the model having fewer high wavenumber updates (reflections), compared to the lower part, resulting in higher resolution in the lower part and a poorer resolution in the upper part. If the passive sources are triggered in a wider range, the illumination of the area would be much better with the same acquisition system. The passive sources in this experiment is quite confined since there are only five sources in a very small area. It leads to the accurate but smooth results for the kinematic components of the model. We compare the vertical profiles of $\alpha$ and $\beta$ at the position of $x=3.0$km as a quality control, shown in Figure~\ref{fig:vprofile}. Apparently, the initial velocities are lower than the true ones. Moreover, in the initial $\beta$, no hydraulic injected area information is given at the arrow-pointed zone in Figure~\ref{fig:vstrue}. After the inversion, we see that the velocities have been updated in the right direction, and the hydraulic injected area (low velocity zone) in $\beta$ is also well retrieved. Certainly, the inverted velocities are more kinematic accurate.
\newline
Although the inverted velocity models are imperfect that the high wave-number part of the model is not well retrieved, they are good enough to locate the micro-seismic sources in space, which reflects their kinematic accuracy (the low wave-number part is accurate). The inverted source images, $\delta \alpha$ and $\delta \beta$, are shown in Figures~\ref{fig:a1inv} to~\ref{fig:b5inv} for the five sources, respectively. Extracting the source coordinates from the source images by searching for the maximum energy point in each source image, where most of the energy is focused at that point, although some unfocused energy caused probably by the remaining data misfit is left over. For verification, the source locations after the first inner loop of the source image update, which is equivalent to the TRI locations, are shown as the blue stars in the figures and the true source locations are shown as the red stars. We may find that both $\delta \alpha$ and $\delta \beta$ focus in the accurate locations, while the initial (TRI) locations, are totally wrong due to the initial velocity errors. Note that $\delta \alpha$ of sources 1 and 2 are zero, because of the reason we discussed in the previous section $(w_{11}+w_{22}=0)$. In the case of an imperfect acquisition and velocity model, the term $(w_{11}+w_{22})$ is not exactly zero, but they are at least sufficiently small. When dealing with the source images, we intended to ignore small values in $\delta \alpha$ as compared to the value of $\delta \beta$. By comparing $\delta \alpha$ with $\delta \beta$, we may find that $\delta \alpha$ exhibits lower resolution than $\delta \beta$. It is reasonable because $\delta \alpha$ simply depends on $V_p$, but $\delta \beta$ depends on both $V_p$ and $V_s$. The fact that $V_p$ is naturally higher than $V_s$ leads to the wavelength (higher resolution) of $\delta \alpha$ being larger (lower) than that of $\delta \beta$. 
\newline
Although the source locations are accurately inverted, we can hardly tell the differences between the source images for the various source types. It is mostly because of the limited acquisition aperture (surface and vertical well receivers) and the inverted velocity imperfections. In addition, there is a leakage of energy in the inverted source images due to the model misfit as well. One can choose to ignore these weak artifacts or use a regularization, for example L1-norm regularization, to suppress the unfocused energy.
\newline
The inverted source wavelets are shown in Figure~\ref{fig:wavelet}. For the first and the second sources on the left (moment tensors $M_3$ and $M_4$), the components $w_{11}$ and $w_{22}$ have opposite signs dictated by the source mechanism. As a result, the summation of $w_{11}$ and $w_{22}$ admits reasonably small values compared to the value of $w_{12}$, as we discussed above. In theory, the $w_{11}$ and $w_{22}$ components of $M_3$ should be both zero but the inversion seems to fail in inverting their correct values, though the result does not affect the source image inversion. For the third and the fourth sources on the left (moment tensors $M_1$ and $M_2$), the components $w_{11}$ and $w_{22}$ show similar features due to the isotropic source mechanism.
\newline
The computational cost for each inner loop of the proposed method is equivalent to a standard FWI iteration. In our inversion scheme, however, besides the nonlinear velocity update, a minor extra cost of the linear source term update in every iteration is required. Thus, the total computational cost is still challenging. When processing with the field monitored data, especially with low S/N data, we could apply some pre-processing of the data to enhance the S/N so that we may obtain clearer subsurface features.
\section{Conclusions}
We introduced an equivalent source term for micro-seismic events, under a similar scheme to that of an elastic reflection full waveform inversion. We decoupled the micro-seismic source term into spatial and temporal components. We then invert for the decoupled source and the velocity models with a nested approach. In the proposed method, the equivalent decoupled source term represents the source mechanisms with the different features of the source images. However, the sensitivity of the source images to the acquisition system affects the inversion results. To mitigate the cycle skipping problem caused by an unknown source function, a modified objective function, that is independent of the source time function, is applied by convolving the ground truth and the predicted data with reference traces, through which the necessity of a source origin time information is mitigated. The inversion for all three parameters yields reasonably good results and the low wavenumber part of the model is retrieved well enough to accurately locate the passive sources. The low resolution in the velocity model is due to the nature of the passive seismic experiment setup. For the modified Marmousi model, the proposed method is successful in finding all the source locations, as well as in updating a kinematically equivalent velocity model.



\section{Appendix A: The Gradient calculation with the conventional objective function}
Here we derive the gradients of the objective function, Equation~\ref{eq:obj_func}, with respect to the model squared velocity $\bf m(\alpha,\beta)$, source images $ \bf {\delta} \bf m(\delta \alpha,\delta \beta)$ and source function $\bf w$:
\begin{equation}
\begin{array}{cl}
dE(\bf u) & \displaystyle = \frac{\partial E}{\partial \bf m}d\bf m+\frac{\partial E}{\partial \delta \bf {m}}d\delta \bf {m}+\frac{\partial E }{\partial \left[ w_{ij} \right]}d\bf w \\& \displaystyle = \int_t dt (\bf u - \bf d) d \bf u,
\end{array}
\label{eq:dE}
\end{equation}
The elastic displacement field when density is constant satisfies
\begin{equation}
\bf u_{tt}-\nabla \cdot (\bf m : \nabla u)=\nabla \cdot (\delta \bf m : \bf w),
\label{eq:utt}
\end{equation}
where $:$ is the Frobenius inner product operator. The total derivatives of the displacement vector $\bf u$ is given by
\begin{equation}
d \bf u_{tt}=\nabla \cdot \nabla d \bf u : \bf m+\nabla \cdot d \bf m : \nabla \bf u +\nabla \cdot d\bf w:\delta \bf m+\nabla d\delta \bf m : \bf w.
\label{eq:dutt}
\end{equation}
The perturbation $d\bf u$ can be written as
\begin{equation}
d \bf u= (\frac{\partial^2}{\partial t^2}-\nabla \cdot(\bf m : \nabla))^{-1}(\nabla \cdot d \bf m : \nabla \bf u +\nabla \cdot d\bf w:\delta \bf m+\nabla d\delta \bf m : \bf w).
\label{eq:du}
\end{equation}
Equation~\ref{eq:du} contains three terms representing the partial derivatives of the displacement vector with respect to $d \bf m$, $d \bf w$ and $d \delta \bf m$. Define the adjoint wavefield for back propagating the data residual into the media as
\begin{equation}
\bf \hat u= (\frac{\partial^2}{\partial t^2}-\nabla \cdot(\bf m : \nabla))^{-T}(\bf u - \bf d),
\label{eq:u_hat}
\end{equation}
where $T$ is the adjoint operation.
\newline
To calculate the gradient for $\bf m(\alpha,\beta)$, $\delta \bf m$ or $\bf w$, we substitute $d\bf u$ into Equation~\ref{eq:dE}
\begin{equation}
\begin{array}{cl}
& \displaystyle dE=(\frac{\partial^2}{\partial t^2}-\nabla \cdot(\bf m : \nabla))^{-1}(\bf u - \bf d)(\nabla \cdot d \bf m : \nabla \bf u +\nabla \cdot d\bf w:\delta \bf m+\nabla d\delta \bf m : \bf w) \\& \displaystyle = \nabla \cdot (\frac{\partial^2}{\partial t^2}-\nabla \cdot(\bf m : \nabla))^{-1}(\bf u - \bf d) : \nabla \bf u d\bf m + \nabla \cdot (\frac{\partial^2}{\partial t^2}-\nabla \cdot(\bf m : \nabla))^{-1}(\bf u - \bf d) : \delta \bf m d\bf w \\& \displaystyle  + \nabla \cdot (\frac{\partial^2}{\partial t^2}-\nabla \cdot(\bf m : \nabla))^{-1}(\bf u - \bf d) : \bf w d\delta \bf m.
\end{array}
\label{eq:total_grd}
\end{equation}
To calculate the gradients for each parameter, we set the other two parameters' total derivatives to be zero (do not change in the current iteration). For example, to calculate the gradient for $\bf m(\alpha,\beta)$, set $d\delta \bf m=0$ and $d\bf w=0$
\begin{equation}
\nabla E_{\bf m}=\frac{dE}{d\bf m}=\int_t dt \nabla \cdot (\frac{\partial^2}{\partial t^2}-\nabla \cdot(\bf m : \nabla))^{-1}(\bf u - \bf d) : \nabla \bf u;
\label{eq:E_model-a}
\end{equation}
Similarly, by setting $d\bf m=0$ and $d\bf w=0$, we can obtain the gradient for $\delta \bf m$
\begin{equation}
\nabla { E }_{ \delta \bf m  }=\frac{dE}{d\delta \bf m}=\int_t dt \nabla \cdot (\frac{\partial^2}{\partial t^2}-\nabla \cdot(\bf m : \nabla))^{-1}(\bf u - \bf d) : \bf w ,
\label{eq:E_img-a}
\end{equation}
and we get the gradient for $\bf w$ by setting $d\bf m=0$ and $d\bf m=0$
\begin{equation}
\nabla { E }_{ \bf w  }=\frac{dE}{d \bf w}=\int_{\bf x} d{\bf x} \nabla \cdot (\frac{\partial^2}{\partial t^2}-\nabla \cdot(\bf m : \nabla))^{-1}(\bf u - \bf d) : \delta \bf m
\label{eq:E_wt-a}
\end{equation}
\section{Appendix B: The Gradient for The Source Independent Objective Function}
We derive the gradient of the source function independent objective function, Equation~\ref{eq:indp_obj}, with respect to elastic velocities $c(x,z)=[V_p;V_s]$, under the constant density assumption. In this case, the objective function without the optional total variation term is given by
\begin{equation}
\ E_{ind}=\sum _{ i }^{ nr }{ \left\| r_i \right\|  ^2} = \sum _{ i }^{ nr }{ \left\| { d }_{ i }^{ pred }*{ d }_{ ref }^{ obs }-{ d }_{ i }^{ obs }*{ d }_{ ref }^{ pred } \right\|  ^2},
\label{eq:misfit2}
\end{equation}
Taking the derivative of Equation~\ref{eq:misfit2} with respect to $c(x,z)$. The gradient can be written as
\begin{equation}
\frac { \partial E_{ind} }{ \partial c } =\sum _{ i }^{ nr }{ \left[ \left( \frac { \partial { d }_{ i }^{ pred } }{ \partial c } *{ d }_{ ref }^{ obs } \right) \cdot r_{ i }-\left( { d }_{ i }^{ obs }*\frac { \partial { d }_{ ref }^{ pred } }{ \partial c }  \right) \cdot r_{ i } \right]  } .
\label{eq:vel_grd_cnv-a}
\end{equation}
\newline
The first convolution term $(\frac { \partial { d }_{ i }^{ pred } }{ \partial c } *{ d }_{ ref }^{ obs })\cdot r_i$ in Equation~\ref{eq:vel_grd_cnv-a} can be rewritten in an integral form,
\begin{equation}
\left( \frac { \partial { d }_{ i }^{ pred } }{ \partial c } *{ d }_{ ref }^{ obs } \right) \cdot { r }_{ i }=\int _{ -\infty  }^{ +\infty  }{ \int _{ -\infty  }^{ +\infty  }{ \frac { \partial { d }_{ i }^{ pred } }{ \partial c } \left( t-\tau  \right)  } { d }_{ ref }^{ obs }\left( \tau  \right) r_{ i }\left( t \right)  } d \tau dt .
\label{eq:first_term}
\end{equation}
Let $\eta = t - \tau$, so that, equation~\ref{eq:first_term} can be written as,
\begin{equation}
\left( \frac { \partial { d }_{ i }^{ pred } }{ \partial c } *{ d }_{ ref }^{ obs } \right) \cdot { r }_{ i }=-\int _{ -\infty  }^{ +\infty  }{ \int _{ -\infty  }^{ +\infty  }{ \frac { \partial { d }_{ i }^{ pred } }{ \partial c } \left( \eta  \right)  } { d }_{ ref }^{ obs }\left( t - \eta  \right) r_{ i }\left( t \right)  } d\eta dt  .
\label{eq:first_term2}
\end{equation}
The partial derivative $\frac { \partial { d }_{ i }^{ pred } }{ \partial c } \left( \eta  \right)$ is now independent of $t$ so we move it outside of the integration of $t$, which gives
\begin{equation}
\left( \frac { \partial { d }_{ i }^{ pred } }{ \partial c } *{ d }_{ ref }^{ obs } \right) \cdot { r }_{ i }=-\int _{ -\infty  }^{ +\infty  }{ \frac { \partial { d }_{ i }^{ pred } }{ \partial c } \left( \eta  \right)  } \left[ \int _{ -\infty  }^{ +\infty  }{ { d }_{ ref }^{ obs }\left( t-\eta  \right) r_{ i }\left( t \right)  } dt \right] d\eta .
\label{eq:first_term3}
\end{equation}
This derivative $\frac { \partial { d }_{ i }^{ pred } }{ \partial c } \left( \eta  \right) $ represents the convolution between $U _ { x , i }$ and $S _ x$, where $U _ { x , i }$ is the adjoint wavefield of the $ith$ backpropagated residual, $S _ x$ is the source wavefield ignited at a virtual source location and $x$ represents the model space location. Using the convolution form to rewrite Equation~\ref{eq:first_term}, we obtain
\begin{equation}
\left( \frac { \partial { d }_{ i }^{ pred } }{ \partial c } *{ d }_{ ref }^{ obs } \right) \cdot { r }_{ i }= -\int _{ -\infty  }^{ +\infty  }{ \left[ \int _{ -\infty  }^{ +\infty  }{ S_{ x }\left( \eta -\tau  \right) U_{ { x,i } } } \left( \tau  \right) d\tau  \right]  } \left[ \int _{ -\infty  }^{ +\infty  }{ { d }_{ ref }^{ obs }\left( t-\eta  \right) r_{ i }\left( t \right)  } dt \right] d\eta 
\label{eq:first_term4}
\end{equation}
Again let $\eta - \tau = t$, then equation~\ref{eq:first_term4} becomes
\begin{equation}
\begin{array}{lcl}
\left( \frac { \partial { d }_{ i }^{ pred } }{ \partial c } *{ d }_{ ref }^{ obs } \right) \cdot { r }_{ i }&=& \displaystyle \int _{ -\infty  }^{ +\infty  }{ \left[ \int _{ -\infty  }^{ +\infty  }{ S_{ x }\left( t \right) U_{ { x,i } } } \left( \eta -t \right) dt \right]  } \left[ \int _{ -\infty  }^{ +\infty  }{ { d }_{ ref }^{ obs }\left( t-\eta  \right) r_{ i }\left( t \right)  } dt \right] d\eta \\
&=& \displaystyle \int _{ -\infty  }^{ +\infty  }{ S_{ x }\left( t \right)  } \left[ \int _{ -\infty  }^{ +\infty  }{ { U_{ { x,i } }\left( \eta -t \right) d }_{ ref }^{ obs }\left( t-\eta  \right) r_{ i }\left( t \right) d\eta  }  \right] dt
\end{array}
\label{eq:first_term5}
\end{equation}
Because of the reciprocity of $U _ { x , i }$, the inner integration $\int _{ -\infty  }^{ +\infty  }{ U_ { x,i }\left( \eta -t \right) } d _{ ref }^{ obs }\left( t-\eta  \right) r_{ i }\left( t \right) d\eta$ can be considered as a back propagation of the new data residual $R_i^{(1)}=\int _{-\infty} ^{+\infty}d_{ ref }^{ obs }\left( t-\eta  \right) r_{ i }\left( t \right)dt$. Thus, the first term of the gradient can be obtained by back-propagating $R_i^{(1)}$ as adjoint sources.
\newline
A similar derivation can be applied to the second term of Equation~\ref{eq:vel_grd_cnv-a}, $-\left( { d }_{ i }^{ obs }*\frac { \partial { d }_{ ref }^{ pred } }{ \partial c }  \right) { r }_{ i }$. This yields
\begin{equation}
-\left( { d }_{ i }^{ obs }*\frac { \partial { d }_{ ref }^{ pred } }{ \partial c }  \right) \cdot { r }_{ i }=\int _{ -\infty  }^{ +\infty  }{ S_{ x }\left( t \right)  } \left[ \int _{ -\infty  }^{ +\infty  }{ { U_{ { x,ref } }\left( \eta -t \right) d }_{ i }^{ obs }\left( t-\eta  \right) r_{ i }\left( t \right) d\eta  }  \right] dt.
\label{eq:second_term}
\end{equation}
In Equation~\ref{eq:second_term}, the second term of the gradient can be obtained by back-propagating the second residual $R_{ i }^{ (2) }=-\int _{ -\infty  }^{ +\infty  } d_{ i }^{ obs }\left( t-\eta  \right) r_{ i }\left( t \right) dt$.
\newline
Thus, the gradient involves calculating the adjoint wavefields with the adjoint sources:
\begin{equation}
\ R^{(1)}_i = d^{obs}_{ref} \bigotimes \left( { d }_{ i }^{ pred } * { d }_{ ref }^{ obs } - { d }_{ i }^{ obs } * { d }_{ ref }^{ pred } \right),
\label{eq:resi1}
\end{equation}
at the $i^{th}$ receiver position, and 
\begin{equation}
\ R^{(2)}_i = - d^{obs}_{i} \bigotimes \left( { d }_{ i }^{ pred } * { d }_{ ref }^{ obs } - { d }_{ i }^{ obs } * { d }_{ ref }^{ pred } \right),
\label{eq:resi2}
\end{equation}
at the reference receiver position.

\end{document}